\documentclass[fleqn,usenatbib]{mnras}

\usepackage{newtxtext,newtxmath}

\usepackage[T1]{fontenc}

\DeclareRobustCommand{\VAN}[3]{#2}
\let\VANthebibliography\thebibliography
\def\thebibliography{\DeclareRobustCommand{\VAN}[3]{##3}\VANthebibliography}

\usepackage{hyperref}
\usepackage{graphicx}	
\usepackage{amsmath}	
\usepackage{threeparttable}
\usepackage{subfig}
\usepackage{longtable}
\usepackage{color}
\usepackage{ulem}

\title[{High-energy tail in gamma-ray globular clusters}]
{{Evidence for a high-energy tail in the gamma-ray spectra of globular clusters}
}

\author[Song et al.] {
\parbox[t]{\textwidth}{
Deheng Song,$^{1}$\thanks{dhsong@vt.edu}
Oscar Macias,$^{1,2,3}$\thanks{o.a.maciasramirez@uva.nl}
Shunsaku Horiuchi,$^{1}$\thanks{horiuchi@vt.edu}
Roland M. Crocker$^{4}$ and David M. Nataf$^{5}$\\}
\\
$^{1}$Center for Neutrino Physics, Department of Physics, Virginia Tech, Blacksburg, VA 24061, USA\\
$^{2}$Kavli Institute for the Physics and Mathematics of the Universe (WPI),University of Tokyo, Kashiwa, Chiba 277-8583, Japan\\
$^{3}$GRAPPA Institute, Institute of Physics, University of Amsterdam, 1098 XH Amsterdam, The Netherlands\\
$^{4}$Research School of Astronomy and Astrophysics, Australian National University, Canberra, ACT 2611, Australia\\
$^{5}$Center for Astrophysical Sciences and Department of Physics and Astronomy, The Johns Hopkins University, Baltimore, MD 21218, USA\\
}

\date{Accepted XXX. Received YYY; in original form ZZZ}

\pubyear{2021}

\begin{document}
\label{firstpage}
\pagerange{\pageref{firstpage}--\pageref{lastpage}}

\maketitle

\begin{abstract}
Millisecond pulsars are very likely the main source of gamma-ray emission from globular clusters. However, the relative contributions of two separate emission processes--curvature radiation from millisecond pulsar magnetospheres vs. inverse Compton emission from relativistic pairs launched into the globular cluster environment by millisecond pulsars--have long been unclear. To address this, we search for evidence of inverse Compton emission in 8-year \textit{Fermi}-LAT data from the directions of 157 Milky Way globular clusters. We find a mildly statistically significant (3.8$\sigma$) correlation between the measured globular cluster gamma-ray luminosities and their photon field energy densities. However, this may also be explained by a hidden correlation between the photon field densities and the stellar encounter rates of globular clusters. Analysed {\it in toto},  we demonstrate that the gamma-ray emission of globular clusters can be resolved spectrally into two components:  i) an exponentially cut-off power law and ii) a pure power law. The latter component--which we uncover at a significance of 8.2$\sigma$--{has a power index of 2.79 $\pm$ 0.25. It} is most naturally interpreted as inverse Compton emission by cosmic-ray electrons and positrons injected by millisecond pulsars. We find the luminosity of this {power-law} component is comparable to, or slightly smaller than, the luminosity of the curved component, suggesting the fraction of millisecond pulsar spin-down luminosity into relativistic leptons is similar to the fraction of the spin-down luminosity into prompt magnetospheric radiation.
\end{abstract}




\section{Introduction}\label{sec:intro}

Over two dozen globular clusters (GCs) have been detected in $\gamma$ rays in {\it Fermi} Large Area Telescope (LAT) data~\citep{2009Sci...325..845A,2010A&A...524A..75A,2010ApJ...712L..36K,2011ApJ...729...90T,2015MNRAS.448.3215Z,2016MNRAS.459...99Z}. The millisecond pulsar (MSP) populations of those GCs are believed to be the main source of this $\gamma$-ray emission. In particular, MSPs have been firmly established as $\gamma$-ray sources~\citep{1996A&A...311L...9V,2009ApJ...699.1171A,2013MNRAS.430..571E,2013ApJS..208...17A} and a large fraction of them have been discovered in GCs~\citep{2005ASPC..328..147C}. Recently, \textit{Fermi}-LAT detected $\gamma$-ray pulsations in two GCs~\citep{2011Sci...334.1107F,2013ApJ...778..106J}, providing further support for this scenario.

The high-energy emission from MSPs emerges from the primary electrons accelerated by them and  subsequent radiation by secondary, relativistic electrons and positrons ($e^\pm$) pair created in their magnetospheres. In particular, \citet{2005ApJ...622..531H} studied the curvature radiation (CR) of primary electrons within MSP magnetospheres with a focus on GeV-scale emission. \citet{2007MNRAS.377..920B} then considered a scenario in which $e^\pm$, injected by MSPs, gradually diffuse through a GC, up-scattering ambient photons, and thus producing inverse Compton (IC) $\gamma$-ray emission in the GeV$-$TeV energy range.~\citet{2009ApJ...696L..52V} calculated the CR and IC spectra for an ensemble of MSPs in the GCs 47 Tucanae and Terzan 5. \citet{2010ApJ...723.1219C} found that the spectra of 47 Tucanae and seven other GCs can be explained by IC alone, invoking background photons from the cosmic microwave background (CMB) or Galactic infrared/stellar radiation. {For a review of the observations and models about the $\gamma$-ray emission from globular clusters, see \citet{2016JASS...33....1T}.} In general, the GeV emission mechanism of MSPs remains in contention with CR, IC, and synchrotron radiation all  proposed~\citep{2021arXiv210105751H}.

Here, motivated by the increasing number of GCs detected in $\gamma$ rays, we perform a collective statistical study of their properties in order to gain insight into the nature of their high-energy emission. Our particular aim is to investigate the importance of the contribution of IC emission to the overall $\gamma$-ray emission of GCs.

Relations between the detected GC gamma-ray luminosities and properties of GC can be used to probe the origins of $\gamma$-ray emission and their underlying sources. For example, correlations with the photon field energy density of GCs could unveil the potential contribution from IC, and correlations with the stellar encounter rate and metallicity could provide insight into the dynamical formation of MPSs in GCs. Previous work here includes a study by \citet{2010A&A...524A..75A} that reported a correlation between the $\gamma$-ray luminosity $L_\gamma$ and the stellar encounter rate of eight GCs. \citet{2011ApJ...726..100H} studied a group of 15 $\gamma$-ray emitting GCs with 2 years of Fermi data and found a positive correlation between $L_\gamma$ and, respectively, encounter rate, metallicity, and Galactic photon energy density. \citet{2016JCAP...08..018H} studied 25 GCs using 85 months of Fermi data, and found that the $\gamma$-ray luminosity function of MSPs in GCs is consistent with that applying to MSPs detected in the field. \citet{2018MNRAS.480.4782L} studied $\gamma$-ray emission from high-latitude GCs and its connection to their X-ray emission. \citet{2019MNRAS.486..851D} reanalysed 9 years of Fermi data and found 23 $\gamma$-ray emitting GCs; they found that the metallicity only mildly contributes to $L_\gamma$ while a very high encounter rate seemed to {\it reduce} the $L_\gamma$ from GCs.

In parallel, modeling of GCs' observed broadband spectral energy distributions provides a handle on their CR and IC emissions. Recently,~\citet{2013ApJ...779..126K} and~\citet{2019ApJ...880...53N} modelled the multiwavelength emission from MSPs considering a potential CR origin for GeV and IC emissions for TeV $\gamma$ rays, as well as synchrotron radiation for the radio and X-ray wavebands. These models are successful in explaining the multiwavelength spectra of Terzan 5. However, Terzan 5 is the only GC (perhaps) detected above TeV energies~\citep{2011A&A...531L..18H}. Detailed spectral modelling similar to that presented by~\citet{2013ApJ...779..126K} and~\citet{2019ApJ...880...53N} is difficult for other GCs at present due to a lack of TeV $\gamma$-ray data. Although \textit{Fermi}-LAT is sensitive to $\gamma$-rays of up to $\approx 1$ TeV, the photon count statistics at the highest energies are very low.

In the recently published \textit{Fermi}-LAT fourth source catalog~\citep{2020ApJS..247...33A} (4FGL)\footnote{The Fermi collaboration has recently released an incremental update of the fourth source catalog~\citep{2020arXiv200511208B} (4FGL-DR2, for Data Release 2). The new catalog uses 10 years of data, a 25\% increase with respect to the 4FGL. However, only 1 new GC (NGC 362) has been detected. Given this marginal change, we retain use of the 4FGL catalog constructed with the 8-year data set.}, 30 GCs have been detected in GeV $\gamma$ rays. With such a number, we can begin to carefully study the nature of the $\gamma$-ray emission from GCs through a population study. In this paper, we repeat the bin-by-bin analysis of the 4FGL data for the 157 known Milky Way GCs in the \citet{1996AJ....112.1487H} catalog\footnote{2010  edition: \url{https://www.physics.mcmaster.ca/~harris/mwgc.dat}}. We search for correlations between the $\gamma$-ray luminosity of the GCs and other parameters of the GCs to probe which are good proxies for the $\gamma$-ray luminosity and study potential IC contributions; to this end, we consider the photon field energy densities, the stellar encounter rate, and the metallicity of the GCs. Unlike previous studies of correlations of the GC $\gamma$-ray emissions, we consider also the upper limits placed by null detections, which we implement via a Kendall $\tau$ coefficient test statistic. Furthermore, we also look for evidence of IC from the spectra of GCs. For the first time, we implement a universal two-component model to study the spectra of $\gamma$-ray-detected GCs. The two-component model comprises a CR component, which is spectrally curved, plus an IC component modeled as a power law in the energy range of interest.

The remainder of the paper is as follows: In Section~\ref{sec:gamma_GCs} we discuss the choice of GC samples and the data analysis procedure. Section~\ref{sec:correlation} presents the methodology and results of our correlation analysis. Section~\ref{sec:spectra} describes the spectral analysis method and reports the $e^\pm$ injection efficiency in the GCs. We discuss the implications of our results in Section~\ref{sec:discussion} and conclude in Section~\ref{sec:conclusion}.

\section{$\gamma$-ray emission from globular clusters}\label{sec:gamma_GCs}

In this section, we describe our choice of GC sample and the GCs' $\gamma$-ray-related parameters. The Fermi data analysis process is reported as well. For GCs with a $\gamma$-ray counterpart in the 4FGL, we update their spectral parameters through a maximum likelihood procedure. For those not detected in the 4FGL, we estimate their 95\% C.L. $\gamma$-ray upper limits.

\subsection{Globular cluster sample}\label{sec:samples}

\begin{figure}
    \centering
    \includegraphics[width=\columnwidth]{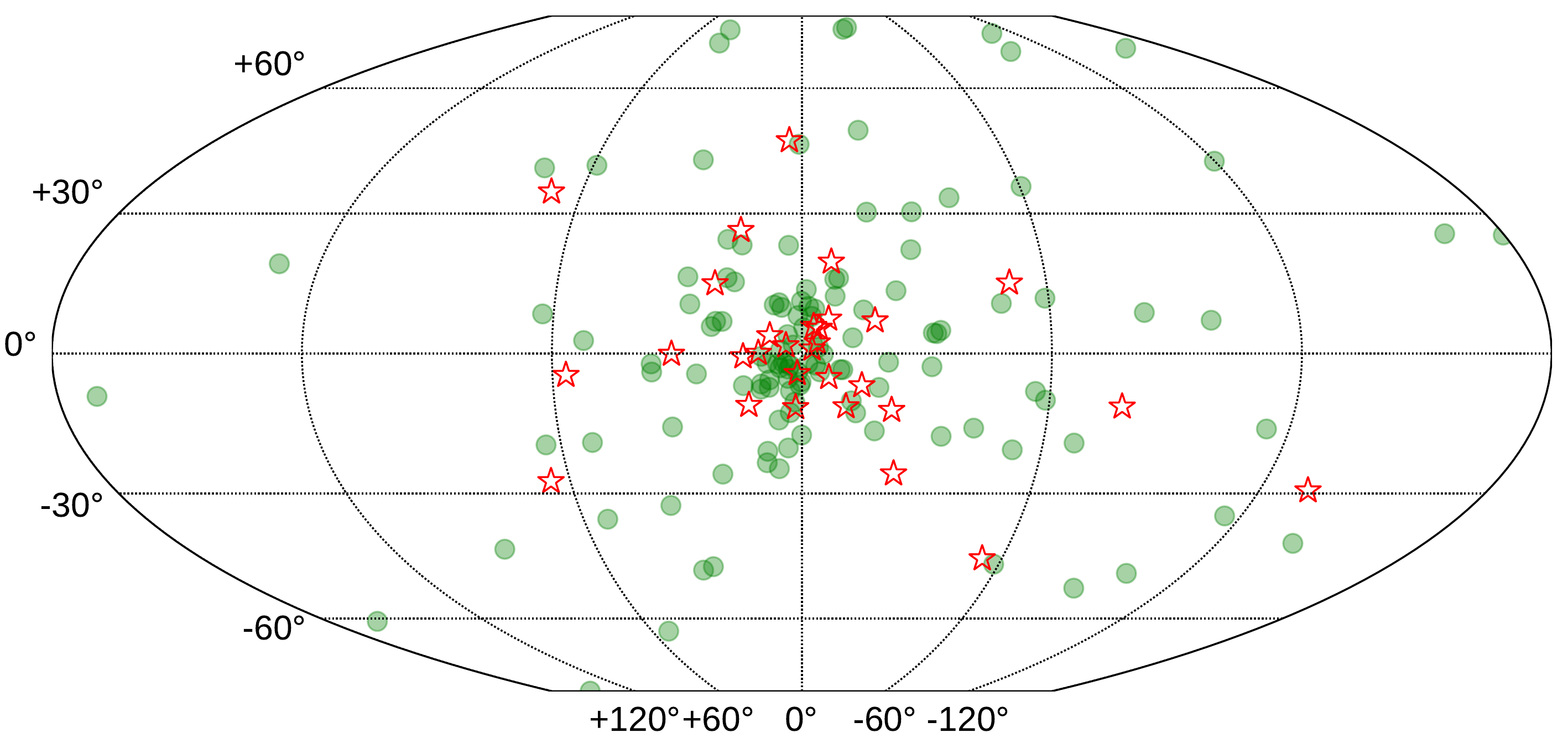} \\
    \includegraphics[width=\columnwidth]{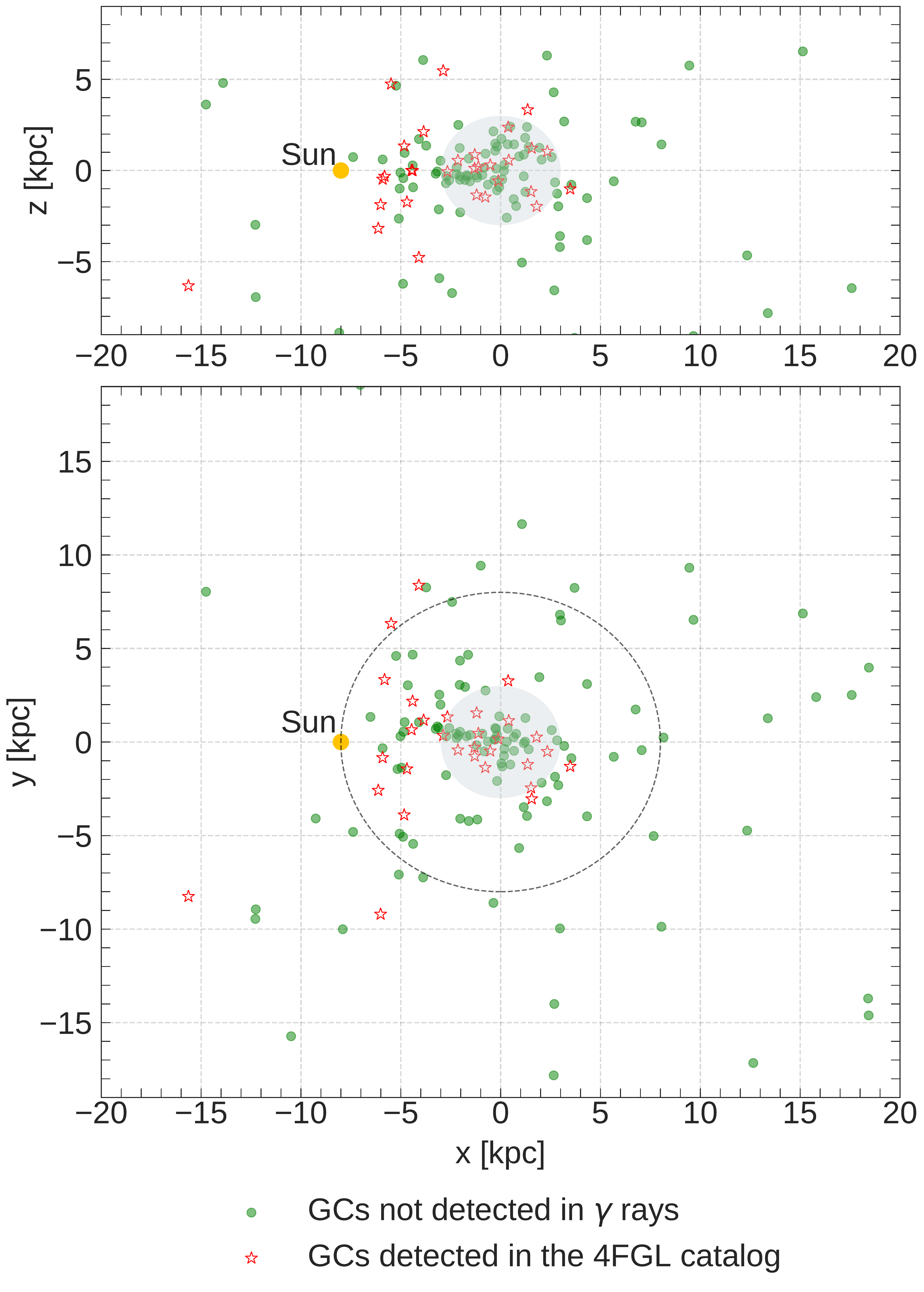}
    \caption{Spatial distribution of the 157 Milky Way GCs in the \citet{1996AJ....112.1487H} catalog. The top panel shows  the all sky spatial distribution in Galactic coordinate. The middle and bottom panel display the three-dimensional (3D) Cartesian 
    coordinates of the GCs, in which 
    the Sun (yellow circle) is located in the negative x-axis. The $\gamma$-ray-detected GCs in the 4FGL catalog are shown as red stars, 
    and the GCs not detected in $\gamma$ rays are shown as green circles. Most $\gamma$-ray-detected GCs are located near the Sun or in the Galactic bulge (grey shaded area in the middle and bottom panel).}
    \label{fig:all_sky_distribution}
\end{figure}

We consider the \citet{1996AJ....112.1487H} catalog (2010 edition), which contains identifications and basic parameters for 157 GCs in the Milky Way. Here, we reanalyse publicly available {\it Fermi}-LAT data from the direction of all GCs in the \citet{1996AJ....112.1487H} catalog. Figure~\ref{fig:all_sky_distribution} shows the spatial distribution of the GCs. The top panel shows the projected direction of the GCs on the celestial plane while the bottom two panels display their 3D coordinates. The GCs which are detected in the 4FGL are marked by red stars while null detections are indicated by green circles. Most $\gamma$-ray GCs are near the Sun (yellow circle) or located in the Galactic bulge (assumed to be sphere of 3 kpc radius, grey circular area).

The origin of the $\gamma$-ray emission from GCs can be studied by comparing its dependency on GC properties. IC emission is sensitive to the ambient photon field on which the $e^\pm$ scatter. \citet{2011ApJ...726..100H} reported a positive correlation between the $\gamma$-ray luminosity $L_\gamma$ and the photon field energy density at the cluster location, indicating an IC contribution. In the present work, we improve upon the Galactic radiation field model used by \citet{2011ApJ...726..100H} by extracting the energy density of the interstellar radiation at the locations of the GCs from the three-dimensional interstellar radiation model in \texttt{GALPROP v56}\footnote{\url{http://galprop.stanford.edu/}}~\citep{2017ApJ...846...67P,2018ApJ...856...45J}. This is a fully 3D model that combines the CMB, infrared, and optical photons of the Milky Way, denoted as $u_\text{MW}$. In addition, photons from stars in the GCs are expected to make a dominant contribution to the total, ambient radiation field. We estimate this component by
\begin{equation}
    u_{\text{GC}} = \dfrac{L_*}{4\pi c r_h^2},
    \label{eq:GCRF}
\end{equation}
where $L_*$ and $r_h$ are the stellar luminosity and the half-light radius of the GC. The total photon field energy density is $u_\text{Total} = u_\text{MW} + u_\text{GC}$.

A potential correlation between the $\gamma$-ray luminosity $L_\gamma$ and the stellar encounter rate has been studied as a way to probe the dynamic formation of MSPs in GCs~\citep{2010A&A...524A..75A,2011ApJ...726..100H,2019MNRAS.486..851D}. In the present work, we adopt the stellar encounter rate estimated by \citet{2013ApJ...766..136B}, which is defined as
\begin{equation}
    \Gamma_c = \frac{4\pi}{\sigma_c}\int\rho(r)^2 r^2dr,
	\label{eq:encounter}
\end{equation}
where $\sigma_c$ is the velocity dispersion at the core radius, $\rho(r)$ is the stellar density profile of the cluster, and the line-of-sight integration is performed along the half-light radius. 

Additionally, it has been argued~\citep{2011ApJ...726..100H,2019MNRAS.486..851D} that high metallicity [Fe/H] in the GCs could enhance the dynamical formation of MSP. The outer convective zone of metal-rich stars enables magnetic braking, which assists orbital shrinking during binary formation. In this analysis, we use the GCs metallicities  reported in the~\citet{1996AJ....112.1487H} catalog, which summarizes  spectroscopic or photometric estimates in the literature.

In summary, we consider the empirical dependence of the inferred $\gamma$-ray luminosity of GCs on four parameters, namely, $u_\text{MW}$, $u_\text{Total}$, $\Gamma_c$, and [Fe/H]. We summarize the values of these parameters for the 30 $\gamma$-ray-detected GCs in Table~\ref{tab:pars}. Also included are the stellar masses M$_*$ of the GCs adopted from~\citet{2017MNRAS.464.2174B},~\cite{2017MNRAS.471.3668S}, and~\citet{2018MNRAS.478.1520B}. They are estimated from N-body modelling of the velocity dispersion and surface density profiles. In Section~\ref{sec:correlation}, we study the correlations between the $\gamma$-ray emission of GCs and these parameters.

\begin{table*}
\centering
\caption{Parameters and data analysis results for 30 $\gamma$-ray-detected GCs.}\label{tab:pars}
\begin{threeparttable}
\begin{tabular}{lccccccccr}
\hline
Name &  $\Gamma_c$\tnote{a} & [Fe/H]\tnote{b}  & M$_*$\tnote{c} & $u_\text{MW}$\tnote{d} & $u_\text{Total}$\tnote{e} & $R_\odot$\tnote{f} & Flux\tnote{g} & $L_\gamma$\tnote{g} & TS\tnote{h}\\
 & &  & ($10^5 M_\odot$) & (eV cm$^{-3}$) & (eV cm$^{-3}$) & (kpc) & (10$^{-8}$ ph cm$^{-2}$ s$^{-1}$)  & (10$^{34}$ erg s$^{-1}$)\\
\hline
2MS-GC01 & ... & ... & ... & 1.79 & 7.14 & 3.60 & $1.96 \pm 0.26$ & $3.88 \pm 0.81$ & 153.82\\
GLIMPSE01 & ... & ... & ... & 1.55 & 30.23 & 4.20 & $2.55 \pm 0.28$ & $8.79 \pm 0.94$ & 535.61\\
GLIMPSE02 & ... & -0.33 & ... & 2.61 & >2.61 & 5.50 & $2.73 \pm 0.25$ & $11.55 \pm 1.57$ & 318.41\\
M 62 & 2470.00 & -1.18 & 6.76 & 2.14 & 293.14 & 6.80 & $0.98 \pm 0.09$ & $9.16 \pm 0.89$ & 1012.19\\
M 80 & 937.00 & -1.75 & 2.82 & 0.92 & 276.86 & 10.00 & $0.17 \pm 0.07$ & $4.26 \pm 1.39$ & 94.83\\
NGC 104 & 1000.00 & -0.72 & 8.13 & 0.55 & 31.12 & 4.50 & $1.34 \pm 0.07$ & $5.61 \pm 0.34$ & 4853.63\\
NGC 1904 & 126.00 & -1.60 & 1.66 & 0.29 & 173.13 & 12.90 & $0.11 \pm 0.04$ & $2.32 \pm 0.98$ & 23.84\\
NGC 2808 & 1210.00 & -1.14 & 8.13 & 0.38 & 467.35 & 9.60 & $0.19 \pm 0.06$ & $3.43 \pm 1.03$ & 90.30\\
NGC 5904 & 120.00 & -1.29 & 3.63 & 0.55 & 56.46 & 7.50 & $0.11 \pm 0.04$ & $1.10 \pm 0.47$ & 39.07\\
NGC 6139 & 407.00 & -1.65 & 3.47 & 1.15 & 161.34 & 10.10 & $0.29 \pm 0.09$ & $5.82 \pm 2.19$ & 59.29\\
NGC 6218 & 18.10 & -1.37 & 0.83 & 0.90 & 14.94 & 4.80 & $0.07 \pm 0.05$ & $0.38 \pm 0.20$ & 33.92\\
NGC 6304 & 150.00 & -0.45 & 1.62 & 2.33 & 23.95 & 5.90 & $0.10 \pm 0.03$ & $1.09 \pm 0.42$ & 21.71\\
NGC 6316 & 131.00 & -0.45 & 3.63 & 1.88 & 270.82 & 10.40 & $0.48 \pm 0.11$ & $10.91 \pm 2.13$ & 207.99\\
NGC 6341 & 265.00 & -2.31 & 3.09 & 0.42 & 97.31 & 8.30 & $0.05 \pm 0.04$ & $0.62 \pm 0.37$ & 15.84\\
NGC 6388 & 1770.00 & -0.55 & 10.47 & 1.30 & 1127.11 & 9.90 & $0.98 \pm 0.09$ & $18.41 \pm 1.63$ & 970.86\\
NGC 6397 & 146.00 & -2.02 & 0.89 & 0.92 & 3.75 & 2.30 & $0.11 \pm 0.06$ & $0.09 \pm 0.05$ & 17.21\\
NGC 6402 & 106.00 & -1.28 & 7.41 & 0.87 & 136.26 & 9.30 & $0.19 \pm 0.08$ & $3.13 \pm 1.18$ & 51.16\\
NGC 6440 & 1750.00 & -0.36 & 5.01 & 2.50 & 721.93 & 8.50 & $0.76 \pm 0.13$ & $10.34 \pm 1.97$ & 259.55\\
NGC 6441 & 3150.00 & -0.46 & 11.75 & 1.36 & 1148.78 & 11.60 & $0.74 \pm 0.10$ & $18.08 \pm 2.62$ & 363.50\\
NGC 6528 & 233.00 & -0.11 & 0.59 & 4.00 & 158.13 & 7.90 & $0.11 \pm 0.03$ & $2.06 \pm 0.75$ & 31.27\\
NGC 6541 & 567.00 & -1.81 & 2.51 & 1.48 & 120.84 & 7.50 & $0.21 \pm 0.06$ & $2.10 \pm 0.63$ & 77.12\\
NGC 6652 & 805.00 & -0.81 & 0.47 & 1.22 & 106.17 & 10.00 & $0.22 \pm 0.06$ & $4.42 \pm 1.03$ & 120.53\\
NGC 6717 & 46.10 & -1.26 & 0.36 & 1.56 & 22.38 & 7.10 & $0.21 \pm 0.07$ & $2.04 \pm 0.63$ & 70.85\\
NGC 6752 & 374.00 & -1.54 & 2.29 & 0.83 & 18.59 & 4.00 & $0.21 \pm 0.05$ & $0.58 \pm 0.12$ & 157.19\\
NGC 6838 & 2.05 & -0.78 & 0.54 & 0.85 & 4.14 & 4.00 & $0.20 \pm 0.08$ & $0.50 \pm 0.18$ & 40.13\\
NGC 7078 & 6460.00 & -2.37 & 4.90 & 0.39 & 248.98 & 10.40 & $0.17 \pm 0.05$ & $2.55 \pm 0.78$ & 46.55\\
Omega Cen & 144.00 & -1.53 & 33.11 & 0.71 & 27.35 & 5.20 & $0.59 \pm 0.07$ & $3.46 \pm 0.42$ & 747.94\\
Terzan 1 & 0.63 & -1.03 & 2.95 & 4.06 & 4.27 & 6.70 & $0.10 \pm 0.02$ & $2.93 \pm 0.73$ & 62.48\\
Terzan 2 & 19.60 & -0.69 & 0.33 & 4.23 & 9.33 & 7.50 & $0.15 \pm 0.05$ & $3.00 \pm 1.00$ & 42.65\\
Terzan 5 & 1400.00 & -0.23 & 6.17 & 4.80 & 98.73 & 6.90 & $3.93 \pm 0.20$ & $38.65 \pm 2.51$ & 3740.32\\
\hline
\end{tabular}
\begin{tablenotes}
\item [a] Stellar encounter rate computed using equation~(\ref{eq:encounter}).
The numerical values are normalized by the encounter rate of NGC 104, which is set to 1000.
\item [b] Metallicity.
\item [c] Stellar mass.
\item [d] Galactic photon field energy density.
\item [e] Total photon field energy density, defined as the sum of the Galactic photon field and the photons from stars in the GC.
\item [f] Distance from the Sun.
\item [g] $\gamma$-ray flux and luminosity between 300 MeV to 500 GeV.
\item [h] Test statistic.
\end{tablenotes}
\end{threeparttable}
\end{table*}

\subsection{Data analysis}\label{sec:data}

We use 8 years of \textit{Fermi}-LAT data, from 2008 August 4 to 2016 August 2. This constitutes the same data as the 4FGL. The newest Pass 8 data release is applied. As recommended by the \textit{Fermi}-LAT data analysis documentation\footnote{\url{http://fermi.gsfc.nasa.gov/ssc/data/analysis/}}, the event class for the analysis is "P8 Source" class (evclass=128) and the event type is "FRONT+BACK" (evtype=3). We use a 90$^{\circ}$ zenith angle cut to remove Earth limb events and filter the data by (DATA\_QUAL>0)\&\&(LAT\_CONFIG==1). The corresponding instrument response function is $\texttt{P8R3\_SOURCE\_V2}$. For our analysis, the \textit{Fermipy} software version \textit{0.18.0} is used, together with the \textit{Fermi Science Tools} version \textit{1.2.21}.

For 30 GCs detected by the 4FGL, we simply reanalyse the 4FGL $\gamma$-ray source. We use a 10$^{\circ}$ by 10$^{\circ}$ Region-of-Interest around the source with a 0.1$^{\circ}$ by 0.1$^{\circ}$ bin size. Photons from 300 MeV to 500 GeV are analysed using 9 logarithmic bins. Given that we use a different Region-of-Interest size and photon class compared to those adopted in the construction of the 4FGL, additional point sources might emerge in our Region-of-Interest. However, since we use the same observation time as in the 4FGL, the impact of those potential new sources is expected to be minimal. Therefore, we only include known 4FGL sources in our analysis. As recommended by the Fermi team, we re-run a maximum likelihood analysis that starts from the best-fit parameter values found in the 4FGL and updating accordingly. The most recent Galactic interstellar emission model \texttt{gll\_iem\_v07} and the isotropic component \texttt{iso\_P8R3\_SOURCE\_V2\_v1} are employed as fore/backgrounds with free-floating normalization. We have followed the \textit{Fermipy} recommended procedure\footnote{\url{http://fermipy.readthedocs.io/en/latest/quickstart.html}} and fixed the spectral parameters of the sources with TS < 10 and 10 < Npred < 100 to their 4FGL values. However, the spectral parameters of the 4FGL sources lying within 5$^{\circ}$ of the Region-of-Interest center are allowed to float freely. The \texttt{MINUIT} algorithm is used to determine the best-fit parameters of the sources for each energy bin independently.

For the 127 additional GCs in the \citet{1996AJ....112.1487H} catalog without 4FGL detections, we estimate the 95\% C.L. $\gamma$-ray upper limits from their locations. More specifically, we place a point source at the coordinates of those GCs. The point source is assumed to have a power-law spectrum $dN/dE\sim E^{-\Gamma}$ with fixed index $\Gamma = 2$. We applied the same pipeline used on the set of detected GCs and obtained the 95\% C.L. flux upper limits on the putative point sources placed at the GCs locations.

Table~\ref{tab:pars} summarizes the Fermi data analysis results. We report the photon flux and luminosity $L_\gamma$ for 30 $\gamma$-emitting GCs. For each GC, the total photon flux is summed over the bin-by-bin fluxes from the Fermi analysis. The statistical error of the total flux is added quadratically from the bin-by-bin flux errors. The energy flux is estimated similarly, then $L_\gamma = 4\pi R_\odot^2 \times (\mathrm{energy\ flux})$. We ignore the uncertainties on $R_\odot$ for the GCs since they are either unavailable in the \citet{1996AJ....112.1487H} catalog or estimated only at percentage level~\citep{2017MNRAS.464.2174B} and so make a negligible contribution to the overall error of $L_\gamma$. For the parameters and flux upper limits of 127 additional GCs, see Appendix~\ref{appx:nodetect}.

\section{Correlation analysis}\label{sec:correlation}

In this section, we investigate the correlation between $L_\gamma$'s and other GC observables. However, GCs not yet detected in $\gamma$ rays and sample selection effects must be taken into account to properly determine the significance of any apparent correlations. We use the Kendall $\tau$ coefficient as the test statistic for estimating the significance of the correlations, and the expectation-maximization (EM) algorithm for the linear regression of the correlations. Both methods allow us to properly incorporate the luminosity upper limits$-$implied by GCs not detected in the 4FGL$-$into our statistical analysis.

\subsection{Linear regression with the expectation–maximization algorithm}\label{sec:EM}

To study the correlations between the $L_\gamma$'s and the other GC observables, we assume a linear relation in logarithmic space of the form:
\begin{equation}
    \log(L_\gamma) = a\log(X)+b,
\end{equation}
where $L_\gamma$ is the gamma-ray luminosity of the GC, $X$ is the independent observable considered, and $a$ and $b$ are parameters to be determined. 

We use an EM algorithm~\citep{1985ApJ...293..192F,1986ApJ...306..490I,1992BAAS...24..839L} to find the maximum likelihood estimates of the parameters $a$ and $b$. In contrast to the standard maximum likelihood method, the EM algorithm is designed to be used with censored data, i.e., data consisting of both measurements and limits. Upper limits must be properly incorporated in the correlation analyses so as to obtain statistically robust results. Briefly, the implementation of the EM algorithm is done as follows: first, the expected values of the censored data are estimated based on the regression parameters and the variance of the uncensored data. Second, a least-squares fit is performed and the variance is updated. Lastly, the procedure is repeated until convergence is achieved on $a$, $b$, and the variance. Using the EM algorithm, we are able to utilize the complete data set (including upper limits) in estimating relations between the $L_\gamma$ and the other observables.

\subsection{Kendall $\tau$ coefficient and significance}\label{sec:kendall}

While the EM algorithm allows us to estimate the linear relations between the $L_\gamma$'s and other GC observables, we are also interested in determining the statistical significance of those relations. To that end, we apply the generalized Kendall $\tau$ rank correlation test and perform Monte Carlo (MC) simulations to determine the significance of each correlation studied with the EM algorithm. 

The Kendall $\tau$ rank correlation coefficient (also referred to as the Kendall $\tau$ coefficient) is a non-parametric statistical test that has been used to study multi-wavelength correlations of star-forming galaxies~\citep{2012ApJ...755..164A,2020ApJ...894...88A}, and misaligned active galactic nuclei~\citep{2014ApJ...780..161D}. It has been generalized to include upper limits in the statistical procedure~\citep{2012ApJ...755..164A}. Therefore, we can calculate the Kendall $\tau$ coefficient using all available information concerning GCs (measurements and upper limits).

To estimate the significance of the correlations, we adopt a similar procedure as advanced previously in the literature~\citep{2012ApJ...755..164A}. Namely, the null hypothesis assumes no correlation between $L_\gamma$ and $X$. A set of null hypothesis samples is generated by repeating the following steps: (1) randomly exchange $L_\gamma$ of two GCs while preserving their locations; (2) if the energy fluxes of the GCs after exchanging the $L_\gamma$ are above the detection threshold of \textit{Fermi}-LAT, the exchange is kept\footnote{This step guarantees the detectability of the null hypothesis samples. It is crucial to apply realistic estimates of the detection threshold so that the null hypothesis samples are valid. \citet{2012ApJ...755..164A} and \citet{2020ApJ...894...88A} have used the minimum fluxes in their data. Since we are using the same amount of data as the 4FGL, we take advantage of the spatial map of the 8-year LAT detection threshold published with the 4FGL. We expect this to be a more rigorous way of generating the samples since the map includes the spatial dependence of the LAT threshold.}; and (3) we perform a large number of exchanges, until obtaining a nearly uniform $L_\gamma$ sample (including corrections from applying the detection threshold) over $X$, as required by the null hypothesis. In Appendix~\ref{appx:MC}, we discuss the number of exchanges needed to generate the null hypothesis sample.

\begin{figure}
    \centering
    \includegraphics[width=\columnwidth]{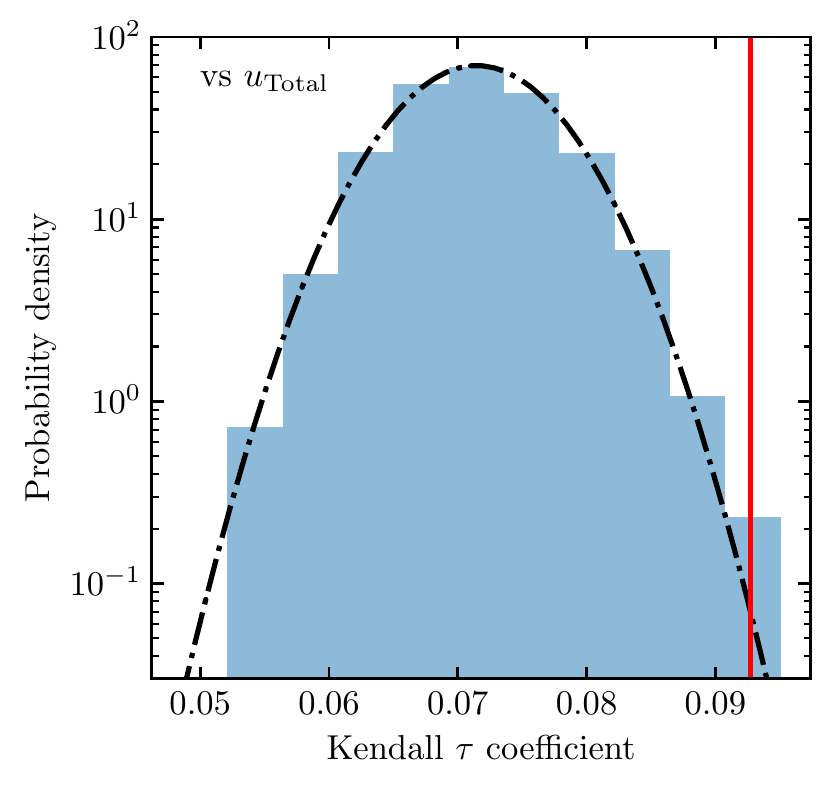}
    \caption{ Probability density distribution of the Kendall $\tau$ coefficients for the $L_\gamma$--$u_\mathrm{Total}$ data set. The blue histogram 
    corresponds to the density of the Kendall $\tau$ coefficients of the null hypothesis samples. The dash-dotted line shows the best-fit normally-distributed probability density function for the null hypothesis. The red vertical line indicates the Kendall $\tau$ coefficient for the 
    actual data.}
    \label{fig:hist}
\end{figure}

For each correlation, we generate 10$^4$ null hypothesis samples and calculate their Kendall $\tau$ coefficients. For a large number of samples, the coefficients can be fitted to a normal distribution~\citep{10.2307/2669997},
\begin{equation}
    \{\hat{\tau}_i\} \sim N(\nu_0,\sigma_0),
\end{equation}
where $\{\hat{\tau}_i\}$ represents the distribution of the $\tau$ coefficients from the null hypothesis sample, and $N(\nu_0,\sigma_0)$ is a normal distribution with mean $\nu_0$ and standard deviation $\sigma_0$. For each correlation, we can compare the observed value of $\tau$ with the corresponding normal distribution from the MC results and compute the significance,
\begin{equation}
    \sigma = \frac{\tau - \nu_0}{\sigma_0}.
\end{equation}
Figure~\ref{fig:hist} shows an example of the $L_\gamma$--$u_\mathrm{Total}$ data set. The blue histogram shows the probability density of Kendall $\tau$ coefficients of the null hypothesis samples. The dash-dotted line is the best fit normal distribution of the probability density, which has $\nu_0 = 0.071$ and $\sigma_0 = 0.0057$. The Kendall $\tau$ coefficient of real data is 0.093, shown by the red vertical line. The real data is about 3.8$\sigma$ away from the center of the null hypothesis distribution.

\begin{table}
\centering
\caption{Summary of correlations between $L_\gamma$ and four astrophysical parameters of the GCs. The best-fit parameters $a, $b, and the corresponding variance of $L_\gamma$ are found using the EM algorithm. The significance of the correlations is found by MC simulations with Kendall $\tau$ coefficients.}
\label{tab:correlation}
\begin{tabular}{lcccr}
\hline
 Correlation & $a$ & $b$ & $\sqrt{\text{Variance}}$ & Significance\\
\hline
vs $\Gamma_c$ & 0.39 $\pm$ 0.10 & 32.99 $\pm$ 0.26 & 0.59 $\pm$ 0.08 & 6.4$\sigma$\\
vs $u_{\mathrm{Total}}$ & 0.59 $\pm$ 0.09 & 32.97 $\pm$ 0.19 & 0.47 $\pm$ 0.06 & 3.8$\sigma$\\
vs [Fe/H] & 0.35 $\pm$ 0.14 & 34.18 $\pm$ 0.19 & 0.64 $\pm$ 0.08 & 1.8$\sigma$\\
vs $u_\mathrm{MW}$ & 0.29 $\pm$ 0.26 & 33.75 $\pm$ 0.12 & 0.68 $\pm$ 0.09 & 1.5$\sigma$\\
\hline
\end{tabular}
\end{table}

\subsection{Correlation results}

\begin{figure}
    \centering
    \includegraphics[width=\columnwidth]{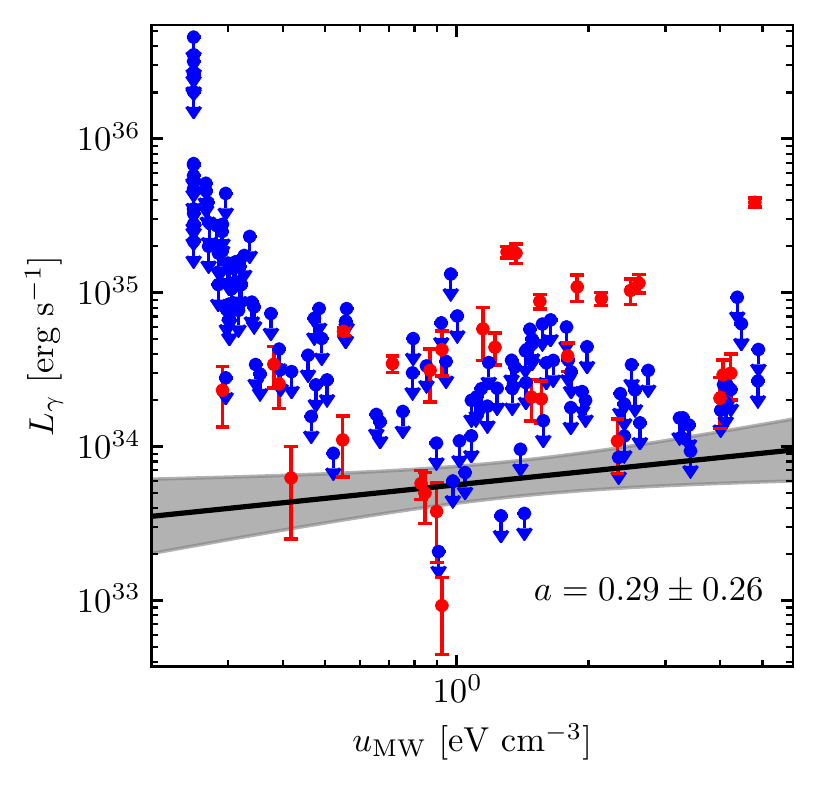}
    \includegraphics[width=\columnwidth]{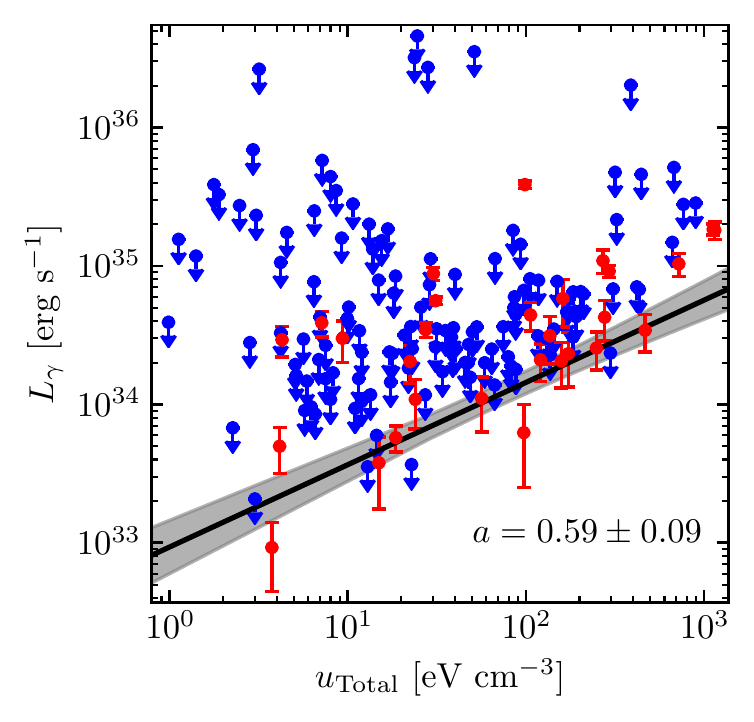}
    \caption{\label{fig:correlation_urad} Correlations between $L_\gamma$ and the photon field energy densities. The top panel shows the $L_\gamma$- $u_\mathrm{MW}$ correlation and the bottom panel shows the $L_\gamma$- $u_\mathrm{Total}$ correlation. GCs with measured $\gamma$ rays are shown in red, while GCs with upper limits are shown in blue. The best-fit correlations (black solid lines) are calculated using the EM algorithm discussed in Section~\ref{sec:EM}, with 1$\sigma$ uncertainties included as the gray shaded bands. We find a shallow correlation between $L_\gamma$ and $u_\mathrm{MW}$ with $a = 0.29 \pm 0.26$. The correlation between $L_\gamma$ and $u_\mathrm{Total}$ is more significant, with $a = 0.59 \pm 0.09$. Numerical values of correlations are summarized in Table~\ref{tab:correlation}, along with their significance.
    }
\end{figure}

The top (bottom) panel of Figure~\ref{fig:correlation_urad} shows the correlations between $L_\gamma$ and $u_\mathrm{MW}$ ($u_\mathrm{Total}$). GCs with measured $\gamma$-ray luminosity are shown in red, while GCs with upper limits are shown in blue. We find a very small slope for the $L_\gamma$-$u_\mathrm{MW}$ correlation, with $a=0.29 \pm 0.26$, which is almost consistent with 0 considering the large statistical error. The significance of the $L_\gamma$-$u_\mathrm{MW}$ correlation is found to be 1.5$\sigma$. When the total photon field is considered, we find a $L_\gamma$--$u_\mathrm{Total}$ correlation with $a = 0.59 \pm 0.09$. In this case, the significance increases to 3.8$\sigma$. The $L_\gamma$--$u_\mathrm{Total}$ correlation is mostly driven by $u_\mathrm{GC}$, the photon field from the starlight in the GCs (see equation~(\ref{eq:GCRF})). As shown by~Table~\ref{tab:pars}, $u_\mathrm{Total}$ is much greater than $u_\mathrm{MW}$ due to the dominant contribution from $u_\mathrm{GC}$. 

\begin{figure}
    \centering
    \includegraphics[width=\columnwidth]{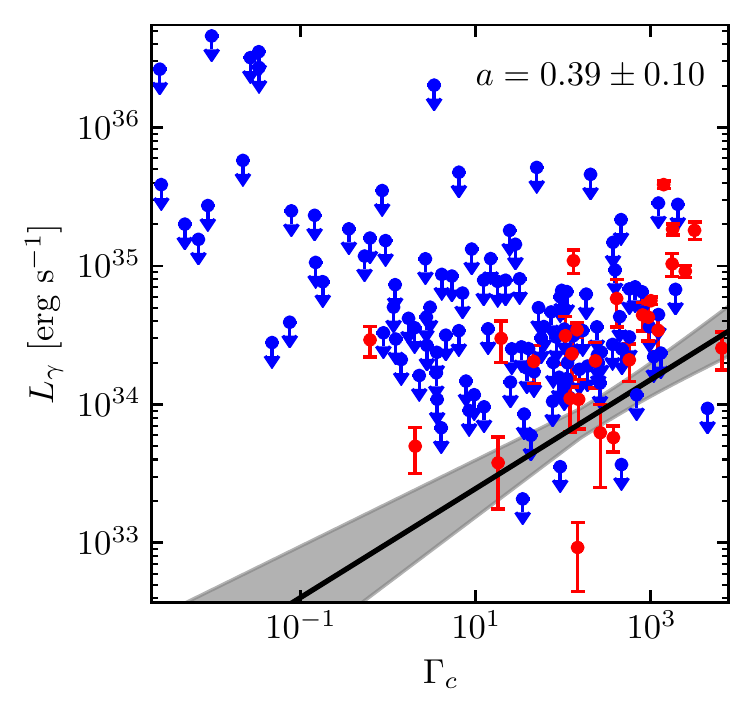}
    \includegraphics[width=\columnwidth]{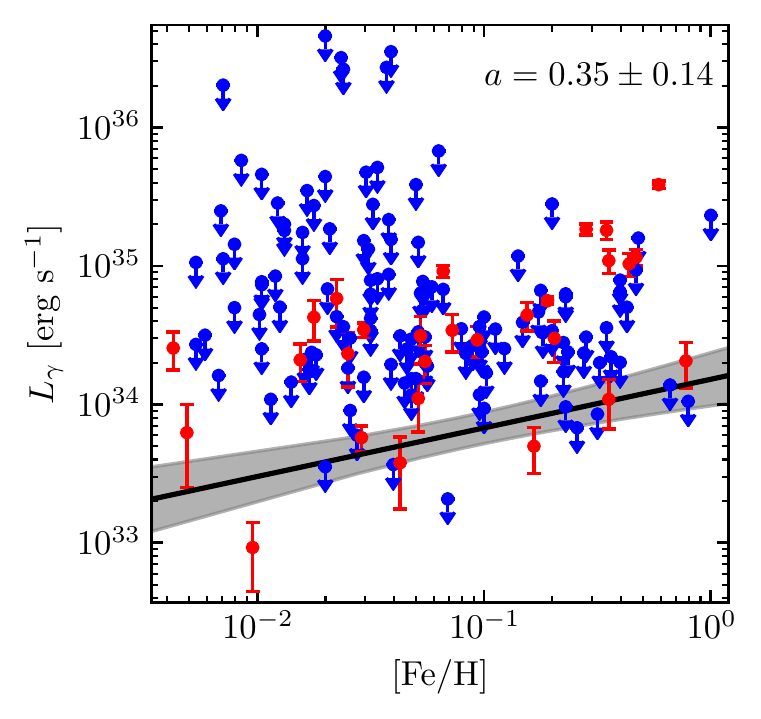}
    \caption{\label{fig:correlation_other} Same as Figure~\ref{fig:correlation_urad}, but correlated with the encounter rate (top panel) and metallicity (bottom panel).}
\end{figure}

We also investigate the correlation of the $L_\gamma$'s with the stellar encounter rate ($\Gamma_c$) and GC metallicities ([Fe/H]). These observables are argued to berelated to the formation of MSPs and may provide a proxy for the total number of MSPs in GCs. Figure~\ref{fig:correlation_other} shows the $L_\gamma$--$\Gamma_c$ correlation (top panel) and the $L_\gamma$--[Fe/H] correlation (bottom pannel) obtained with the EM algorithm. We find a positive correlation between the $L_\gamma$ and $\Gamma_c$, with $a = 0.39 \pm 0.10$, for which the Kendall $\tau$ test yields a 6.4$\sigma$ statistical significance. Similarly, we find a correlation between $L_\gamma$ and [Fe/H] with the best-fit value $a = 0.35 \pm 0.14$. However, the statistical significance of the correlation is only 1.8$\sigma$. 

We summarize the best-fit correlation results and their respective statistical significance in Table~\ref{tab:correlation}.

\subsection{Hidden correlation and interpretations}\label{sec:hidden}

Positive and statistically significant correlations are obtained in both the $L_\gamma$--$u_\mathrm{Total}$ and the $L_\gamma$--$\Gamma_c$ space. The positive $L_\gamma$--$u_\mathrm{Total}$ correlation could indicate a  significant contribution from IC emission. If the $e^\pm$ injected by MSPs lose energy through multiple comparable processes, e.g., IC and synchrotron radiaton, the $L_\gamma$ is proportional to the IC energy loss rates, which is linearly proportional to the $u_\mathrm{Total}$. In the extreme limit where all the $e^\pm$ injected by MSPs lose their energy through IC, the $L_\gamma$ is constrained by the energy injection rate of $e^\pm$ by MSPs and the $u_\mathrm{Total}$ would have less impact. Since we find a preference for a non-linear correlation ($a= 0.59 \pm 0.09$), the $\gamma$ rays are unlikely all originated from IC radiation.

\begin{figure}
    \centering
    \includegraphics[width=\columnwidth]{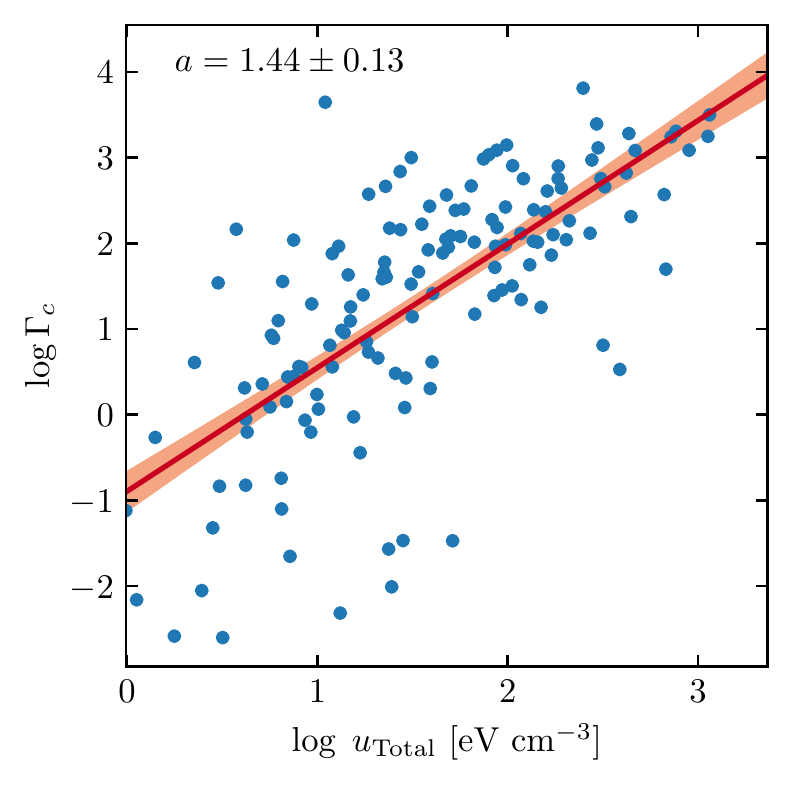}
    \caption{Hidden correlation between $u_\mathrm{Total}$ and $\Gamma_c$. GCs with higher encounter rates tend to have higher total photon field energy densities. The red line shows the relation between $u_\mathrm{Total}$ and $\Gamma_c$ based on a least-squires method in logarithmic space.}
    \label{fig:hidden_0}
\end{figure}

However, the positive correlation between $L_\gamma$ and $u_\mathrm{Total}$ could alternatively be driven by the $L_\gamma$--$\Gamma_c$ correlation. Here, we investigate a potential hidden correlation between $u_\mathrm{Total}$ and $\Gamma_c$ in order to better understand the nature of our detections. Figure~\ref{fig:hidden_0} shows the $u_\mathrm{Total}$ and $\Gamma_c$ values for our sample of GCs. It is apparent that the $u_\mathrm{Total}$ tends to be higher for GCs with higher encounter rates. Since these data are uncensored, we simply estimate the correlation using the Spearman coefficient: we find 0.72, confirming a strong correlation. This result is not surprising because a higher photon density implies higher stellar density which implies higher encounter rates (see also equation~(\ref{eq:encounter})).

An important implication of this result is that the $L_\gamma$--$u_\mathrm{Total}$ and the $L_\gamma$--$\Gamma_c$ correlations are not necessarily independent. Using a simple least squares method in the logarithmic space, we find the relation between $\Gamma_c$ and $u_\mathrm{Total}$ to be
\begin{equation}
    \Gamma_c \propto u_\mathrm{Total}^{1.44 \pm 0.13}.
\end{equation}
As reported in Table~\ref{tab:correlation}, the correlation between $L_\gamma$ and $\Gamma_c$ has a power index $a=0.39 \pm 0.10$. Based on the hidden relation between $\Gamma_c$ and $u_\mathrm{Total}$, the projected correlation between $L_\gamma$ and $u_\mathrm{Total}$ would have an index $a = 0.56 \pm 0.15$. Within the uncertainty, this projected result is consistent with the directly measured correlation between $L_\gamma$ and $u_\mathrm{Total}$ found in real data, $a = 0.59 \pm 0.09$. Therefore, the positive correlation between $L_\gamma$ and $u_\mathrm{Total}$ could be evidence for IC, or alternatively, an indirect effect of the $L_\gamma$--$\Gamma_c$ correlation which is connected to the dynamic formation of MSPs. The correlation found between $L_\gamma$ and $u_\mathrm{Total}$ cannot be considered as concrete evidence for IC due to this ambiguity implicated by the hidden correlation. However, as we discuss next, evidence for IC emission in GCs may still be revealed from the detailed spectral properties of these objects.

\section{Spectral analysis}\label{sec:spectra}

Motivated by the correlations detected in the previous section, we perform a spectral analysis of the 30 GCs detected in the 4FGL catalog, with the aim of finding further evidence for IC emission. First, we model the spectra of the GCs individually and compare their spectral parameters with those describing the field MSPs. Second, we fit the GCs spectra with universal spectral models which phenomenologically describe possible IC emission. Lastly, we use the detected IC component to constrain the $e^\pm$ injection efficiency in the GCs.

\subsection{Individual spectral fits}\label{sec:spectra_individual}

We consider two possible mechanisms of $\gamma$-ray emission, not mutually exclusive: CR, and IC up-scattering of starlight.

Detailed theoretical models predict that the maximum energy of the $e^\pm$ accelerated by the MSPs is limited by the CR in the pulsar magnetosphere. The predicted CR spectrum exhibits an energy cut-off which is related to the $e^\pm$ Lorentz factor~\citep{2005ApJ...622..531H}. For this reason, we model the GC $\gamma$-ray spectrum--as predicted by CR models--using a power law with an exponential cut-off (PLE) of the form:
\begin{equation}\label{eq:PLE}
    \left[ \frac{dN}{dE} \right]_{\mathrm{CR}} = N_0 \left(\frac{E}{E_0} \right)^{-\Gamma}\exp{ \left( -\frac{E}{E_\text{cut}} \right) },
\end{equation}
where $N_0$ is the normalization factor, $\Gamma$ is the spectral index, $E_0$ is the scaling energy, and $E_\mathrm{cut}$ is the energy cutoff. 

The $e^\pm$ may also leave the MSPs through open magnetic field lines and diffuse into the GC medium. Escaping pairs may up-scatter ambient photons and produce IC emission. The spectrum of the IC is determined by the $e^\pm$ spectrum and the ambient photon field. Theoretical studies~\citep{2011ApJ...743..181H} show that the MSPs can inject $e^\pm$ with Lorentz factors $\gamma_{e^\pm}$ $>$ 10$^6$ efficiently. Given ambient photons of $ E_0\sim$ 1 eV energy, the up-scattered IC photons can reach to above $\gamma_{e^\pm}^2 E_0 = 1$ TeV. Thus, in the Fermi GeV energy range, we assume a power law (PL) injection distribution for $e^\pm$. In the Thomson regime, the IC spectrum resulting from the interaction of power-law-like $e^\pm$ with ambient photons following a black-body radiation distribution~\citep{1970RvMP...42..237B} is still a power law in $\gamma$-ray energy. We consider this spectral form as a phenomenological description of the IC model. Specifically\footnote{For the maximum $\gamma$-ray energy (hundreds of GeV) and the photon field (starlight) we considered, the IC is in transition from the Thomson regime to the Klein-Nishina regime with the Thomson regime still an adequate approximation.},
\begin{equation}\label{eq:PL}
    \left[ \frac{dN}{dE} \right]_{\mathrm{IC}} = N_0 \left(\frac{E}{E_0} \right)^{-\Gamma}.
\end{equation}

We first estimate the GCs' spectral parameters using a maximum likelihood method. For this, we use the CR model and the IC model separately. We perform a $\chi^2$ test using the bin-by-bin $\gamma$-ray fluxes (9 energy bins from 300 MeV to 500 GeV) of each GC and the CR and IC emission models. Therefore, we define 
\begin{equation}
    \chi^2=\sum_i\frac{(F_\text{data}^i-F_\text{model}^i)^2}{(\Delta F_\text{data}^i)^2+(f_\text{ref}^i F_\text{data}^i)^2},
\end{equation}
where $F_\mathrm{data}^i$ and $\Delta F_\mathrm{data}^i$ are the measured fluxes and flux uncertainties obtained at each independent energy bin,  $F_\mathrm{model}^i$ are the predicted fluxes (either the CR or IC models). We allow all model parameters to be free (i.e., normalization, power-law index, and cut-off energy for CR, and normalization and power-law index for IC). The $f_\text{ref}^i$ values encapsulate the systematic uncertainties on the effective area of the LAT. We follow the values reported in the 4FGL catalog~\citep{2020ApJS..247...33A}, and set $f_\text{ref}$ to 0.05 for the first three energy bins, 0.06 for the fourth bin, and 0.1 for the last five bins.

\begin{figure}
    \centering
    \includegraphics[width=1\columnwidth]{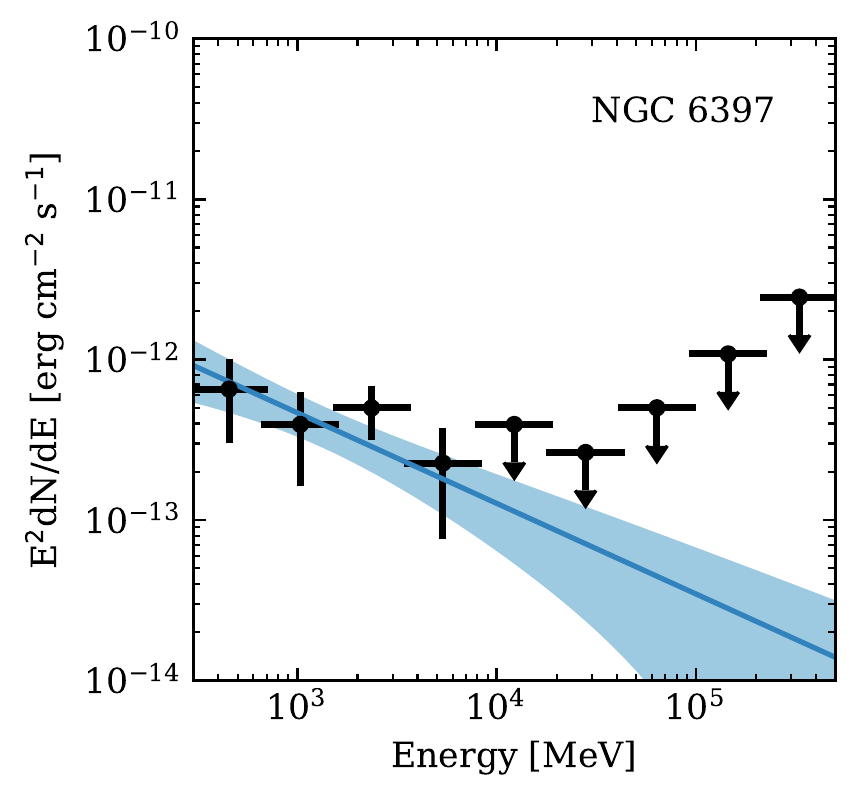}
    \includegraphics[width=1\columnwidth]{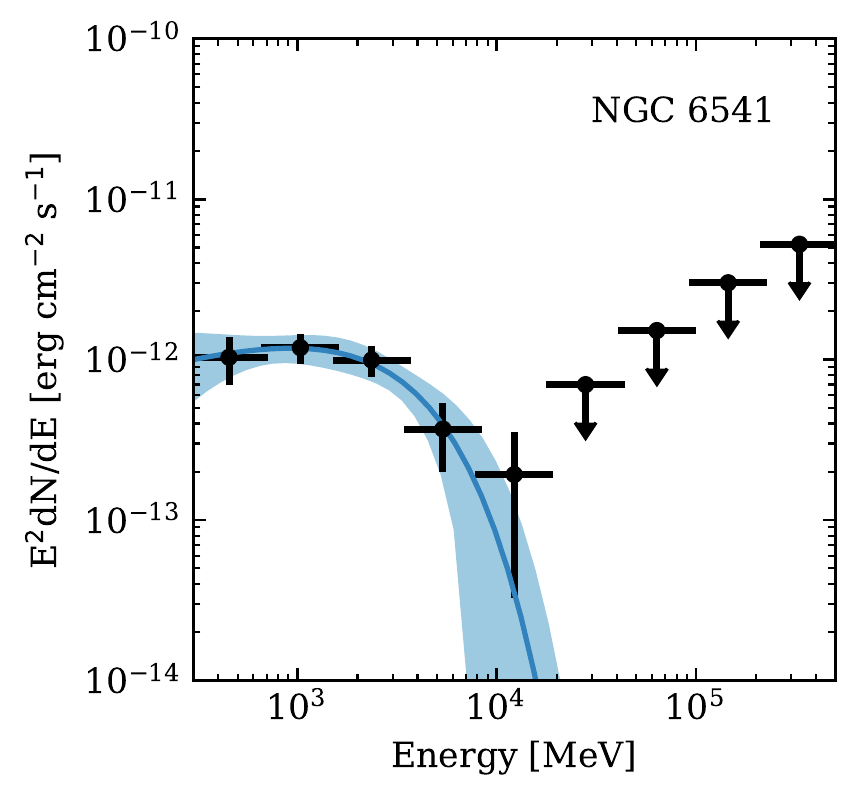}
    \caption{Best-fit spectra (blue solid line) and the 1$\sigma$ uncertainties (blue band) for two GCs. The bin-by-bin fluxes from the Fermi data analysis are included as black points. The top panel shows the spectrum of NGC 6397 as a simple power law because the PLE model is only slightly favored (TS$_\mathrm{curve} = 2$). The bottom panel shows the spectrum for NGC 6541, which prefers an exponential cutoff $\sim$ GeV with TS$_\mathrm{curve} = 4$.}
    \label{fig:spectra_example}
\end{figure}

The significance of the spectral curvature is estimated by computing the difference of the best-fit $\chi^2$ between the IC and the CR models, TS$_\mathrm{curve} = \chi^2_\text{IC} - \chi^2_\text{CR}.$ We apply a 2$\sigma$ threshold to determine the type of spectrum: for GCs with TS$_\mathrm{curve} \ge 4$, their PLE spectra are reported. Otherwise, the power-law spectra are reported. Note that this is a lower threshold than the 4FGL, which requires TS$_\mathrm{curve} \ge 9$ before detection of curvature is claimed. We adopt this low threshold because our analysis removes potentially contaminated photons < 300 MeV. Bins encompassing this low energy range usually generate upper limits in the 4FGL analysis and contribute to the detection of curvature. We find, a posteriori, the 2$\sigma$ threshold adequate in our analysis since our fits generate finite $E_\mathrm{cut}$'s within uncertainties for all GCs with TS$_\mathrm{curve} \ge 4$. For those GCs with TS$_\mathrm{curve} < 4$, the fits only generate lower limits for $E_\mathrm{cut}$. Table~\ref{tab:spectra} summarizes the best-fit parameters of the spectra for 30 $\gamma$-ray-detected GCs, sorted by their TS$_\mathrm{curve}$. The majority prefer curved spectra, with only 5 preferring simple power law spectra. Figure~\ref{fig:spectra_example} shows the spectra for 2 GCs as examples. The top panel shows the spectrum of NGC 6397, which is best fit by a simple power law, while the lower panel shows the spectrum for NGC 6541, which prefers an exponential cut-off at $\sim$ 350 MeV with TS$_\mathrm{curve}=4$. 

\begin{table*}
\centering
\caption{Spectral parameters for 30 $\gamma$-ray-detected GCs from the individual fits, ordered from the least curved to the most curved. For GCs with TS$_\mathrm{curve} <$ 4 (2$\sigma$), the best-fit simple power laws (PL) are reported. For the rest GCs, the power laws with an exponential cutoff (PLE) are reported. }\label{tab:spectra}
\begin{threeparttable}
\begin{tabular}{lccccr}
\hline
Name&$\Gamma$ & $\log\left(\frac{E_\mathrm{cut}}{\mathrm{MeV}}\right)$ & $\chi^2$/d.o.f. & Type\tnote{a} & TS$_\mathrm{curve}$\\
\hline
2MS-GC01 & {2.68$\pm$0.08} & ... & 1.03 & PL & 0 \\
NGC 1904 & {2.89$\pm$0.28} & ... & 0.63 & PL & 0 \\
NGC 6397 & {2.56$\pm$0.20} & ... & 0.36 & PL & 2 \\
NGC 7078 & {2.74$\pm$0.16} & ... & 0.70 & PL & 2 \\
NGC 5904 & {2.53$\pm$0.15} & ... & 0.54 & PL & 2 \\
\hline
NGC 6341 & 0.94$\pm$1.12 & {3.24$\pm$0.38} & 0.74 & PLE & 4 \\
NGC 6541 & 1.64$\pm$0.57 & {3.41$\pm$0.37} & 0.25 & PLE & 4 \\
NGC 6528 & 1.85$\pm$0.54 & {3.68$\pm$0.53} & 1.24 & PLE & 4 \\
GLIMPSE02 & 2.58$\pm$0.16 & {3.94$\pm$0.35} & 2.54 & PLE & 5 \\
NGC 6717 & 1.85$\pm$0.34 & {3.71$\pm$0.30} & 0.09 & PLE & 6 \\
NGC 6218 & 0.00$\pm$1.61 & {3.42$\pm$0.10} & 0.43 & PLE & 7 \\
NGC 6402 & 1.86$\pm$0.38 & {3.73$\pm$0.32} & 0.10 & PLE & 7 \\
NGC 2808 & 1.83$\pm$0.33 & {3.75$\pm$0.31} & 0.11 & PLE & 7 \\
NGC 6139 & 1.94$\pm$0.32 & {3.77$\pm$0.26} & 0.43 & PLE & 7 \\
NGC 6838 & 1.38$\pm$0.65 & {3.27$\pm$0.23} & 0.29 & PLE & 9 \\
NGC 6304 & 0.86$\pm$0.81 & {3.10$\pm$0.28} & 0.74 & PLE & 12 \\
M 80 & 1.60$\pm$0.32 & {3.71$\pm$0.24} & 0.31 & PLE & 14 \\
Terzan 2 & 0.60$\pm$0.59 & {3.40$\pm$0.18} & 1.00 & PLE & 16 \\
NGC 6440 & 1.88$\pm$0.22 & {3.63$\pm$0.19} & 1.07 & PLE & 17 \\
NGC 6441 & 1.83$\pm$0.23 & {3.59$\pm$0.21} & 0.84 & PLE & 17 \\
NGC 6652 & 1.29$\pm$0.42 & {3.29$\pm$0.20} & 0.71 & PLE & 22 \\
NGC 6316 & 1.60$\pm$0.23 & {3.54$\pm$0.14} & 1.16 & PLE & 24 \\
NGC 6752 & 0.83$\pm$0.58 & {2.99$\pm$0.20} & 0.18 & PLE & 28 \\
Terzan 1 & 0.00$\pm$0.36 & {3.28$\pm$0.06} & 0.77 & PLE & 37 \\
GLIMPSE01 & 1.67$\pm$0.14 & {3.57$\pm$0.10} & 0.53 & PLE & 61 \\
M 62 & 1.48$\pm$0.14 & {3.47$\pm$0.08} & 0.74 & PLE & 90 \\
Omega Cen & 1.05$\pm$0.27 & {3.25$\pm$0.12} & 1.11 & PLE & 103 \\
NGC 6388 & 1.33$\pm$0.15 & {3.35$\pm$0.07} & 0.46 & PLE & 137 \\
Terzan 5 & 1.58$\pm$0.09 & {3.54$\pm$0.06} & 2.14 & PLE & 159 \\
NGC 104 & 1.28$\pm$0.11 & {3.37$\pm$0.05} & 0.54 & PLE & 207 \\
\hline
\end{tabular}
\begin{tablenotes}
\item [a] Spectrum type: PL for power law; PLE for power law with an exponential cut-off.
\end{tablenotes}
\end{threeparttable}
\end{table*}

\begin{figure}
    \centering
    \includegraphics[width=\columnwidth]{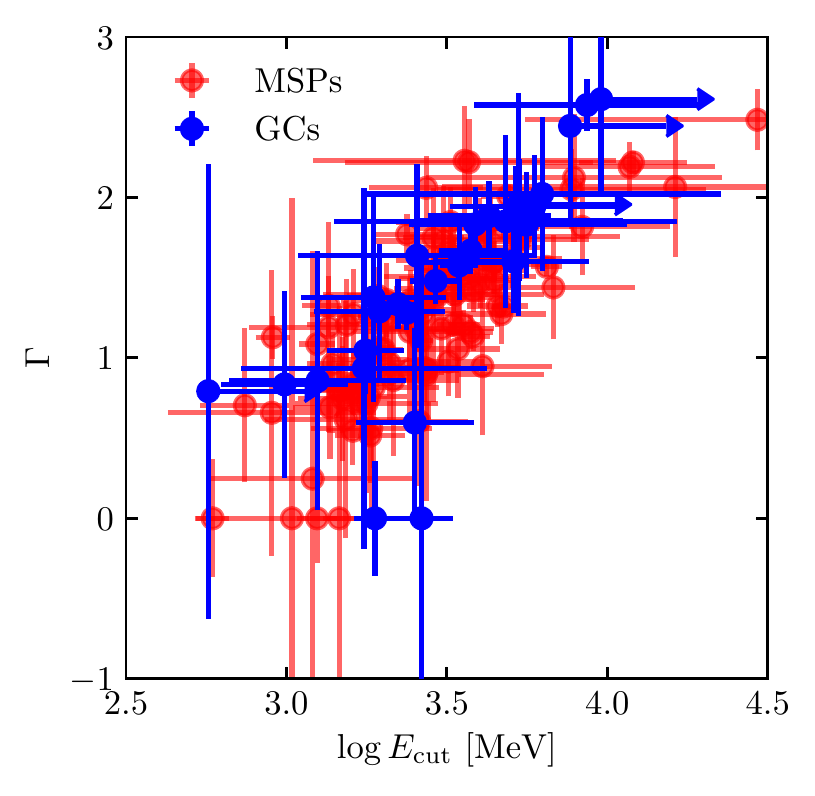}
    \caption{Spectral parameters of the $\gamma$-ray emission. The PLE spectrum is assumed, with $\Gamma$ and $E_\mathrm{cut}$ as free parameters. Both the GCs (blue dots) and the MSPs (red dots) detected in the 4FGL are included. The error bars represent the 1$\sigma$ parameter uncertainties. Within uncertainties, the distribution of the GCs' spectra is very similar to that of the MSPs.}
    \label{fig:msps}
\end{figure}

The \textit{Fermi}-LAT has detected more than 200 pulsars. Most of these have been found to have a curved spectrum with best-fit energy cutoffs of the order of a few GeV. Therefore, their $\gamma$-ray emission is likely dominated by a CR process. {Nevertheless,~\citet{2018PhRvD..98d3005H,2021arXiv210400014H} find} that many MSPs could be surrounded by TeV halos of IC. The IC emission may also extend to the GeV energy range. Figure~\ref{fig:msps} compares the distribution of the spectral parameters of 108 field MSPs in the 4FGL (red dots) with the $\gamma$-ray GCs (blue dots), assuming a PLE spectra. The 1$\sigma$ uncertainties of the best-fit parameters are also shown. We find that within uncertainties, the spectral distribution of the GCs and the field MSPs are very similar. However, given the starlight in GCs typically contributes a much larger photon field energy density than for field MSPs, IC emission may still provide a sizeable contribution to the overall GC $\gamma$-ray emission. The results from individual spectral fit cannot rule out the presence of IC for the following reasons: (1) There are 5 GCs for which the spectra shows no obvious energy cutoffs. This is hard to explain using the CR emission model alone. (2) Many GCs have energy bins above 10 GeV detected even though their spectra have cutoffs of order a GeV (see Appendix~\ref{appx:spectra}). These high-energy measurements may be indicative of an IC component.

\subsection{Fit assuming universal spectral components}\label{sec:spectra_global}

The curvature of the GCs' spectra at around a few GeV's, as well as their similarity to the field MSPs spectra, support the hypothesis that the GeV $\gamma$-ray emission from most GCs is due to mainly local CR emission from MSPs within GCs. However, IC may still contribute sub-dominantly, especially at the high-energy end. To probe this possibility, we perform a reduced $\chi^2$ analysis in which we fit, bin-by-bin, the  GCs' spectra using a linear combination of the spectral components introduced in equation~\ref{eq:PLE} and \ref{eq:PL}. Specifically;
\begin{align}\label{eq:PLE+PL}
    \frac{dN}{dE} &= \left[ \frac{dN}{dE} \right]_\mathrm{CR} + \left[ \frac{dN}{dE} \right]_\mathrm{IC} \\\nonumber
    &= N_1\left( \frac{E}{E_0} \right)^{-\Gamma_1}\exp\left(-\frac{E}{E_\mathrm{cut}}\right) + N_2\left(\frac{E}{E_0}\right)^{-\Gamma_2}.
\end{align}

Fitting such a two-component model to each GC's bin-by-bin data is difficult since the GC spectra only contains 9 energy bins, and many high energy bins only provide upper limits. On the other hand, typical GCs can host close to $\sim$ 20 MSPs~\citep{2019ApJ...877..122Y} each. So, as a simplifying approximation, we hypothesise that the $\gamma$-ray and $e^\pm$ injection from the collection of MSPs in each GC to be similar to one another. Then, we can fit a common or universal spectrum to all the $\gamma$-ray detected GCs, i.e., one set of spectral shape parameters in the two component model in equation~(\ref{eq:PLE+PL}) for all the GCs. More specifically, we tie the $\Gamma_1$, $\Gamma_2$ and $E_\mathrm{cut}$ across all GCs considered (hereafter referred to as the ``universal model''). The normalization factors $N_1$ and $N_2$ are allowed to float for each GC as these should depend on the number of MSPs and the photon field energy density in the GCs.

\begin{figure*}
    \includegraphics[width=0.8\textwidth]{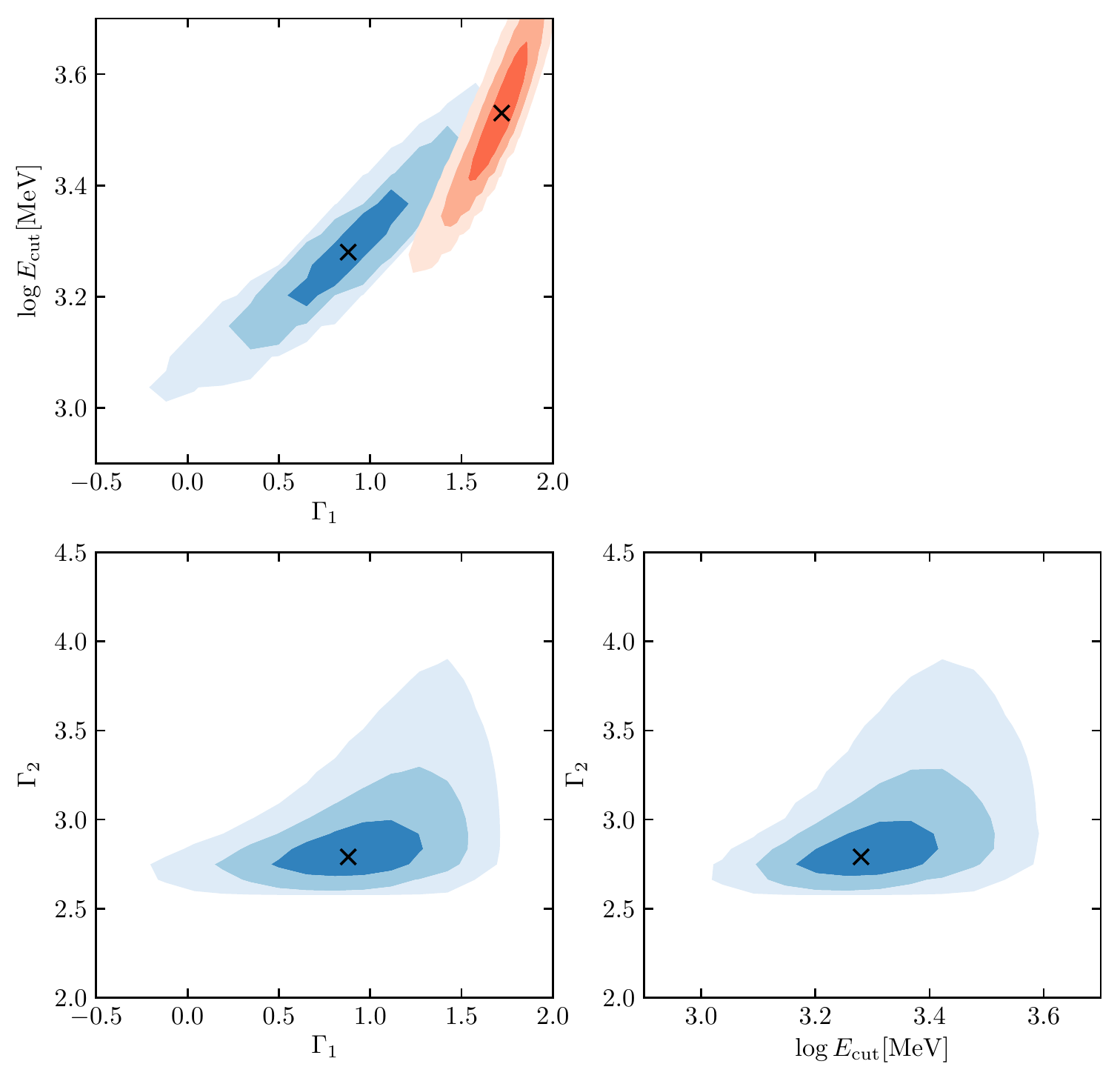}
    \caption{The projected parameter space of the universal model fit, as illustrated in equation~(\ref{eq:PLE+PL}). The blue shaded contours show the 1$\sigma$, 2$\sigma$, and 3$\sigma$ confidence levels for the two-component model. The crosses indicate the best fit values for $\Gamma_1$, $\Gamma_2$, and $E_\mathrm{cut}$. On the top-left panel, the red shaded region shows the best-fit value and 3$\sigma$ confidence levels for the null hypothesis, which includes only the CR model.}
    \label{fig:global}
\end{figure*}

We perform the universal fit by minimizing the total $\chi^2$ of 30 $\gamma$-ray-detected GCs,
\begin{equation}
    \chi^2_\mathrm{total}(\Gamma_1,\;\Gamma_2,\;E_\mathrm{cut}) = \sum_i\chi_i^2(\Gamma_1,\;\Gamma_2,\;E_\mathrm{cut},\;N_1^i,\;N_2^i).
\end{equation}
In practice, we assign the same $\Gamma_1$, $\Gamma_2$, and $E_\mathrm{cut}$ to all $\gamma$-ray GCs and perform a minimum $\chi^2$ for each different object. However, during the fit, we free the $N_1^i$ and $N_2^i$ parameters. By scanning the parameter space of $\Gamma_1$, $\Gamma_2$, and $E_\mathrm{cut}$, we find the values that minimize the total $\chi^2$ for the two-component model. These are, 
\begin{align}\nonumber
    \Gamma_1 = 0.88 \pm 0.44,\\\nonumber
    \Gamma_2 = 2.79 \pm 0.25,\\\nonumber
    \log\left(\frac{E_\mathrm{cut}}{\mathrm{MeV}}\right) = 3.28 \pm 0.16,
\end{align}
for which we find a $\chi^2_\mathrm{total} / \mathrm{d.o.f} = 204/206 = 0.99$ (we have $30\times 9$ data points, and the number of free parameters is $60+3$ as there are 2 normalization factors for each GC, and 3 global parameters. So, we have $\mathrm{d.o.f}=30\times 9 - 60 - 3 -1 = 206$). In Figure~\ref{fig:global}, we show the associated 3$\sigma$ contours and correlated uncertainties for the parameters $\Gamma_1$, $\Gamma_2$, and $E_\mathrm{cut}$ as found in this procedure.

\begin{figure}
    \centering
    \includegraphics[width=\columnwidth]{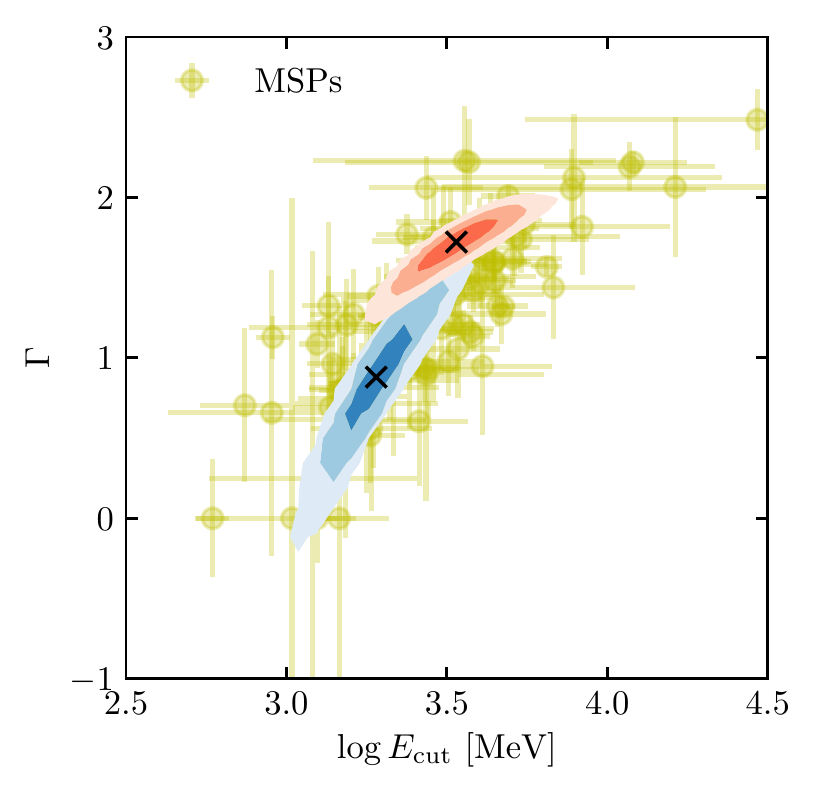}
    \caption{Same as Figure~\ref{fig:msps}, but with best-fit parameters of GC replaced by those obtained from the universal models. The 3$\sigma$ contours are shown for the CR-only model (red) and the CR component from the two-component model (blue), as in Figure~\ref{fig:global}. The MSP parameters are included in the background (yellow dots).}
    \label{fig:msps_global}
\end{figure}

In order to compute the statistical significance of the {PL} component, it is necessary to define the null hypothesis. This corresponds to the universal model containing only the CR component (see equation~\ref{eq:PLE}). Again, we tie $\Gamma_1$ and $E_\mathrm{cut}$ across all GCs and allow the normalization factors to individually vary. We find that the best-fit parameters for the CR-only model are:
\begin{align}\nonumber
    \Gamma_1 = 1.72 \pm 0.21, \\\nonumber
    \log\left(\frac{E_\mathrm{cut}}{\mathrm{MeV}}\right) = 3.53 \pm 0.19.
\end{align}
In this case, we find a $\chi^2_\mathrm{total} / \mathrm{d.o.f} = 349/237=1.47$ (the null hypothesis has 30 + 2 free parameters so the $\mathrm{d.o.f}=30\times 9 - 30 - 2 - 1 = 237$). This implies that the two-component model is preferred at the 8.2$\sigma$ level ($\Delta \chi^2=349-204$ for 31 d.o.f [1 power-law index plus 30 normalization factors]). It is useful to compare the best-fit spectral results of the CR component for the universal models with the best-fit spectral parameters of the MSPs in the 4FGL catalog. As seen in Figure~\ref{fig:msps_global}, although the CR-only model (null hypothesis) has larger $\Gamma$ and higher $E_\mathrm{cut}$ than the CR component from the two-component model, our results for both models are compatible with the field MSPs, up to statistical uncertainties.

The universal fitting procedure used in this section is similar to a stacking analysis. This method is usually applied to explore the characteristics of an astrophysical population, especially one that is undetected. Numerous studies have shown that this technique can increase the detection sensitivity to such population characteristics. So, even though there is good statistical evidence for the {PL} component in the universal fit,  this might not be apparent from individual fitting of the two-component model.

\begin{figure}
    \centering
    \includegraphics[width=1\columnwidth]{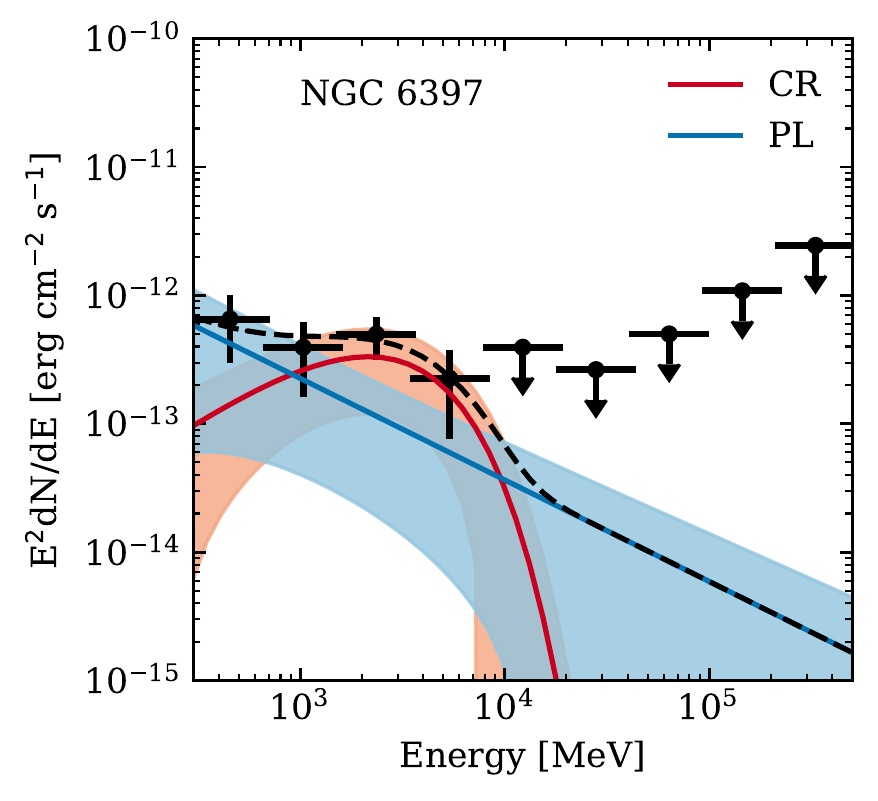}
    \includegraphics[width=1\columnwidth]{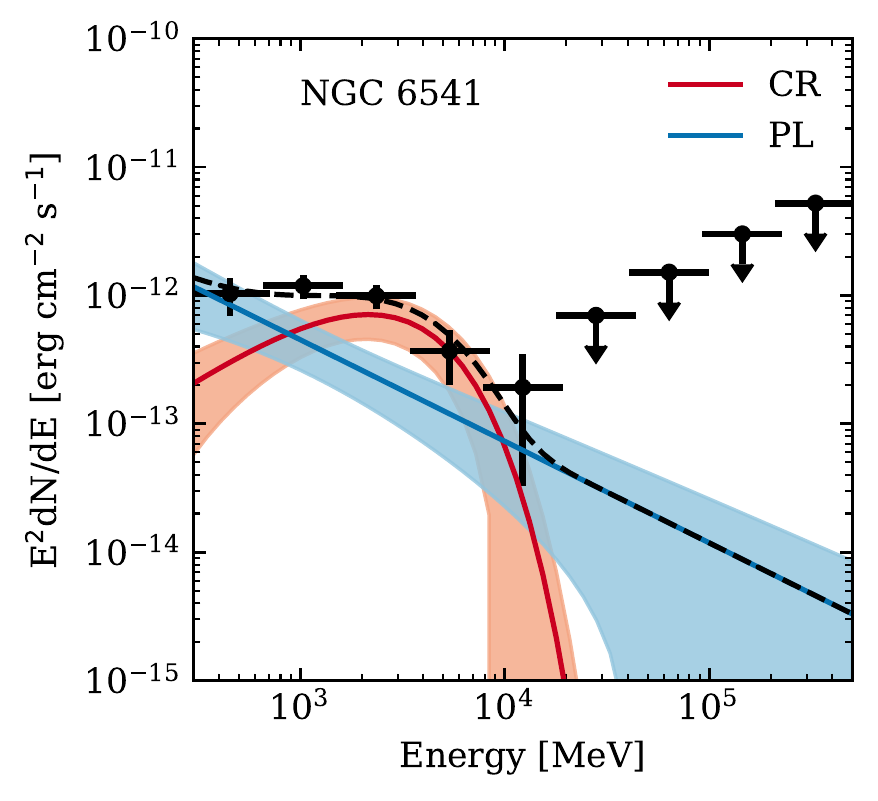}
    \caption{The best-fit two-component spectra for NGC 6397 (top panel) and NGC 6541 (bottom panel). The spectra are fit to a universal shape with same $\Gamma_1$, $\Gamma_2$, and $E_\mathrm{cut}$ for all GCs.  Only the normalizations of the two components are allowed to vary between GCs. The best-fit parameters for the CR component (red line with shaded band) is $\Gamma_1 = 0.88 \pm 0.44$ and $\log(E_\mathrm{cut}/\mathrm{MeV})=3.28 \pm 0.16$. The best fit parameter for the {PL} component (blue line with shaded band) is $\Gamma_2 = 2.79 \pm 0.25$. The black dashed line indicates the total of the two components.}
    \label{fig:global_spectra}
\end{figure}

We show examples of the spectra obtained in the universal fit of the two-component model for NGC 6397 and NGC 6541 in Figure~\ref{fig:global_spectra}. As can be seen, the solutions for the CR and {PL} components look physically plausible. The spectra also include 1$\sigma$ bow-tie errors, which immediately reveal the level of statistical support for the CR and {PL} components, respectively. For comparison, the results shown in Figure~\ref{fig:spectra_example} presented a {single-component (e.g.,~\citet{2020ApJS..247...33A})} spectral curvature analysis applied to NGC 6397 and NGC 6541, individually. 

\begin{figure}
    \centering
    \includegraphics[width=1\columnwidth]{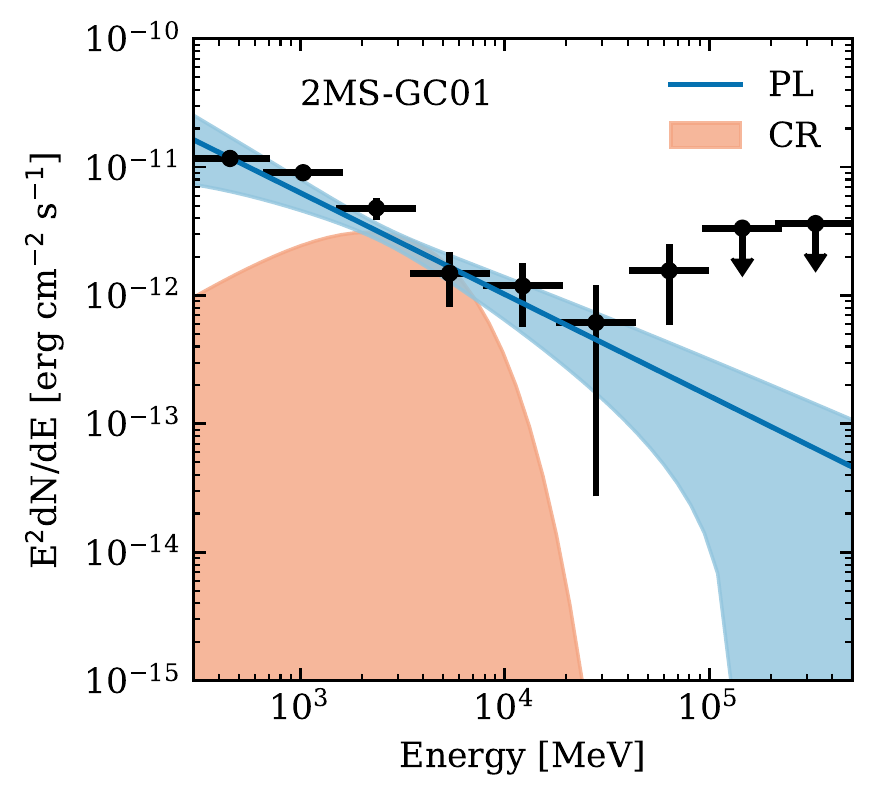}
    \includegraphics[width=1\columnwidth]{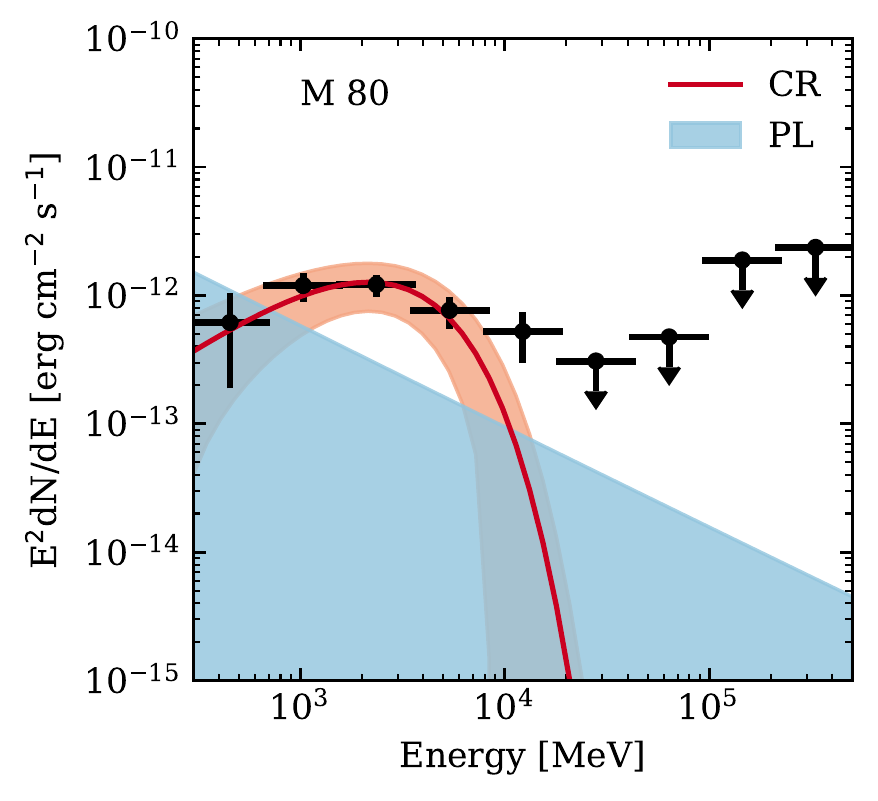}
    \caption{Spectra of 2MS-GC01 (top panel) and M 80 (bottom panel) from the two-component fit. The 2MS-GC01 prefers only the {PL} model (blue solid line). The 95\% upper limit on the normalization of the CR model is shown by the red shaded region. In contrast, only the CR model is detected for M 80 (red solid curve). The 95\% upper limit on the normalization of the {PL} model is shown by the blue shaded region.}
    \label{fig:global_spectra_ul}
\end{figure}

We show some additional noteworthy results of the universal fit in Figure~\ref{fig:global_spectra_ul}. Here, we see that in the case of GC 2MS-GC01, the {PL} model is sufficient to explain the bin-by-bin spectrum over the full energy range, but we also display the estimated 95\% C.L. upper limit for the normalization of the CR model. By contrast, in the case of GC M 80, we find that the data is best described by the CR model alone, and we show the 95\% C.L. upper limit for the normalization of the {PL} component. These examples might indicate special conditions of the environment of the GC.

For 19 GCs (out of the 30 GCs included in the universal fit), we find good statistical support for both the CR and {PL} models. For the remaining 11 GCs we find that only one component is sufficient to explain the spectrum: 7 GCs require only the CR model and the other 4 GCs require only the {PL} model. The two-component spectral results for all 30 GCs are shown in Appendix~\ref{appx:spectra}.

To explain the best-fit index of the {PL} component ($\Gamma_2 = 2.79 \pm 0.25$) {as IC emission}, the implied emitting $e^\pm$ spectrum would have an index of $4.58 \pm 0.50$. The minimum $e^\pm$ energy required to maintain a power law IC in the energy range of our analysis (300 MeV) is $\lesssim$ 10 GeV given that the upscattered photon field has energies $\sim$ 1 eV~\citep{1970RvMP...42..237B}. Interestingly,~\citet{2011ApJ...743..181H} has simulated $e^{\pm}$ pair cascades from pulsar polar caps. For typical MSP parameters, they show that the injected $e^\pm$ flux decreases by $\sim 5 - 10$ orders of magnitude when the $e^\pm$ energy increases from $\sim$10 GeV to $\sim$ 1 TeV. The soft $e^\pm$ spectrum we found is in line with their results.

\subsection{Leptonic injection efficiency within the globular clusters}\label{spectra_fe}

The relative normalization of the CR and IC components probes an important property of the MSPs: the $\gamma$-ray production and $e^\pm$ injection efficiencies, respectively. Indeed, the spin-down energy of MSPs can be injected into $\gamma$ rays and $e^\pm$. While prompt $\gamma$ rays are mainly produced by CR in the magnetosphere, the $e^\pm$ can propagate into the interstellar environment. We can write down the following empirical relations,
\begin{align}
    L_\mathrm{CR} &= f_\gamma L_\mathrm{sd},\\
    L_{e^\pm} &= f_{e^\pm} L_\mathrm{sd},
\end{align}
where $L_\mathrm{sd}$ is the spin-down luminosity and the $f$'s are efficiency parameters. 

Assuming that the $\gamma$-ray emission is the superposition of CR and IC processes, we have that
\begin{equation}
    L_\gamma = L_\mathrm{CR} + L_\mathrm{IC}.
\end{equation}
However, $e^\pm$ can also lose energy via synchrotron radiation. We can compare the relative strength of the sychrotron radiation vs the IC emission through
\begin{equation}
    \frac{\dot{E}_\mathrm{SR}}{\dot{E}_\mathrm{IC}} = \frac{u_\mathrm{B}^2}{u_\mathrm{rad}^2},
\end{equation}
where $\dot{E}_\mathrm{SR}$ and $\dot{E}_\mathrm{IC}$ are the synchrotron and IC energy loss rates, respectively, $u_\mathrm{B}$ is the magnetic energy density, and $u_\mathrm{rad}$ is the radiation field energy density. We note that this relation assumes that the $e^\pm$ lose all their energy within the GCs. We provide justifications for this assumption in Appendix~\ref{appx:system_fe}. For a typical GC the magnetic field is estimated to be $B\lesssim \mathrm{10\;\mu G}$~\citep{2007MNRAS.377..920B}, so we expect to have a $u_\mathrm{B} = {\mathrm{(10\;\mu G)}^2}/{(2\mu_0)} = 2.5$ eV cm$^{-3}$, which is much smaller than the total radiation field of most GCs shown in Table~\ref{tab:pars}. Thus in the usual instance when IC is the leading energy loss process, we have that
\begin{equation}
    L_\mathrm{IC} \simeq L_{e^\pm}.
\end{equation}

Since no GC is detected as an extended source by the \textit{Fermi}-LAT, the energy carried away to the interstellar medium by $e^\pm$ propagation is expected to be small. Thus, we can use the following approximate scaling relation,
\begin{equation}\label{eq:fefgamma}
    \dfrac{f_{e^\pm}}{f_\gamma} \simeq \dfrac{L_\mathrm{IC}}{L_\mathrm{CR}}.
\end{equation}
Using this\footnote{We discuss caveats to the approximation used in equation~(\ref{eq:fefgamma}) in Appendix~\ref{appx:system_fe}.}, we estimate the ratios $f_{e^\pm}/f_\gamma$ for all $\gamma$-ray emitting GCs in Table~\ref{tab:ratio}. These are found to be in the range $\approx 0.17 - 1.04$. Note that for some GCs we present only upper or lower limits as only one component is detected. The measurement of pulsars by \textit{Fermi}-LAT estimated the $f_{\gamma}$ efficiency from observations of pulsars and found that on average, $f_\gamma \sim \mathrm{10\%}$. Furthermore, the $e^\pm$ efficiency $f_{e^\pm}$ was also estimated to be around 10\% from TeV observations of {nearby pulsars~\citep{2017PhRvD..96j3013H, 2018PhRvD..98d3005H, 2021arXiv210400014H} and} the Galactic center~\citep{2013MNRAS.435L..14B}, although \citet{2019MNRAS.484.2876M} claims $f_{e^\pm}$ is at the percentage level for one GC they observed (NGC 7078), and \citet{2020arXiv200508982S} suggest $f_{e^\pm} \sim 90\%$ on the basis of the  radio continuum emission detected from galaxies with low specific star formation rates.

For the CR and IC luminosities, we integrate the best-fit two-component spectra from 300 MeV to 500 GeV, the same energy range used in the Fermi data analysis. For the IC emission, the minimum $e^\pm$ injection energy probed by this energy range is $\lesssim$ 10 GeV assuming the ambient photon field is starlight. We note that \citet{2011ApJ...743..181H} investigated the $e^\pm$ pair cascades from MSPs and proposed several theoretical models. Their Figure 10 shows that their predicted pair spectra peak at $\sim$ GeV and extend to $\gtrsim$ TeV. This roughly corresponds to the Fermi energy range we assume. If the $e^\pm$ injection spectra extend to lower energy, they will lead to higher $L_\mathrm{IC}$. Therefore, the choice of $\gamma$-ray energy range will contribute as systematic uncertainties on the estimated $f_{e^\pm}/f_\gamma$. For example, we verify that the $f_{e^\pm}/f_\gamma$ would be $\sim$ 5 times larger if the minimum $\gamma$-ray energy is assumed to be 30 MeV.

\begin{table}
    \centering
    \caption{$\gamma$-ray luminosity for the IC and CR  components and the ratios between $f_{e^\pm}$ and $f_\gamma$. For GCs with only one component detected, the 95\% C.L. upper limits are reported for another component.}
    \begin{tabular}{lccr}
\hline
Name & $L_\mathrm{{IC}}$ & $L_\mathrm{CR}$ & $f_{e^\pm}/f_\gamma$ \\
 & (10$^{34}$ erg s$^{-1}$) & (10$^{34}$ erg s$^{-1}$) &  \\
\hline
GLIMPSE02 & 10.90 $\pm$ 1.06 & < 1.70 & > 6.40 \\
2MS-GC01 & 3.20 $\pm$ 0.44 & < 1.08 & > 2.95 \\
NGC 7078 & 2.38 $\pm$ 0.62 & < 1.14 & > 2.08 \\
NGC 1904 & 1.75 $\pm$ 0.75 & < 1.62 & > 1.08 \\
NGC 5904 & 0.54 $\pm$ 0.31 & 0.52 $\pm$ 0.24 & 1.04 $\pm$ 0.77 \\
NGC 6397 & 0.05 $\pm$ 0.04 & 0.05 $\pm$ 0.03 & 0.97 $\pm$ 1.00 \\
NGC 6440 & 4.83 $\pm$ 1.20 & 5.12 $\pm$ 0.97 & 0.94 $\pm$ 0.29 \\
NGC 6541 & 0.99 $\pm$ 0.42 & 1.09 $\pm$ 0.35 & 0.92 $\pm$ 0.48 \\
NGC 6139 & 2.33 $\pm$ 1.14 & 2.64 $\pm$ 0.90 & 0.88 $\pm$ 0.52 \\
NGC 6441 & 8.36 $\pm$ 1.80 & 9.57 $\pm$ 1.57 & 0.87 $\pm$ 0.24 \\
NGC 6752 & 0.25 $\pm$ 0.09 & 0.33 $\pm$ 0.08 & 0.76 $\pm$ 0.33 \\
NGC 6717 & 0.80 $\pm$ 0.42 & 1.19 $\pm$ 0.35 & 0.67 $\pm$ 0.40 \\
NGC 6402 & 1.18 $\pm$ 0.79 & 1.83 $\pm$ 0.62 & 0.65 $\pm$ 0.48 \\
NGC 2808 & 1.18 $\pm$ 0.63 & 2.09 $\pm$ 0.56 & 0.56 $\pm$ 0.34 \\
NGC 6838 & 0.16 $\pm$ 0.14 & 0.28 $\pm$ 0.11 & 0.55 $\pm$ 0.54 \\
GLIMPSE01 & 2.77 $\pm$ 0.65 & 5.89 $\pm$ 0.62 & 0.47 $\pm$ 0.12 \\
NGC 6316 & 2.99 $\pm$ 1.44 & 7.71 $\pm$ 1.26 & 0.39 $\pm$ 0.20 \\
Terzan 5 & 10.02 $\pm$ 1.68 & 27.46 $\pm$ 1.83 & 0.37 $\pm$ 0.07 \\
NGC 6652 & 1.16 $\pm$ 0.77 & 3.35 $\pm$ 0.68 & 0.35 $\pm$ 0.24 \\
M 62 & 2.20 $\pm$ 0.60 & 6.92 $\pm$ 0.63 & 0.32 $\pm$ 0.09 \\
NGC 6388 & 4.37 $\pm$ 1.18 & 14.22 $\pm$ 1.27 & 0.31 $\pm$ 0.09 \\
NGC 104 & 0.96 $\pm$ 0.22 & 4.67 $\pm$ 0.30 & 0.21 $\pm$ 0.05 \\
Omega Cen & 0.50 $\pm$ 0.24 & 2.90 $\pm$ 0.27 & 0.17 $\pm$ 0.08 \\
NGC 6528 & < 2.41 & 1.38 $\pm$ 0.58 & < 1.74 \\
NGC 6218 & < 0.39  &  0.23 $\pm$ 0.12 & < 1.67 \\
NGC 6341 & < 0.71  &  0.49 $\pm$ 0.23 & < 1.45 \\
NGC 6304 & < 0.79  &  0.89 $\pm$ 0.30 & < 0.88 \\
M 80 & < 2.31  &  3.45 $\pm$ 0.74 & < 0.67 \\
Terzan 2 & < 0.49  &  2.66 $\pm$ 0.52 & < 0.18 \\
Terzan 1 & < 0.16  &  2.42 $\pm$ 0.49 & < 0.07 \\
\hline
    \end{tabular}
    \label{tab:ratio}
\end{table}

\section{Discussion}\label{sec:discussion}

\subsection{Implications of the correlation analysis}

We have found strong positive correlations between $L_\gamma$, the stellar encounter rate $\Gamma_c$, and the total photon field energy density $u_\mathrm{Total}$ of GCs. The latter correlation may indicate a significant contribution of IC upscattering of ambient starlight to the total $\gamma$-ray emission of GCs. However, we showed in Figure~\ref{fig:hidden_0} that the $u_\mathrm{Total}$ also increases with $\Gamma_c$. So, the detection of the $L_\gamma$--$u_\mathrm{Total}$ correlation alone does not unambiguously demonstrate the presence of IC emission in GCs~\footnote{We also analyzed other potential hidden correlations, but no obvious correlations with other parameters such as the interstellar radiation field and the distance from the Sun were found (see Appendix~\ref{appx:system_fe}).}. On the other hand, corroborating evidence for IC emission was found from the universal two-component fit, wherein we were able to estimate, separately, the luminosities of the CR and IC components of most GCs. The ratios of the luminosities between the CR and IC components were found to be comparable. This implies that MSPs in GCs can potentially inject $e^{\pm}$ as efficiently as they inject prompt magnetospheric $\gamma$ rays.

Overall, our correlation results in the $L_\gamma$-$\Gamma_c$ plane are consistent with those in \citet{2011ApJ...726..100H} and~\citet{2019MNRAS.486..851D}, though it is important to note that our method is more statistically robust since we include GCs with detection limits which were previously neglected. In particular, our high significance ($6.4\sigma$) detection of a $L_\gamma$--$\Gamma_c$ correlation naively supports a dynamic formation scenario for MSPs in GCs. However, as pointed out earlier, this correlation may not be independent due to the hidden correlation of $u_\mathrm{Total}$ and $\Gamma_c$. On the other hand, we have not found an obvious correlation between $f_{e^\pm}/f_\gamma$ and $u_\mathrm{Total}$ (see Appendix~\ref{appx:system_fe}). The lack of this latter correlation may indicate that IC is, in fact, the leading energy loss process for $e^\pm$ in GCs: in the limit of IC dominance, the IC luminosity of GCs already saturates the power going into freshly-injected $e^\pm$ pairs, so ``dialling-up'' the light field energy density has no effect on the IC luminosity. Thus, in this situation of IC dominance, we expect, at most, only a weak correlation between $L_\gamma$ and $u_\mathrm{Total}$ and we would anticipate that the $L_\gamma$-$\Gamma_c$ correlation is the fundamental one (while the $L_\gamma$-$u_\mathrm{Total}$ correlation is caused by the fact that GCs with higher stellar encounter rate naturally have higher stellar density which leads to higher photon field density). With the uncertainties of the data and the number of variables involved, it is challenging to statistically confirm this scenario. Overall, however, our results are consistent with there being both a significant role for dynamical formation of MSPs in GCs and for the presence of a significant contribution of IC to the overall $\gamma$-ray emission of GCs.

Previous studies~\citep{2011ApJ...726..100H} found a positive correlation between $L_\gamma$ and $u_\mathrm{MW}$, and $L_\gamma$ and [Fe/H]. However, our study does not confirm these results. The former discrepancy is possibly due to the different interstellar radiation field models assumed in these works, or it could be due to the more limited sample data used in~\cite{2011ApJ...726..100H}. Specifically, while we have used the most up-to-date interstellar radiation field for the Milky Way--which is the 3D radiation field model in GALPROP v56~\citep{2017ApJ...846...67P}$-$ ~\citet{2011ApJ...726..100H} used the 2D radiation field model in GALPROP v54. Also, as explained above, our correlation study includes 30 $\gamma$-ray-detected GCs , as well as the luminosity upper limits from the 127 non-detected ones, thus covering the entire GC~\citet{1996AJ....112.1487H} catalog. As for the latter discrepancy, similar results for the $L_\gamma$--[Fe/H] correlation were obtained by~\citet{2019MNRAS.486..851D}, which also found low statistical evidence for this correlation.

\subsection{Implications for the Fermi GeV excess}

\begin{figure}
    \centering
    \includegraphics[width=1\columnwidth]{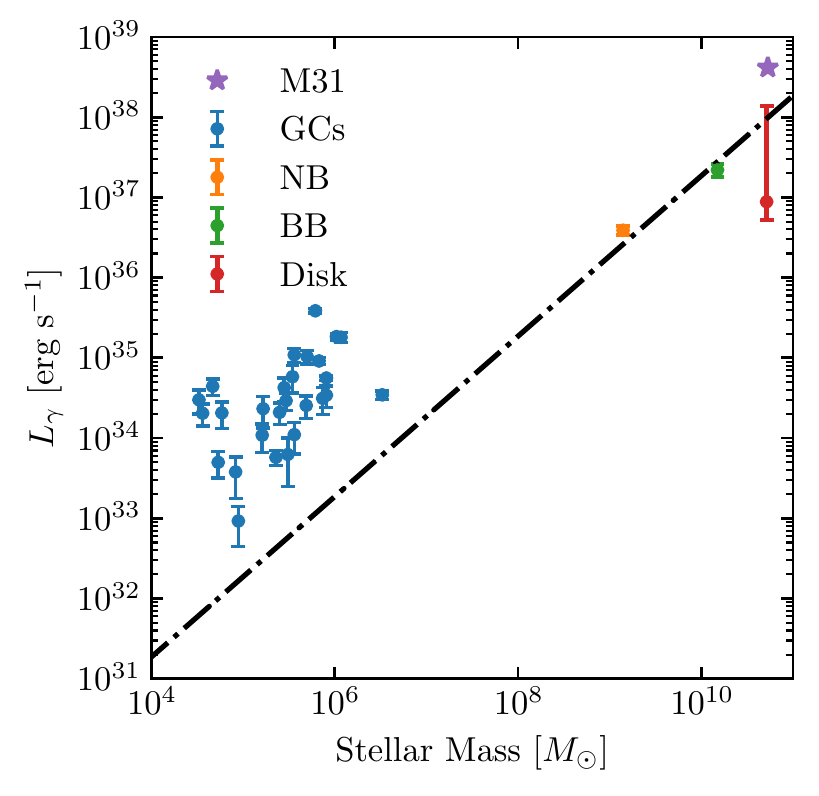}
    \caption{Relations between $L_\gamma$ and the stellar mass for several systems. The results for the nuclear bulge (orange dot) and boxy bulge (green dot) are adopted from~\citet{2019JCAP...09..042M}. The Galactic disk (red dot) luminosity is predicted by~\citet{2018NatAs...2..819B}, and the M31 (purple star) are adopted from~\citet{2017ApJ...836..208A}. The blue dots show the data for 30 $\gamma$-ray detected GCs. The dash-dotted line is the relation implied by the nuclear bulge and boxy bulge.}
    \label{fig:stellar_mass}
\end{figure}

The emission from a putative population of about $40,000$~\citep{2020JCAP...12..035P} unresolved MSPs in the Galactic Center region is currently the preferred explanation for the Fermi GeV excess~\citep{2018NatAs...2..387M,2018NatAs...2..819B,2019JCAP...09..042M,2020PhRvD.102d3012A}. Since GCs also contain large numbers of unresolved MSPs, it is useful to compare the light-to-mass ratios for these two systems so as to obtain additional clues for the physical processes causing the observed high-energy $\gamma$-ray emissions in their directions. In Figure~\ref{fig:stellar_mass}, we show the relation between $L_\gamma$ and the stellar mass for several different systems. The blue dots show the sample of the $\gamma$-ray detected GCs in this work. The nuclear bulge (orange dot) has a stellar mass around $1.4\times 10^9$ M$_\odot$ and a $\gamma$-ray luminosity of $(3.9\pm 0.5)\times 10^{36}$ erg s$^{-1}$ and the boxy bulge (green dot) has $1.5\times 10^{10}$ M$_\odot$ and $(2.2 \pm 0.4)\times 10^{37}$ erg s$^{-1}$~\citep{2019JCAP...09..042M}. The combination of the nuclear bulge and the boxy bulge is responsible for the Galactic center GeV excess. Also included are the Galactic disk (red dot) luminosity predicted by~\citet{2018NatAs...2..819B} and the M31 galaxy (purple star)~\citep{2017ApJ...836..208A}. The dot-dashed line shows the $\gamma$-ray luminosity-to-stellar-mass relation implied for the nuclear bulge and the boxy bulge, which is $2 \times 10^{27}$ erg s$^{-1}$ M$_\odot^{-1}$. 

As can be seen in Figure~\ref{fig:stellar_mass}, the luminosities of the detected sample of GCs exceed the luminosities expected based on the bulge correlations. In total, the GC samples have a stellar mass of $\sim 1.4\times 10^7$ M$_\odot$ and a $\gamma$-ray luminosity of $\sim 1.5\times 10^{36}$ erg s$^{-1}$. This means that GCs systematically emit $\sim$ 50 times more $\gamma$ rays per stellar mass than other objects such as the nuclear bulge and the Galactic bulge. The GCs have long been known for producing MSPs efficiently~\citep{2005ASPC..328..147C}. On average GCs make up $\sim 0.05\%$ of the total number of stars in the Milky Way~\citep{2019ApJ...877..122Y}, but more than one-third of the known MPSs are found in these systems~\citep{2005AJ....129.1993M}. Our observations support this scenario. This is also consistent with the larger stellar densities and larger stellar encounter rates in GCs than in the Galactic bulge. We also note in passing that a large fraction of $\gamma$-ray-detected GCs are located in the Galactic bulge region (see Figure~\ref{fig:all_sky_distribution}), which means that it is possible that at least some of the unresolved MSPs contributing to the Fermi GeV excess are hosted by GCs in the Galactic bulge region.

\subsection{TeV observations of globular clusters}

\begin{figure}
    \centering
    \includegraphics{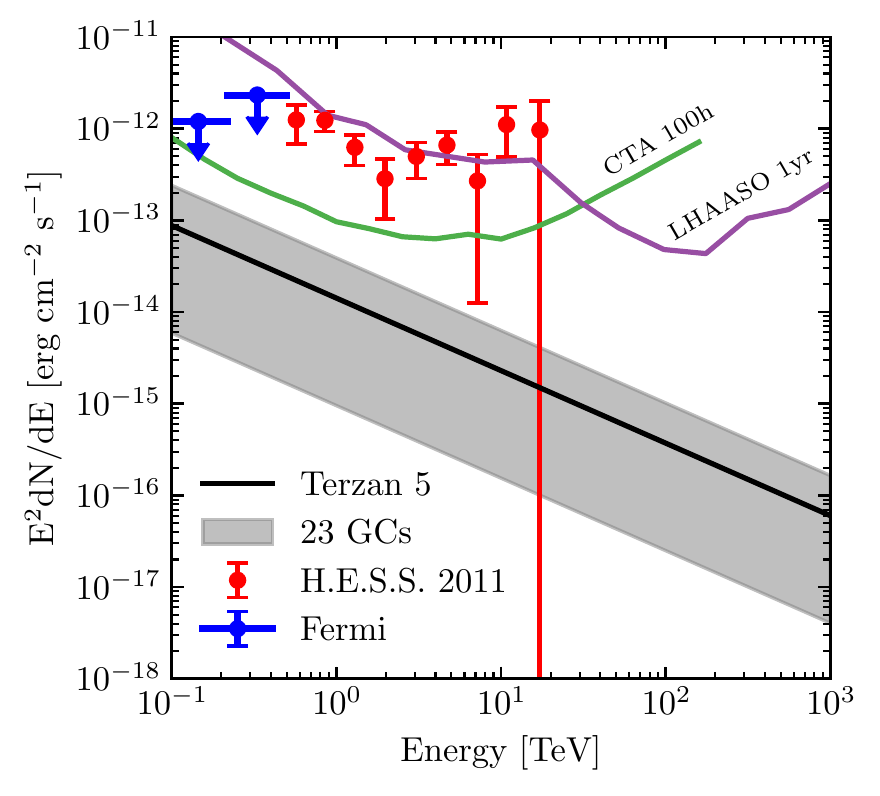}
    \caption{Extrapolated spectra for Terzan 5 (black line) and 23 GCs with the IC component detected (grey region). The H.E.S.S. measurement of Terzan 5 is shown by red dots with error bars. The {\it Fermi}-LAT only find upper limits (blue arrows) for Terzan 5 in this energy range. Also included are the sensitivities for 100-hour CTA South (green line) and 1-year LHAASO (purple line).}
    \label{fig:TeV}
\end{figure}

Using the universal two-component fit, we identified a power law component with a slope of $2.79 \pm  0.25$ from the spectra of $\gamma$-ray-detected GCs. The power law component can be plausibly explained by IC emission from GCs. The fact that this power law component is rather soft may explain why most GCs are not detected in the TeV energy range. In order to explore this more closely, we extrapolate the high energy tail of the GCs spectra to TeV energies in Figure~\ref{fig:TeV}. In this figure, the black line shows the extrapolated fluxes for Terzan 5, and the gray band shows the range of extrapolated fluxes for the other 23 GCs with a detected IC component. Above 100 GeV, {\it Fermi}-LAT only find upper limits (blue arrows) for Terzan 5. The red dots are the H.E.S.S. measurements from the direction of Terzan 5. It is interesting to note that the extrapolated spectrum for Terzan 5 is about one order of magnitude lower than the H.E.S.S. measurements from the same object. This discrepancy might be explained by the fact that the $\gamma$-ray source reported by H.E.S.S is misaligned with the center of Terzan 5 so that this association could be a chance coincidence. However, such a coincidence with known objects has been estimated to be improbable ($\sim 10^{-4}$)~\citep{2011A&A...531L..18H}. If the H.E.S.S. source is indeed associated to Terzan 5, it could be that $e^\pm$ injection spectrum from MSPs has a spectral break at approximately 1 TeV. Note that a substantial fraction of stars in Terzan 5 have been identified as young and centrally concentrated~\citep{2016ApJ...828...75F,2020BAAA..61R...90G}, which could lead to a larger number of younger pulsars. The H.E.S.S. measurements could be explained if these young pulsars have higher energy $e^\pm$ cutoffs. Therefore, Terzan 5 may not be representative compared to other GCs which are dominated by old stellar systems. However, \citet{2019AJ....158...14N} also find that the abundance variations among Terzan 5 is indeed consistent with a regular globular cluster. Alternatively, the TeV $\gamma$ rays could originate from sources other than MSPs (e.g., hadronic emission from supernova remnants). Further investigation of those scenarios, though very interesting, is beyond the scope of this work. 

We also include in Figure~\ref{fig:TeV} the sensitivities to point-like sources for the next generation $\gamma$ ray observatories. The green line shows the sensitivity for the Cherenkov Telescope Array (CTA)$-$South assuming 100 hours of observation time. The purple line shows the 1-year sensitivity of the Large High Altitude Air Shower Observatory (LHAASO). The extrapolated IC fluxes are close to the 100-hour CTA sensitivity. It is clear that it will be difficult for the next-generation TeV $\gamma$-ray telescopes to actually detect each individual GC considered in our study. This might require a much more ambitious observation strategy that increases the sensitivity by factor of a few at the TeV energy range.  Efforts to measure the diffuse IC emission from the putative MSP population responsible for the Fermi GeV excess have been made and are very encouraging; see~\citet{2019PhRvD..99l3020S} and~\citet{2021arXiv210205648M}. Alternatively,~\citet{2016MNRAS.458.1083B} studied TeV $\gamma$-ray emission from MSPs taking into account the advection of $e^\pm$ with the wind from the GC. They showed that CTA can constrain models incorporating such effects. 

\section{Summary and Conclusions}\label{sec:conclusion}

We have reanalyzed \textit{Fermi}-LAT data in the energy range between 300 MeV and 500 GeV from the direction of 157 GCs in the~\citet{1996AJ....112.1487H} catalog. Using the same data cuts adopted in the construction of the 4FGL catalog, we confirmed the detection of 30 GCs in $\gamma$ rays, and updated the $\gamma$-ray spectral parameters for the sample of detected objects. We also estimated the 95\% C.L. luminosity upper limits for the sample of 127 undetected GCs in the 4FGL catalog. The main objective of our reanalysis was to find evidence for IC emission from $e^{\pm}$ injected by MSPs in GCs. This was done using two different methodologies. First, we searched for correlations of the $\gamma$-ray luminosities with other GCs properties. Second, we performed a spectral analysis of the GCs with a universal fit method that enhances the sensitivity to the high energy tail of the spectra. Specifically:

\begin{itemize}
    \item[1.] Using an expectation-maximization algorithm that properly incorporates null detections ($L_\gamma$ upper limits) in the pipeline, we found a correlation between $L_\gamma$ and the GCs' total photon field energy density $u_\mathrm{Total}$ of the form, \begin{equation}
    \log\left(\frac{L_\gamma}{\mathrm{erg}\;\mathrm{s}^{-1}}\right) = (0.59 \pm 0.09)\log\left(\frac{u_\mathrm{Total}}{\mathrm{eV}\;\mathrm{cm}^{-3}}\right) + (32.97 \pm 0.19).
    \end{equation}
    Using the Kendall $\tau$ coefficient as the test statistic we determined this correlation to have a $3.8\sigma$ significance. The total photon field is dominated by the stellar light of the GCs  ($u_{\rm GC}$), and we find a much weaker correlation (below the 2$\sigma$ level) when only the photon field at the location of the GC ($u_{\rm MW}$) is used. In addition, we obtained a strong correlation (at 6.4$\sigma$ significance) between $L_\gamma$ and the stellar encounter rate $\Gamma_c$, which is given by 
    \begin{equation}
    \log\left(\frac{L_\gamma}{\mathrm{erg}\;\mathrm{s}^{-1}}\right) = (0.39 \pm 0.10)\log\left(\Gamma_c\right) + (32.99 \pm 0.26).
    \end{equation}
    Finally, we found only weak evidence (below the 2$\sigma$ level) for a correlations between $L_\gamma$ and the stellar metallicity [Fe/H].
    \item[2.] We revealed a hidden correlation between $u_\mathrm{Total}$ and $\Gamma_c$, which implies that the $L_\gamma$--$u_{\rm Total}$ and $L_\gamma$--$\Gamma_c$ correlations are not entirely independent. However, as described below, we find spectral evidence for IC emission. The correlation results are consistent with there being both a significant role for dynamical formation of MSPs in GCs and the for the presence of a significant contribution of IC to the observed $\gamma$-ray luminosity.
    \item[3.] We applied a universal spectral fit to the sample of 30 GCs in the 4FGL catalog and searched for evidence of an IC component on top of a curvature radiation model--accounting for the MSPs prompt emission in the GCs. We found that the extra power-law IC component is preferred at the 8.2$\sigma$ significance over the curvature radiation model only. The best-fit power law index of the IC component was found to be $2.79 \pm 0.25$. This implies a power-law $e^\pm$ spectrum with an index of $4.58 \pm 0.50$ and a minimum energy as low as $\sim$ 10 GeV. 
    \item[4.] We estimated the $e^\pm$ injection efficiency $f_{e^\pm}$ for MSPs residing in GCs. We determined the IC $\gamma$-ray luminosities over 300 MeV to 500 GeV, which roughly corresponds to $e^\pm$ energies from 10 GeV to 1 TeV. We found the fraction of MSP spin-down energy injected to $e^\pm$ is comparable to or slightly smaller than that injected to $\gamma$ rays, $f_{e^\pm} \lesssim f_\gamma$ and is at $\lesssim 10\%$ level. This parameter has been estimated in different environments, such as {nearby pulsars~\citep{2017PhRvD..96j3013H, 2018PhRvD..98d3005H, 2021arXiv210400014H}}, the Galactic center~\citep{2013MNRAS.435L..14B}, individual GCs~\citep{2019MNRAS.484.2876M}, and galaxies with low specific star formation rate~\citep{2020arXiv200508982S}. Our results provide new insights into the $f_{e^\pm}$ parameter based on the universal properties of $\gamma$-ray-detected GCs in the Milky Way.
\end{itemize}

In summary, our analysis reveals strong evidence for {soft} IC emission in \textit{Fermi}-LAT GCs. This is indicative of $e^\pm$ injected by MSPs hosted by such systems. Although the {\it Fermi}-LAT sensitivity for energies larger than 10 GeV is not sufficiently high to claim a detection in each individual GC, we employed a universal fit method with the bin-by-bin spectra of the sample of detected objects and were able to increase the sensitivity to the IC component. Our results also explain why it is difficult to detect GCs with TeV $\gamma$-ray telescopes: we have obtained a very soft spectra for the high energy tail of the GC population. It is possible that with a more aggressive observation campaign such objects could be detected by forthcoming TeV telescopes [see~\citep{2018MNRAS.473..897N} for a recent sensitivity analysis] such as CTA~\citep{2019scta.book.....C} and LHAASO~\citep{2019arXiv190502773B}. Globular clusters remain some of the most important systems within which to search for and study millisecond pulsars. We have shown the potential of extracting critical knowledge from $\gamma$-ray data of globular clusters with advanced statistical tools and intensive modelling. 

\section*{Acknowledgements}
We thank Shin'ichiro Ando, Holger Baumgardt and Zhe Li for discussions. The work of D.S.\ is supported by the U.S.\ Department of Energy under the award number DE-SC0020262. O.M. acknowledges support by JSPS KAKENHI Grant Numbers JP17H04836, JP18H04340, JP18H04578, and JP20K14463. The work of S.H.\ is supported by the U.S.\ Department of Energy under the award number DE-SC0020262 and NSF Grant numbers AST-1908960 and PHY-1914409. This work was supported by World Premier International Research Center Initiative (WPI Initiative), MEXT, Japan. R.M.C. acknowledges support from the Australian Government through the Australian Research Council for grant DP190101258 shared with Prof.~Mark Krumholz at the ANU. D.M.N. acknowledges support from NASA under award Number 80NSSC19K0589.

\section*{Data Availability}

The {\it Fermi}-LAT data analysed in this article were accessed from \url{https://heasarc.gsfc.nasa.gov/FTP/fermi/data/lat/weekly/}. The derived data generated in this research are available in the article.



\bibliographystyle{mnras}
\bibliography{main} 




\appendix

\section{Globular clusters not detected in the 4FGL catalog}\label{appx:nodetect}

Table~\ref{tab:nondetect} reports the parameters and $\gamma$-ray analysis results for 127 additional GCs in the~\citet{1996AJ....112.1487H} catalog with no counterpart detected in the 4FGL. For fluxes and $L_\gamma$, we report their 95\% C.L. upper limits by placing a putative point source at the sky location of the GC and running a maximum-likelihood procedure in which we assume a power-law spectrum with a spectral slope of $\Gamma = -2$.

Importantly, we included the $L_\gamma$ upper limits in the correlation analysis shown in Section~\ref{sec:correlation}. The EM algorithm uses the $L_\gamma$ upper limits to perform maximum likelihood estimates and find the best-fit parameters for the correlations with the other GC observables. The significance of the correlations is estimated with the generalized Kendall $\tau$ coefficient as the test statistic, which also includes the luminosity upper limits. 

\begin{table*}
\centering
\caption{Parameters and data analysis results of 127 GCs not detected in the 4FGL. Notations same as Table~\ref{tab:pars}.}\label{tab:nondetect}
\begin{tabular}{lcccccccr}
\hline
Name &  $\Gamma_c$& [Fe/H]  & $M_*$ & $u_\text{MW}$ & $u_\text{Total}$ & $R_\odot$ & Flux 95\% UL & $L_\gamma$ 95\% UL \\
 & &  & ($10^5 M_\odot$) & (eV cm$^{-3}$) & (eV cm$^{-3}$) & (kpc) & (10$^{-8}$ ph cm$^{-2}$ s$^{-1}$)  & (10$^{34}$ erg s$^{-1}$)\\
\hline
2MS-GC02 & ... & -1.08 & 0.26 & 2.56 & 17.79 & 4.90 & < 0.57 & < 2.33\\
AM 1 & 0.01 & -1.70 & 0.32 & 0.25 & 24.57 & 123.30 & < 0.04 & < 457.85\\
AM 4 & 0.00 & -1.30 & 0.01 & 0.27 & 1.77 & 32.20 & < 0.09 & < 38.62\\
Arp 2 & 0.01 & -1.75 & 0.37 & 0.28 & 2.47 & 28.60 & < 0.06 & < 27.28\\
BH 176 & 0.15 & 0.00 & 0.48 & 0.34 & 3.06 & 18.90 & < 0.16 & < 23.19\\
BH 261 & ... & -1.30 & 0.31 & 3.29 & 11.51 & 6.50 & < 0.07 & < 1.54\\
Djorg 1 & 9.04 & -1.51 & 0.78 & 0.97 & 13.81 & 13.70 & < 0.16 & < 13.22\\
Djorg 2 & 46.40 & -0.65 & 0.56 & 4.01 & 34.01 & 6.30 & < 0.09 & < 1.72\\
E 3 & 0.08 & -0.83 & 0.03 & 0.46 & 0.99 & 8.10 & < 0.16 & < 3.92\\
Eridanus & 0.03 & -1.43 & 0.11 & 0.25 & 28.17 & 90.10 & < 0.07 & < 271.32\\
ESO280-SC06 & ... & -1.80 & 0.26 & 0.33 & 4.54 & 21.40 & < 0.06 & < 17.45\\
ESO452-SC11 & 1.72 & -1.50 & 0.06 & 1.44 & 9.94 & 8.30 & < 0.10 & < 4.18\\
FSR 1735 & ... & ... & 0.58 & 1.47 & 173.86 & 9.80 & < 0.23 & < 5.79\\
HP 1 & 2.75 & -1.00 & 1.62 & 4.89 & 6.99 & 8.20 & < 0.17 & < 4.27\\
IC 1257 & ... & -1.70 & 0.65 & 0.30 & 8.01 & 25.00 & < 0.21 & < 44.16\\
IC 1276 & 7.78 & -0.75 & 0.72 & 1.58 & 5.88 & 5.40 & < 0.17 & < 1.47\\
IC 4499 & 0.94 & -1.53 & 1.29 & 0.31 & 15.50 & 18.80 & < 0.09 & < 15.22\\
Ko 1 & ... & ... & ... & 0.25 & 2.94 & 48.30 & < 0.03 & < 68.99\\
Ko 2 & ... & ... & ... & 0.25 & 1.89 & 34.70 & < 0.02 & < 32.76\\
Liller 1 & 391.00 & -0.33 & 6.61 & 4.38 & >4.38 & 8.20 & < 0.56 & < 9.34\\
Lynga 7 & ... & -1.01 & 1.02 & 1.24 & 17.13 & 8.00 & < 0.06 & < 2.39\\
NGC 1261 & 17.90 & -1.27 & 1.74 & 0.30 & 149.73 & 16.30 & < 0.03 & < 7.73\\
NGC 1851 & 1910.00 & -1.18 & 2.82 & 0.30 & 433.15 & 12.10 & < 0.14 & < 6.77\\
NGC 2298 & 5.37 & -1.92 & 0.54 & 0.30 & 18.54 & 10.80 & < 0.04 & < 8.44\\
NGC 2419 & 3.37 & -2.15 & 14.45 & 0.25 & 388.12 & 82.60 & < 0.08 & < 202.00\\
NGC 288 & 1.23 & -1.32 & 1.20 & 0.35 & 5.64 & 8.90 & < 0.02 & < 2.96\\
NGC 3201 & 8.45 & -1.59 & 1.45 & 0.52 & 5.73 & 4.90 & < 0.12 & < 0.90\\
NGC 362 & 569.00 & -1.26 & 3.39 & 0.42 & 184.01 & 8.60 & < 0.06 & < 3.07\\
NGC 4147 & 14.90 & -1.80 & 0.29 & 0.28 & 67.12 & 19.30 & < 0.03 & < 11.28\\
NGC 4372 & 2.28 & -2.17 & 2.19 & 0.65 & 5.13 & 5.80 & < 0.14 & < 1.62\\
NGC 4590 & 4.58 & -2.23 & 1.29 & 0.40 & 20.79 & 10.30 & < 0.03 & < 3.16\\
NGC 4833 & 25.00 & -1.85 & 2.04 & 0.67 & 17.40 & 6.60 & < 0.10 & < 1.45\\
NGC 5024 & 28.50 & -2.10 & 4.27 & 0.30 & 93.40 & 17.90 & < 0.06 & < 14.33\\
NGC 5053 & 0.15 & -2.27 & 0.62 & 0.31 & 4.20 & 17.40 & < 0.09 & < 10.59\\
NGC 5272 & 167.00 & -1.50 & 3.63 & 0.35 & 35.36 & 10.20 & < 0.04 & < 3.41\\
NGC 5286 & 569.00 & -1.69 & 3.80 & 0.47 & 308.67 & 11.70 & < 0.13 & < 6.82\\
NGC 5466 & 0.18 & -1.98 & 0.52 & 0.32 & 6.45 & 16.00 & < 0.02 & < 7.68\\
NGC 5634 & 24.50 & -1.88 & 2.00 & 0.29 & 84.71 & 25.20 & < 0.04 & < 18.05\\
NGC 5694 & 205.00 & -1.98 & 3.89 & 0.27 & 444.24 & 35.00 & < 0.10 & < 45.81\\
NGC 5824 & 1220.00 & -1.91 & 8.51 & 0.27 & 897.80 & 32.10 & < 0.03 & < 28.45\\
NGC 5897 & 1.16 & -1.90 & 1.55 & 0.49 & 10.12 & 12.50 & < 0.05 & < 5.04\\
NGC 5927 & 251.00 & -0.49 & 3.47 & 1.08 & 58.71 & 7.70 & < 0.05 & < 2.00\\
NGC 5946 & 122.00 & -1.29 & 1.15 & 0.85 & 50.13 & 10.60 & < 0.06 & < 3.34\\
NGC 5986 & 56.10 & -1.59 & 3.31 & 0.79 & 130.52 & 10.40 & < 0.03 & < 3.01\\
NGC 6101 & 1.21 & -1.98 & 1.29 & 0.38 & 28.76 & 15.40 & < 0.04 & < 7.32\\
NGC 6121 & 34.50 & -1.16 & 0.89 & 0.91 & 3.01 & 2.20 & < 0.07 & < 0.21\\
NGC 6144 & 3.61 & -1.76 & 0.48 & 1.14 & 11.98 & 8.90 & < 0.06 & < 2.38\\
NGC 6171 & 9.65 & -1.02 & 0.78 & 1.08 & 13.42 & 6.40 & < 0.06 & < 1.17\\
NGC 6205 & 89.90 & -1.53 & 4.47 & 0.46 & 48.74 & 7.10 & < 0.04 & < 1.57\\
NGC 6229 & 49.90 & -1.47 & 2.95 & 0.27 & 677.70 & 30.50 & < 0.01 & < 51.41\\
NGC 6235 & 7.11 & -1.28 & 1.12 & 0.92 & 18.12 & 11.50 & < 0.30 & < 6.36\\
NGC 6254 & 42.80 & -1.56 & 1.86 & 0.98 & 14.51 & 4.40 & < 0.05 & < 0.60\\
NGC 6256 & 242.00 & -1.02 & 1.07 & 1.66 & 53.00 & 10.30 & < 0.11 & < 3.63\\
NGC 6273 & 246.00 & -1.74 & 6.46 & 1.93 & 136.93 & 8.80 & < 0.06 & < 2.28\\
NGC 6284 & 797.00 & -1.26 & 2.40 & 0.56 & 184.37 & 15.30 & < 0.03 & < 6.50\\
NGC 6287 & 52.30 & -2.10 & 1.32 & 1.49 & 85.63 & 9.40 & < 0.21 & < 4.99\\
NGC 6293 & 1220.00 & -1.99 & 1.38 & 1.99 & 87.63 & 9.50 & < 0.12 & < 4.45\\
NGC 6325 & 189.00 & -1.25 & 0.72 & 2.42 & 82.73 & 7.80 & < 0.13 & < 1.89\\
NGC 6333 & 153.00 & -1.77 & 3.16 & 1.82 & 87.91 & 7.90 & < 0.04 & < 1.79\\
NGC 6342 & 83.70 & -0.55 & 0.60 & 1.82 & 38.20 & 8.50 & < 0.16 & < 3.06\\
NGC 6352 & 12.50 & -0.64 & 0.55 & 1.40 & 6.23 & 5.60 & < 0.12 & < 0.96\\
NGC 6355 & 130.00 & -1.37 & 1.17 & 2.74 & 117.16 & 9.20 & < 0.13 & < 3.13\\
NGC 6356 & 110.00 & -0.40 & 3.80 & 0.56 & 203.09 & 15.10 & < 0.06 & < 6.52\\
NGC 6362 & 3.57 & -0.99 & 1.07 & 0.75 & 8.27 & 7.60 & < 0.03 & < 1.69\\
\hline
\end{tabular}
\end{table*}

\begin{table*}
\centering
\contcaption{}
\begin{tabular}{lcccccccr}
\hline
Name &  $\Gamma_c$& [Fe/H]  & $M_*$ & $u_\text{MW}$ & $u_\text{Total}$ & $R_\odot$ & Flux & $L_\gamma$ \\
 & &  & ($10^5 M_\odot$) & (eV cm$^{-3}$) & (eV cm$^{-3}$) & (kpc) & (10$^{-8}$ ph cm$^{-2}$ s$^{-1}$)  & (10$^{34}$ erg s$^{-1}$)\\
\hline
NGC 6366 & 4.06 & -0.59 & 0.59 & 1.04 & 2.26 & 3.50 & < 0.11 & < 0.68\\
NGC 6380 & 96.20 & -0.75 & 3.02 & 1.64 & 97.36 & 10.90 & < 0.12 & < 6.66\\
NGC 6401 & 60.20 & -1.02 & 2.75 & 1.80 & 22.57 & 10.60 & < 0.11 & < 3.65\\
NGC 6426 & 2.68 & -2.15 & 0.63 & 0.32 & 29.15 & 20.60 & < 0.04 & < 11.25\\
NGC 6453 & 183.00 & -1.50 & 2.34 & 1.57 & 210.77 & 11.60 & < 0.12 & < 6.25\\
NGC 6496 & 2.02 & -0.46 & 0.83 & 0.94 & 39.16 & 11.30 & < 0.10 & < 3.58\\
NGC 6517 & 661.00 & -1.23 & 3.02 & 1.00 & 419.34 & 10.60 & < 0.22 & < 7.06\\
NGC 6522 & 467.00 & -1.34 & 2.14 & 4.15 & 64.33 & 7.70 & < 0.08 & < 2.51\\
NGC 6535 & 1.42 & -1.79 & 0.13 & 1.11 & 6.87 & 6.80 & < 0.27 & < 2.11\\
NGC 6539 & 271.00 & -0.63 & 2.40 & 1.34 & 38.89 & 7.80 & < 0.24 & < 2.39\\
NGC 6540 & 263.00 & -1.35 & 0.43 & 2.62 & >2.62 & 5.30 & < 0.13 & < 1.42\\
NGC 6544 & 462.00 & -1.40 & 1.15 & 1.43 & 22.80 & 3.00 & < 0.13 & < 0.37\\
NGC 6553 & 103.00 & -0.18 & 3.02 & 3.40 & 66.75 & 6.00 & < 0.06 & < 1.38\\
NGC 6558 & 109.00 & -1.32 & 0.39 & 3.23 & 7.50 & 7.40 & < 0.07 & < 1.53\\
NGC 6569 & 72.80 & -0.76 & 2.40 & 1.49 & 169.48 & 10.90 & < 0.08 & < 4.65\\
NGC 6584 & 22.00 & -1.50 & 1.17 & 0.56 & 117.73 & 13.50 & < 0.13 & < 7.88\\
NGC 6624 & 1080.00 & -0.44 & 0.62 & 2.37 & 79.61 & 7.90 & < 0.07 & < 2.21\\
NGC 6626 & 688.00 & -1.32 & 2.82 & 2.41 & 27.22 & 5.50 & < 0.08 & < 1.17\\
NGC 6637 & 92.40 & -0.64 & 1.48 & 1.78 & 86.29 & 8.80 & < 0.09 & < 5.98\\
NGC 6638 & 103.00 & -0.95 & 1.74 & 1.60 & 143.62 & 9.40 & < 0.07 & < 3.50\\
NGC 6642 & 112.00 & -1.26 & 0.25 & 1.97 & 47.34 & 8.10 & < 0.04 & < 2.00\\
NGC 6656 & 92.40 & -1.70 & 4.07 & 1.26 & 12.92 & 3.20 & < 0.05 & < 0.35\\
NGC 6681 & 964.00 & -1.62 & 1.12 & 1.33 & 74.61 & 9.00 & < 0.12 & < 3.64\\
NGC 6712 & 33.40 & -1.02 & 1.20 & 1.44 & 31.07 & 6.90 & < 0.19 & < 2.60\\
NGC 6715 & 2030.00 & -1.49 & 15.85 & 0.29 & 765.61 & 26.50 & < 0.09 & < 27.82\\
NGC 6723 & 13.90 & -1.10 & 1.74 & 1.18 & 31.53 & 8.70 & < 0.11 & < 3.51\\
NGC 6749 & 38.50 & -1.60 & 0.79 & 1.17 & 21.91 & 7.90 & < 0.09 & < 1.83\\
NGC 6760 & 77.00 & -0.40 & 2.57 & 1.11 & 45.55 & 7.40 & < 0.10 & < 2.01\\
NGC 6779 & 25.90 & -1.98 & 1.66 & 0.48 & 40.35 & 9.40 & < 0.15 & < 2.52\\
NGC 6809 & 3.65 & -1.94 & 1.86 & 1.01 & 7.99 & 5.40 & < 0.10 & < 1.09\\
NGC 6864 & 370.00 & -1.29 & 3.98 & 0.32 & 663.99 & 20.90 & < 0.09 & < 14.80\\
NGC 6934 & 31.80 & -1.47 & 1.41 & 0.34 & 105.49 & 15.60 & < 0.05 & < 8.08\\
NGC 6981 & 4.12 & -1.42 & 0.69 & 0.34 & 40.01 & 17.00 & < 0.06 & < 8.68\\
NGC 7006 & 6.46 & -1.52 & 1.48 & 0.25 & 316.89 & 41.20 & < 0.07 & < 47.47\\
NGC 7089 & 441.00 & -1.65 & 5.01 & 0.39 & 191.32 & 11.50 & < 0.02 & < 4.30\\
NGC 7099 & 366.00 & -2.27 & 1.26 & 0.50 & 47.69 & 8.10 & < 0.14 & < 2.71\\
NGC 7492 & 0.86 & -1.78 & 0.29 & 0.25 & 8.61 & 26.30 & < 0.09 & < 34.99\\
Pal 1 & 0.05 & -0.65 & 0.02 & 0.30 & 2.82 & 11.10 & < 0.04 & < 2.79\\
Pal 10 & 75.80 & -0.10 & 0.55 & 0.90 & 11.97 & 5.90 & < 0.08 & < 1.05\\
Pal 11 & 12.40 & -0.40 & 0.11 & 0.48 & 14.90 & 13.40 & < 0.05 & < 7.89\\
Pal 12 & 0.54 & -0.85 & 0.06 & 0.32 & 1.41 & 19.00 & < 0.04 & < 11.78\\
Pal 13 & 0.00 & -1.88 & 0.03 & 0.27 & 13.18 & 26.00 & < 0.07 & < 20.02\\
Pal 14 & 0.00 & -1.62 & 0.15 & 0.25 & 3.18 & 76.50 & < 0.06 & < 263.46\\
Pal 15 & 0.02 & -2.07 & 0.42 & 0.25 & 7.18 & 45.10 & < 0.06 & < 57.78\\
Pal 2 & 457.00 & -1.42 & 2.29 & 0.25 & 323.49 & 27.20 & < 0.07 & < 21.58\\
Pal 3 & 0.03 & -1.63 & 0.23 & 0.25 & 23.67 & 92.50 & < 0.10 & < 319.08\\
Pal 4 & 0.03 & -1.41 & 0.28 & 0.25 & 51.34 & 108.70 & < 0.04 & < 352.19\\
Pal 5 & 0.01 & -1.41 & 0.18 & 0.30 & 1.12 & 23.20 & < 0.07 & < 15.55\\
Pal 6 & 40.20 & -0.91 & 1.00 & 4.07 & 23.00 & 5.80 & < 0.23 & < 2.53\\
Pal 8 & 3.03 & -0.37 & 0.58 & 0.79 & 25.72 & 12.80 & < 0.15 & < 5.03\\
Pyxis & ... & -1.20 & 0.23 & 0.25 & >0.25 & 39.40 & < 0.17 & < 67.61\\
Rup 106 & 0.36 & -1.68 & 0.35 & 0.29 & 16.78 & 21.20 & < 0.08 & < 18.51\\
Terzan 10 & 4430.00 & -1.00 & 2.45 & 3.42 & 10.99 & 5.80 & < 0.03 & < 0.94\\
Terzan 12 & 35.80 & -0.50 & 0.01 & 2.34 & 6.57 & 4.80 & < 0.12 & < 0.85\\
Terzan 3 & 0.89 & -0.74 & 0.58 & 1.36 & 4.20 & 8.20 & < 0.11 & < 3.29\\
Terzan 4 & ... & -1.41 & 0.76 & 4.12 & 5.07 & 7.20 & < 0.23 & < 1.94\\
Terzan 6 & 1300.00 & -0.56 & 0.89 & 4.23 & 298.38 & 6.80 & < 0.13 & < 2.35\\
Terzan 7 & 0.62 & -0.32 & 0.19 & 0.31 & 9.23 & 22.80 & < 0.04 & < 15.89\\
Terzan 8 & 0.08 & -2.16 & 0.54 & 0.29 & 6.49 & 26.30 & < 0.13 & < 24.96\\
Terzan 9 & 2.79 & -1.05 & 0.04 & 4.88 & 7.51 & 7.10 & < 0.10 & < 2.67\\
Ton 2 & 6.45 & -0.70 & 0.26 & 2.51 & 11.62 & 8.20 & < 0.13 & < 3.40\\
UKS 1 & 100.00 & -0.64 & 0.78 & 4.47 & >4.47 & 7.80 & < 0.45 & < 6.28\\
Whiting 1 & ... & -0.70 & 0.02 & 0.25 & 10.69 & 30.10 & < 0.05 & < 28.00\\
\hline
\end{tabular}
\end{table*}

\section{Kendall $\tau$ test}\label{appx:MC}

\begin{figure*}
    \centering
    \includegraphics[width=0.45\textwidth]{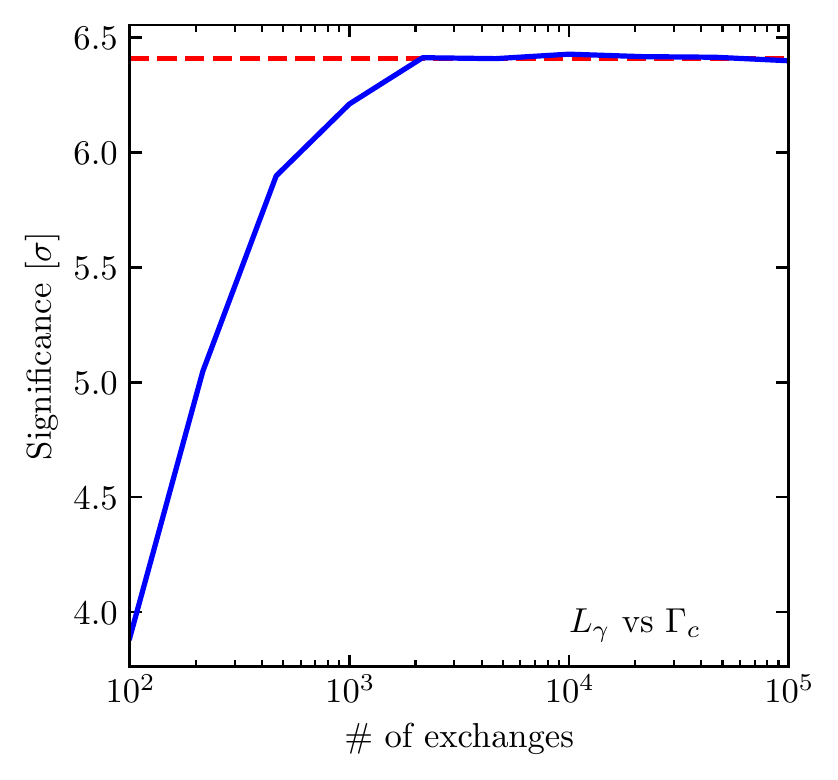}
    \includegraphics[width=0.45\textwidth]{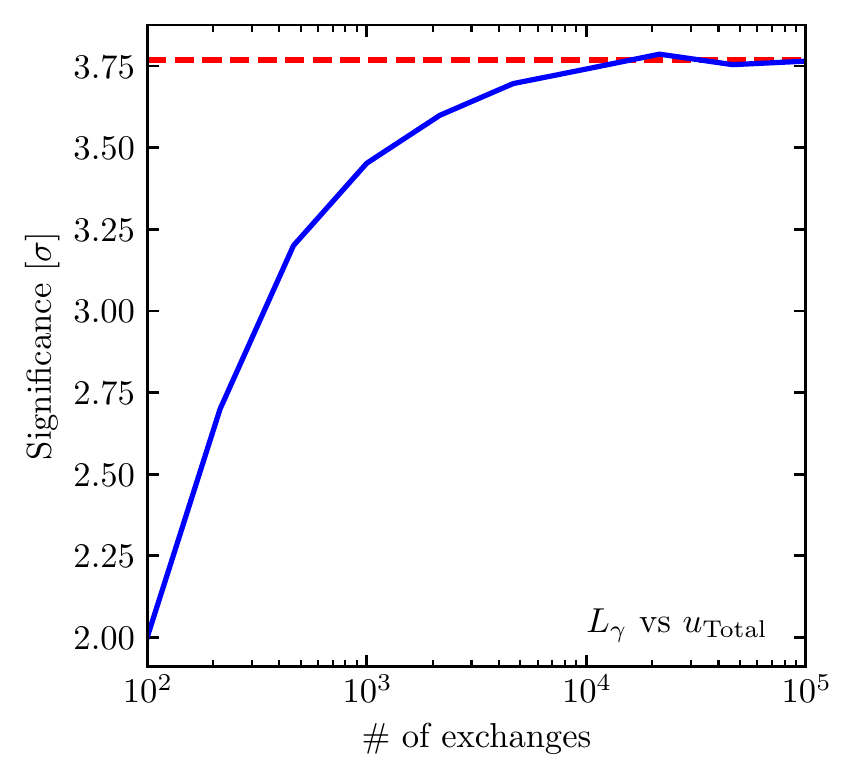}
    \includegraphics[width=0.45\textwidth]{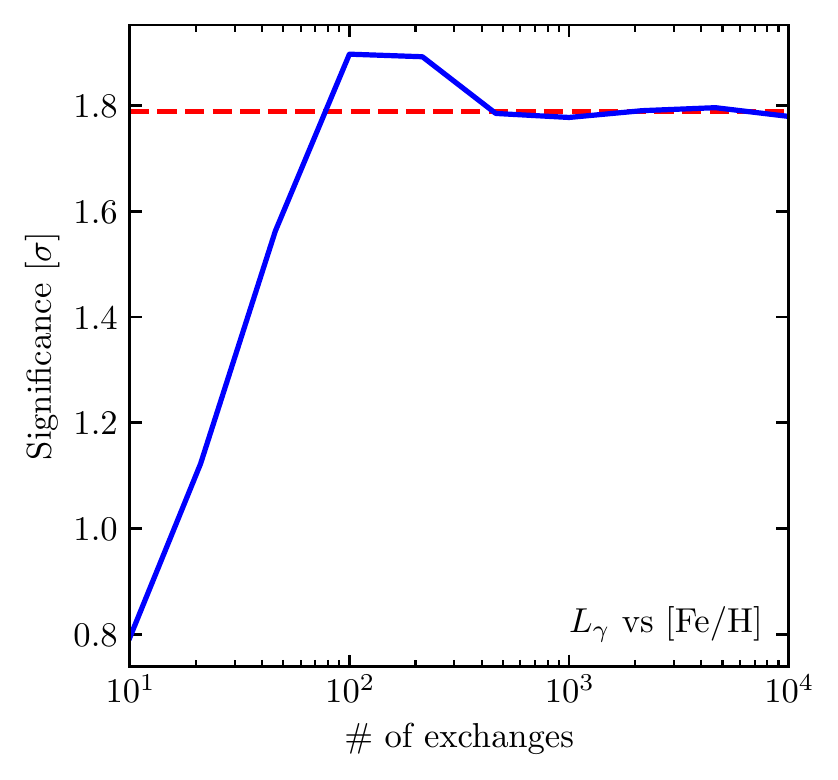}
    \includegraphics[width=0.45\textwidth]{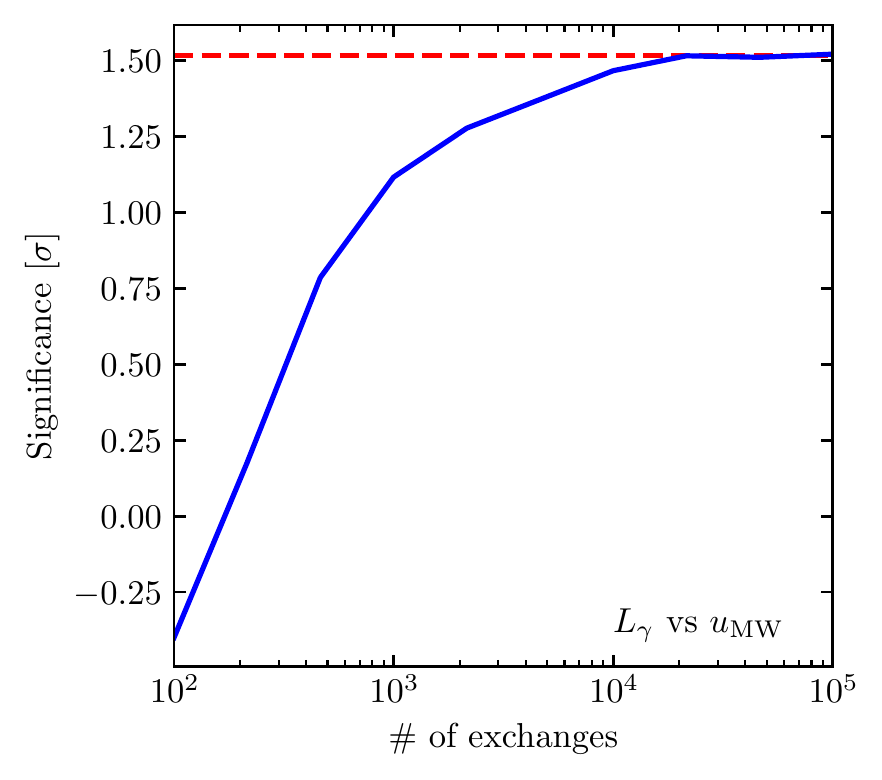}
    \caption{Tests of the number of $L_\gamma$ exchanges during the MC simulations. The significance of correlations increases with the number of exchanges until the number is large enough to generate fully randomized samples. For the four observables investigated, the typical number of exchanges needed to achieve converged significance is $\sim 10^4$.}
    \label{fig:exchange}
\end{figure*}

During the MC simulations that determine the significance of correlations between $L_\gamma$ and other observables, the null hypothesis samples are generated by repetitively exchanging $L_\gamma$ of two GCs while keeping their location fixed. A large number of exchanges is needed to fully randomize the data and fulfill the requirement of the null hypothesis. A previous study~\citep{2012ApJ...755..164A} that used this technique with star forming galaxies found that 1000 exchanges were enough in their case. In this section, we test different numbers of exchanges and their impact on the significance of the correlations. Figure~\ref{fig:exchange} shows the calculated significance of each correlation when different numbers of exchanges are used to generate null hypothesis samples. We find that the significance usually increases with the number of exchanges until some large exchange number. This is expected as a small number of exchanges will generate samples too close to the real data, which is unwanted for the null hypothesis. The dependence of significance on the exchanges disappears from a large number of exchanges when the samples are fully randomized and represents the null hypothesis. We find that about $10^4$ exchanges are needed to generate the null hypothesis samples and achieve converged significance for four correlations estimated.

\section{Spectral energy distributions}\label{appx:spectra}

Figure~\ref{fig:spectra} shows the best-fit spectra for 30 $\gamma$-ray-detected GCs from the individual fits. We report the $\gamma$-ray fluxes of the GCs from 300 MeV to 500 GeV in 9 logarithmic energy bins (black dots). The spectra are sorted by their spectral curvature (TS$_\mathrm{curve}$). For five of them (2MS-GC01, NGC 1904, NGC 6397, NGC 7078, and NGC 5904), we report the best-fit spectra from a power law since they have TS$_\mathrm{curve}$ < 4. For the rest of them, we present the best-fit spectra using a PLE function. In general, we find the cut-off energy $E_\mathrm{cut}$ to be around $\sim 1$ GeV.

We also performed a universal fit with the sample of detected GCs. Figure~\ref{fig:2comp} shows the best-fit spectra from the universal two-component fits described in Section~\ref{sec:spectra_global}. This comprises a curved CR component and a power law IC component. The shape of the two components are tied across all GCs, based on the parameters that minimize the $\chi^2_\mathrm{total}$ in the universal fit. The normalization factors of the two-component model are allowed to float. We report 19 GCs with both components detected (nonzero normalization factors). Four GCs are only fitted to the IC component. The rest of them are fitted to the CR component. For GCs with only one component detected, we include the 95\% upper limit on the normalization of the non-detected component.

\begin{figure*}
\centering
\includegraphics[width=0.24\textwidth]{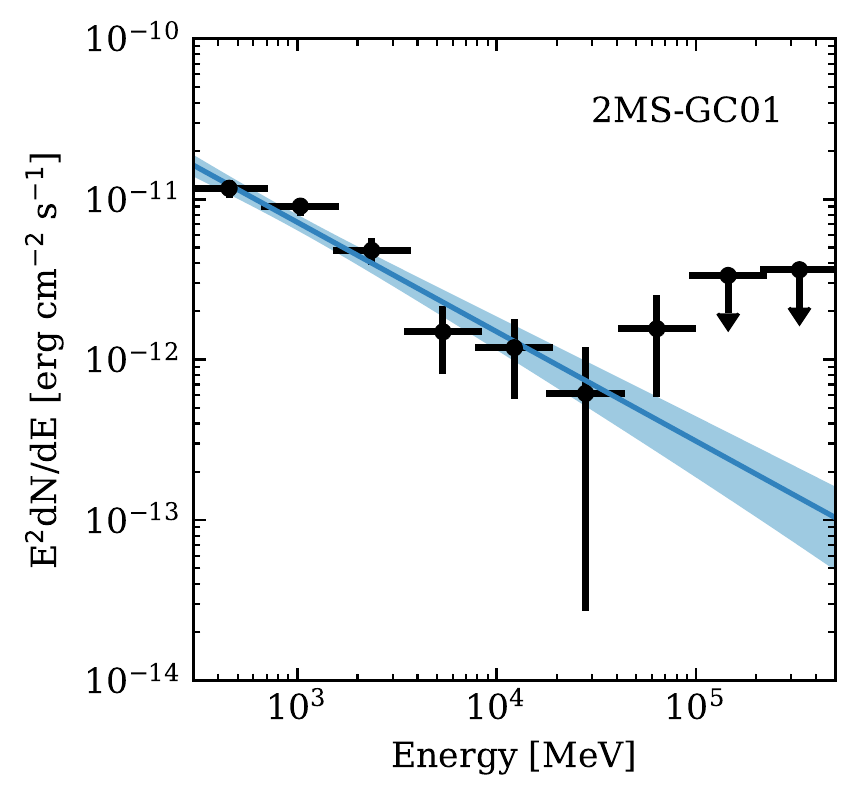}
\includegraphics[width=0.24\textwidth]{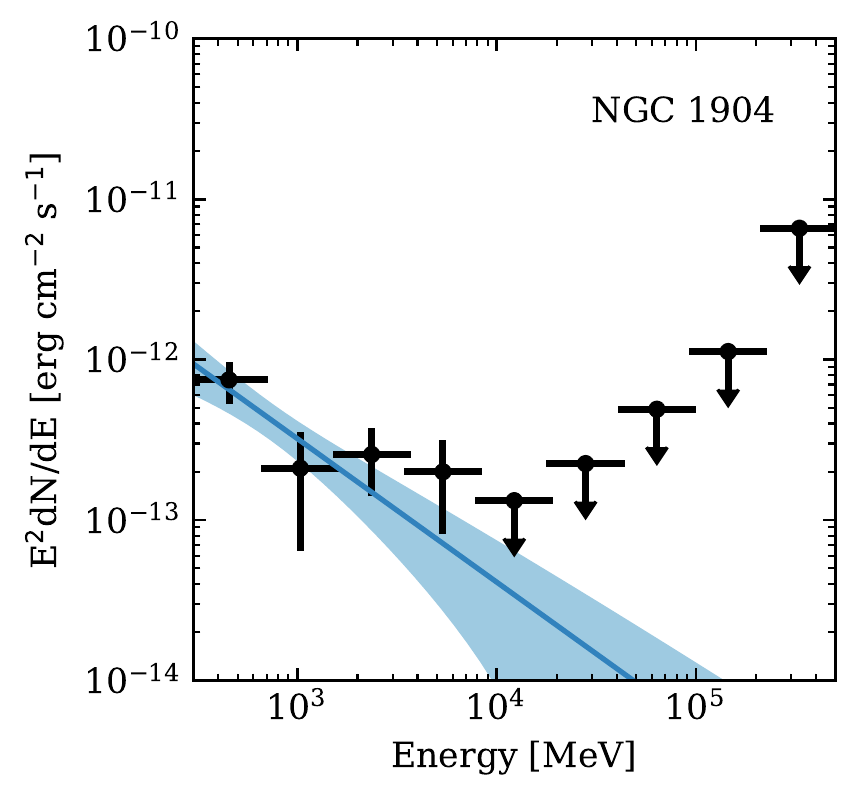}
\includegraphics[width=0.24\textwidth]{figures/spectra/PL_spectrum_16.pdf}
\includegraphics[width=0.24\textwidth]{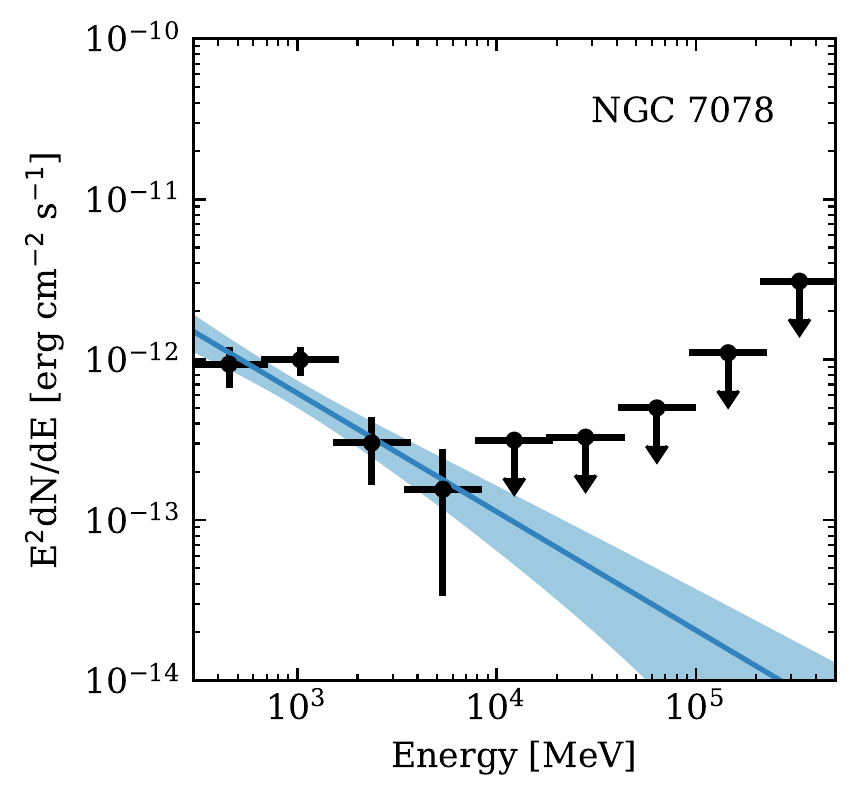}
\includegraphics[width=0.24\textwidth]{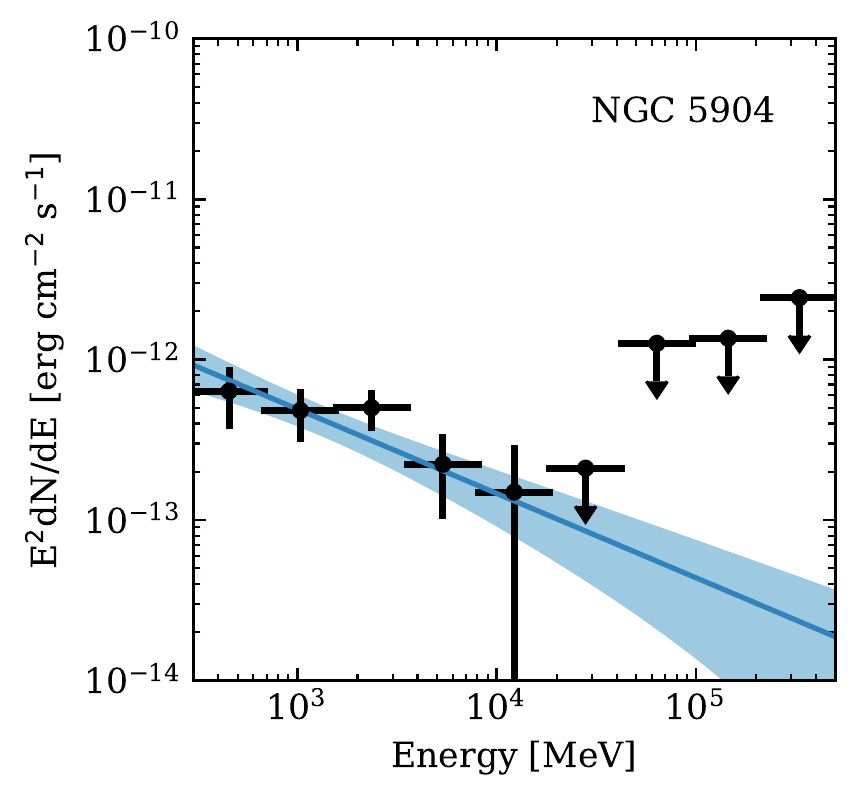}
\includegraphics[width=0.24\textwidth]{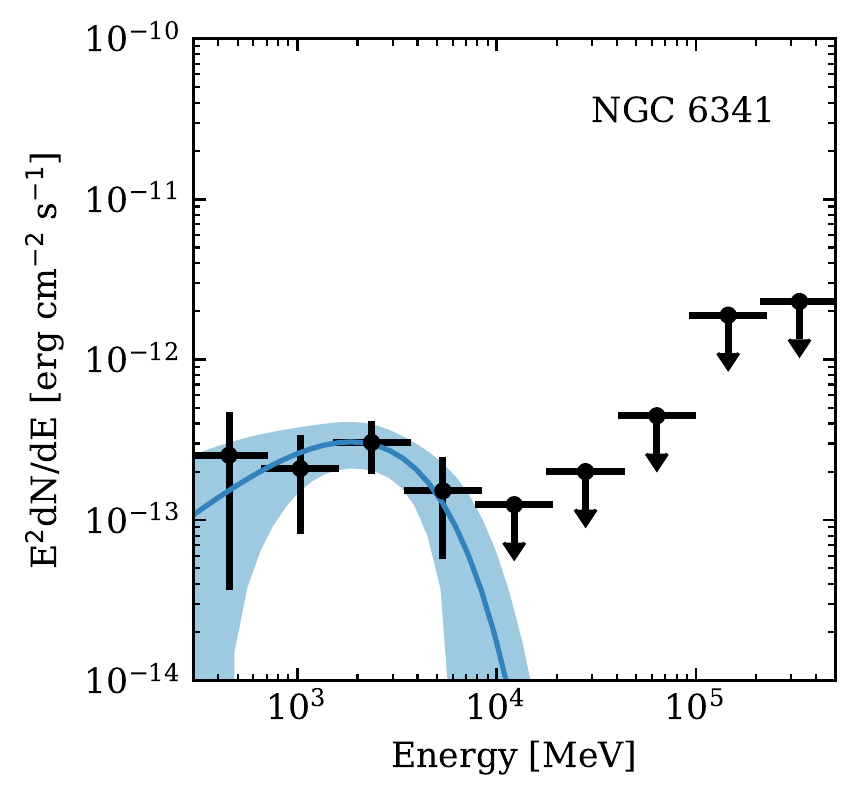}
\includegraphics[width=0.24\textwidth]{figures/spectra/PLE_spectrum_21.pdf}
\includegraphics[width=0.24\textwidth]{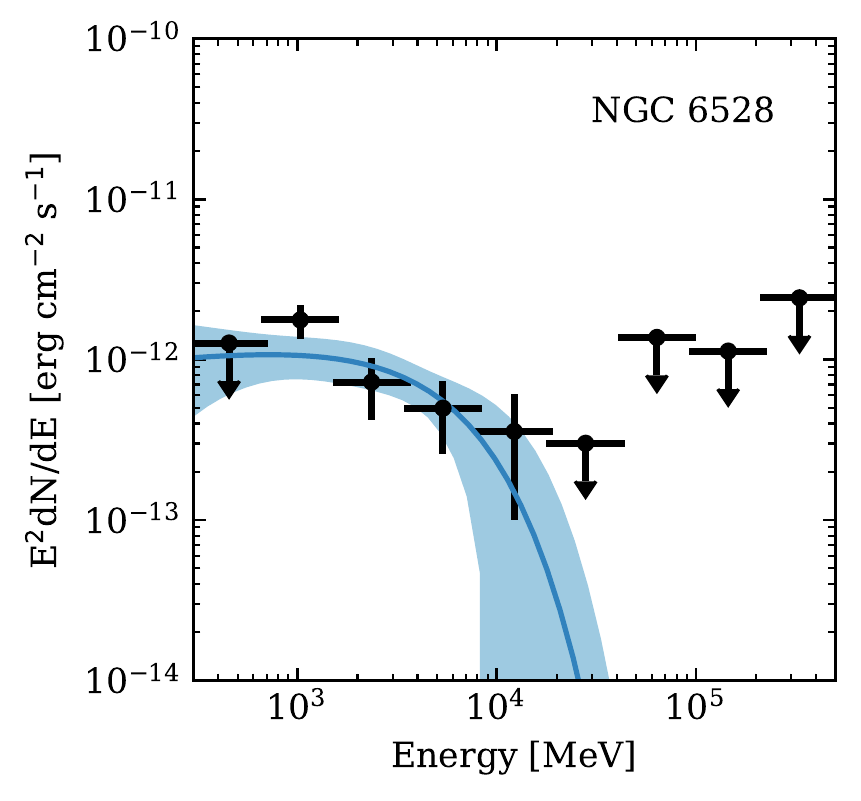}
\includegraphics[width=0.24\textwidth]{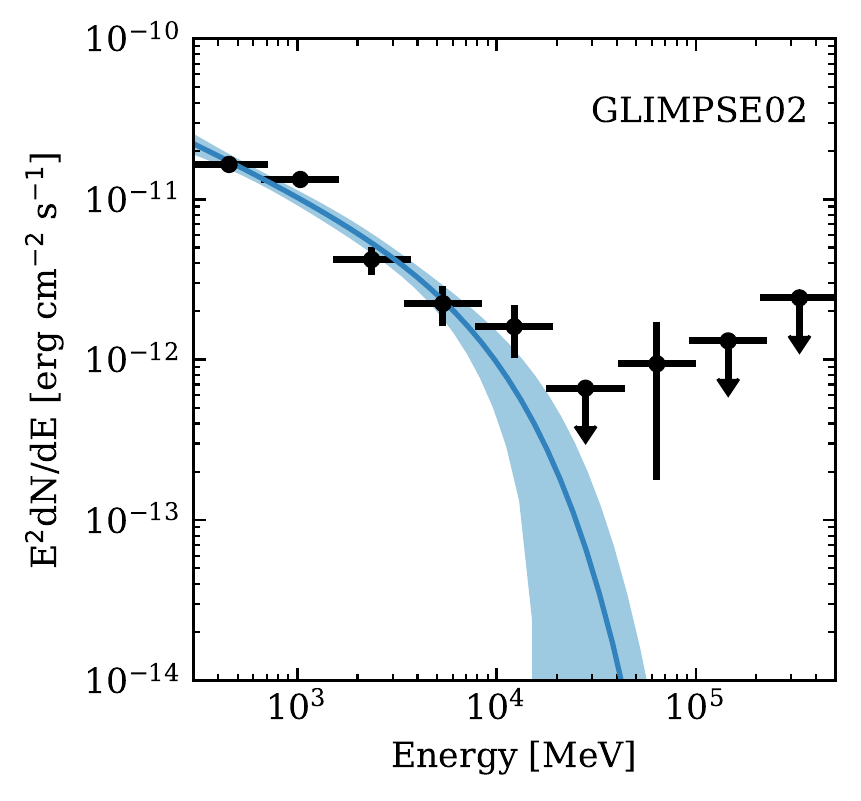}
\includegraphics[width=0.24\textwidth]{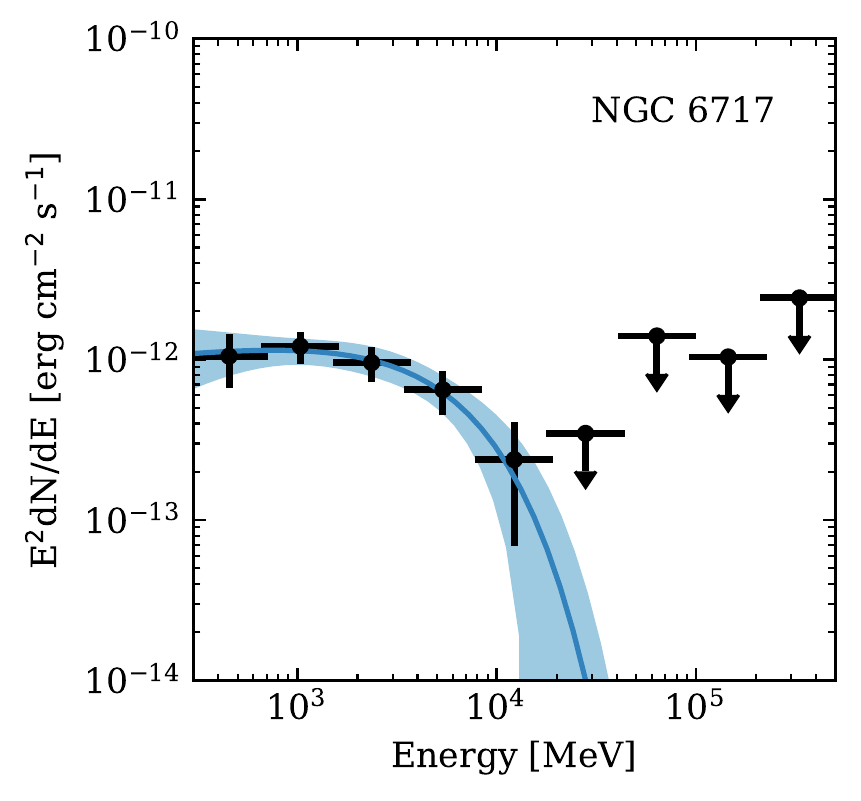}
\includegraphics[width=0.24\textwidth]{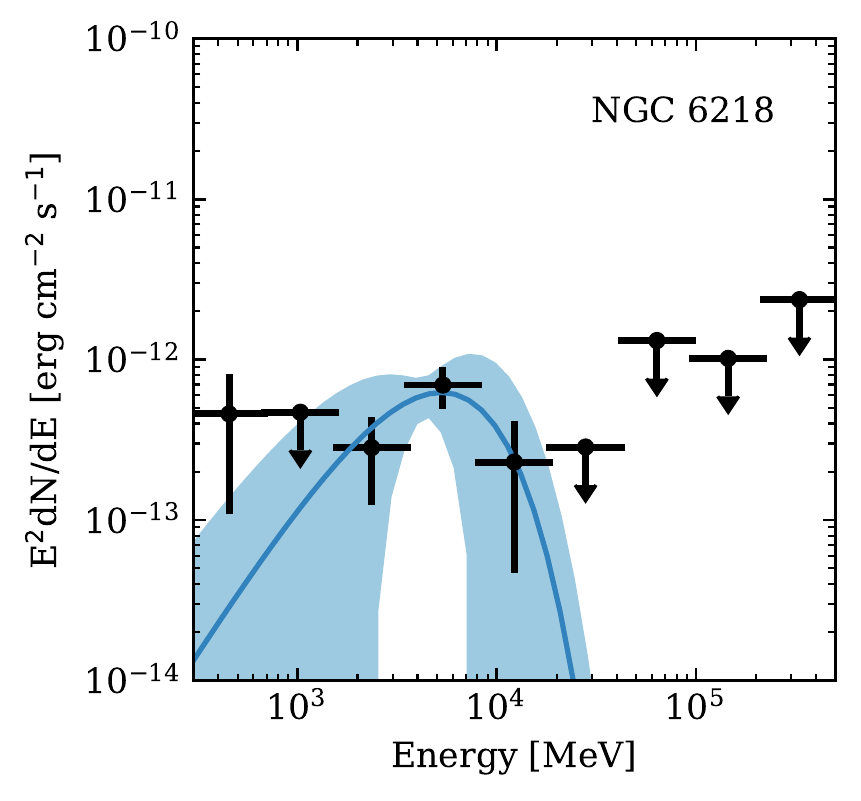}
\includegraphics[width=0.24\textwidth]{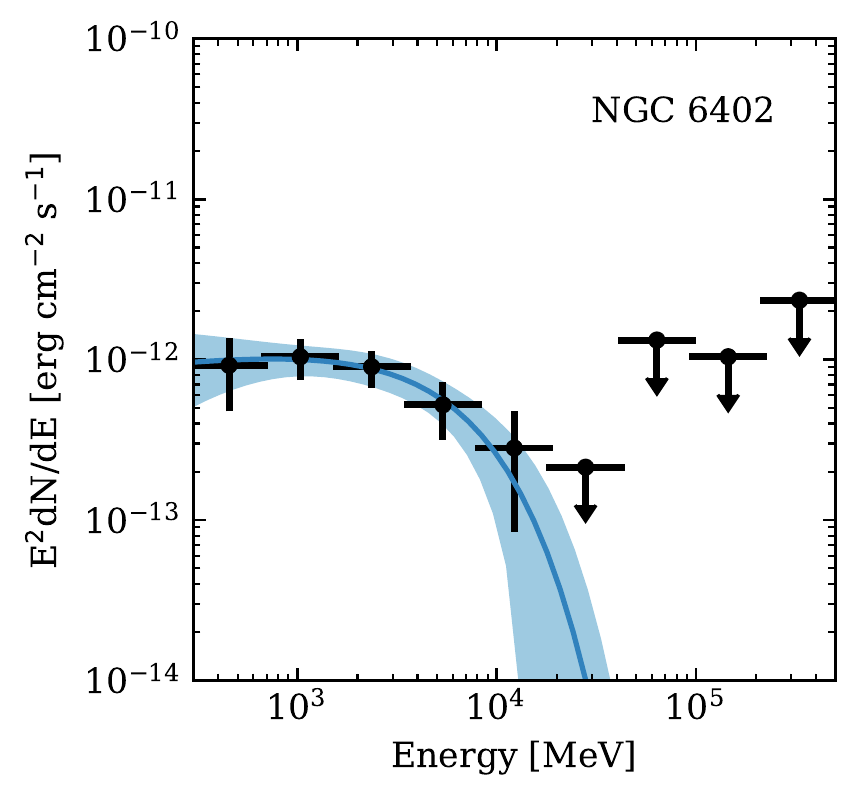}
\includegraphics[width=0.24\textwidth]{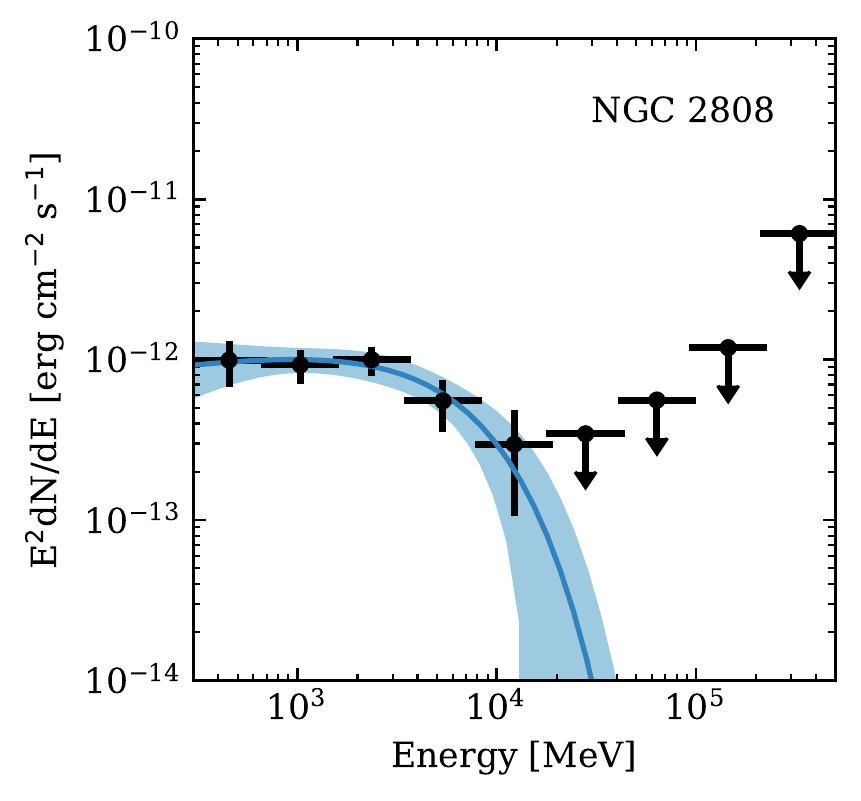}
\includegraphics[width=0.24\textwidth]{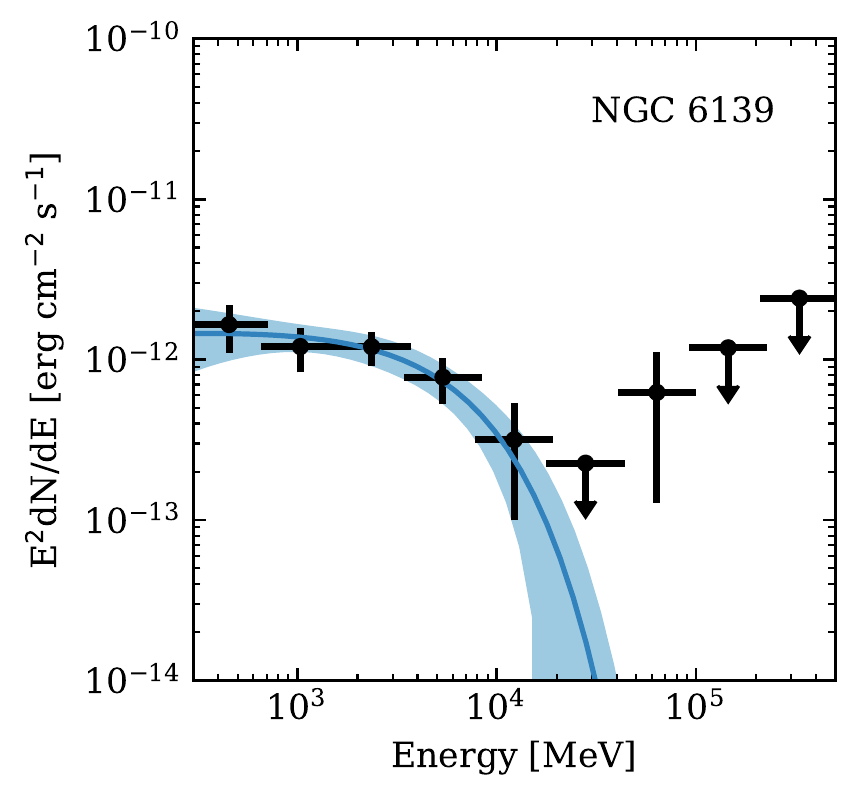}
\includegraphics[width=0.24\textwidth]{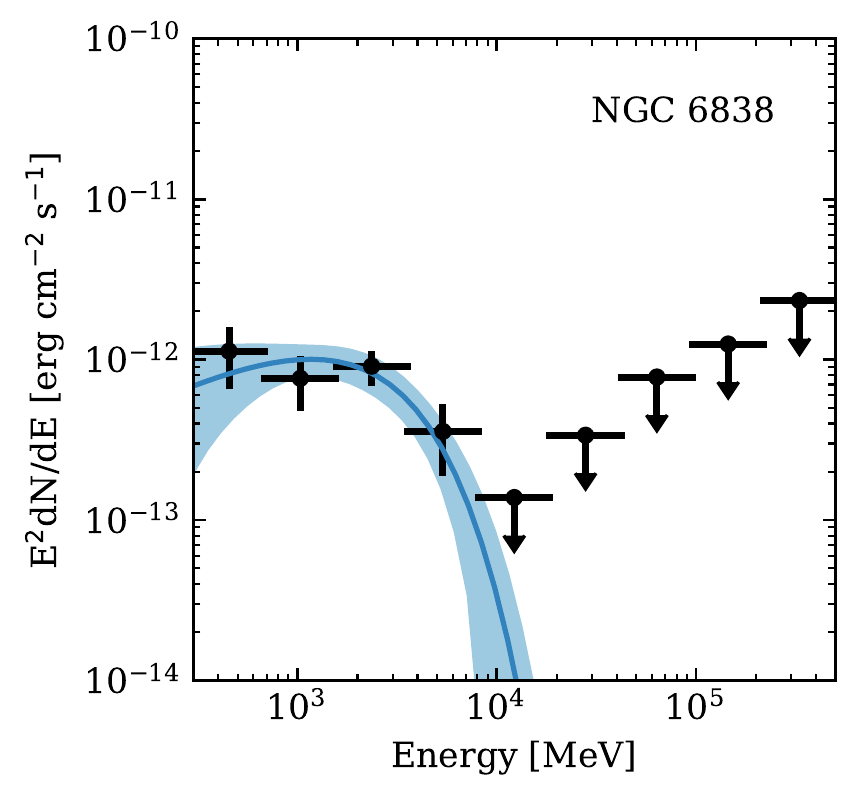}
\includegraphics[width=0.24\textwidth]{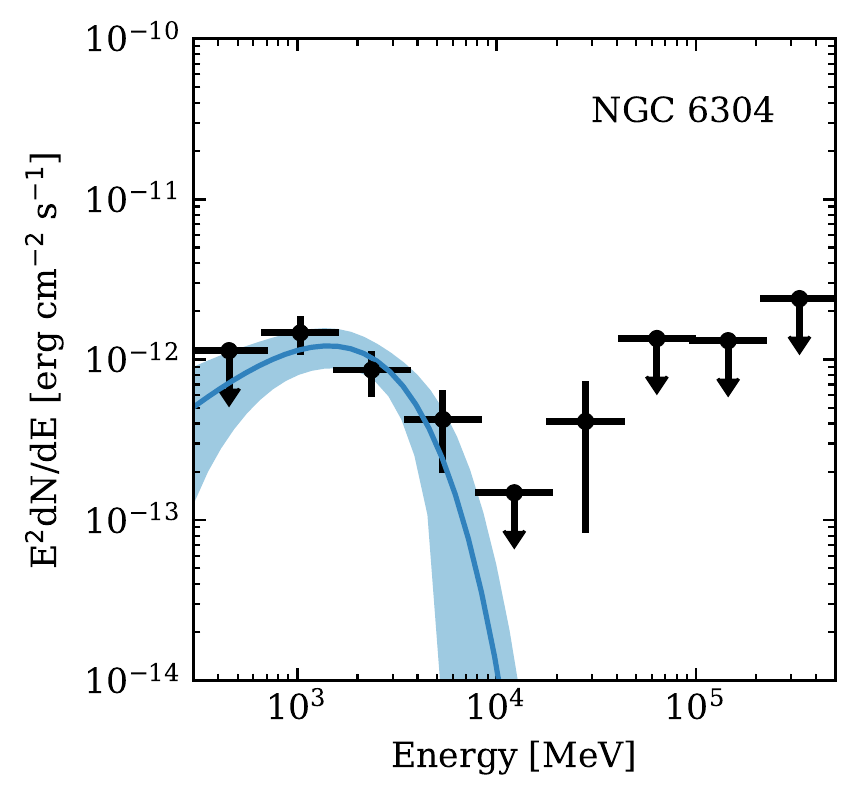}
\caption{Best-fit spectra and 1$\sigma$ uncertainty (blue line and shaded area) for 30 $\gamma$-ray-detected GCs from the individual fits. The bin-by-bin fluxes from Fermi data analysis are include by the black dots with error bars. The spectra are sorted by their curvature. For GCs with TS$_{\rm curve} < 4$, we report their best-fit PL spectrum. For the rest, the best-fit PLE spectra are shown.}
\label{fig:spectra}
\end{figure*}

\begin{figure*}
\centering
\includegraphics[width=0.24\textwidth]{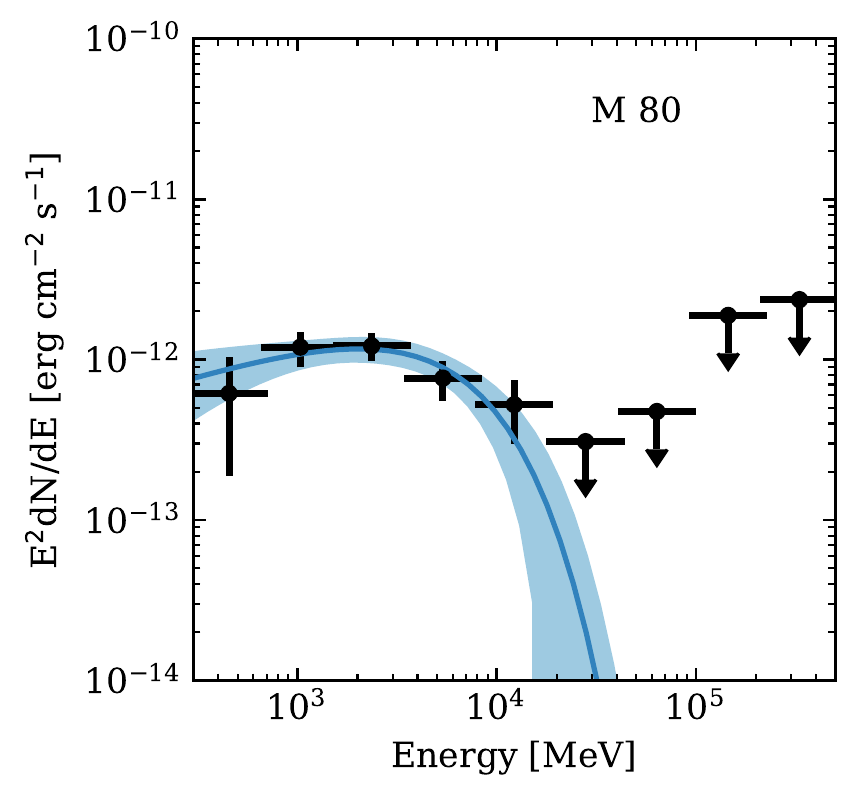}
\includegraphics[width=0.24\textwidth]{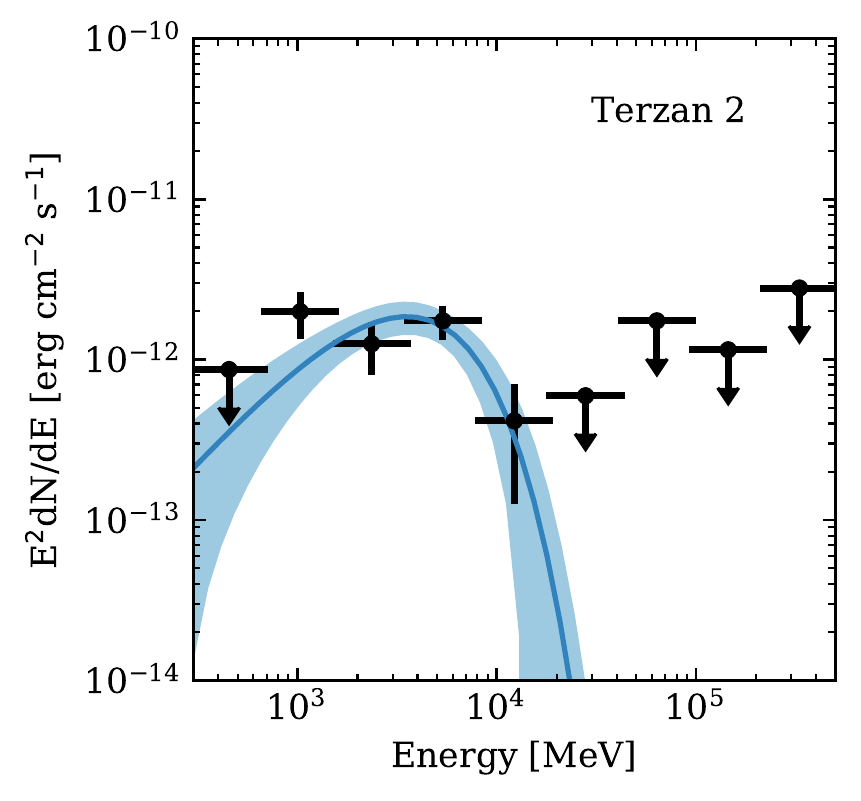}
\includegraphics[width=0.24\textwidth]{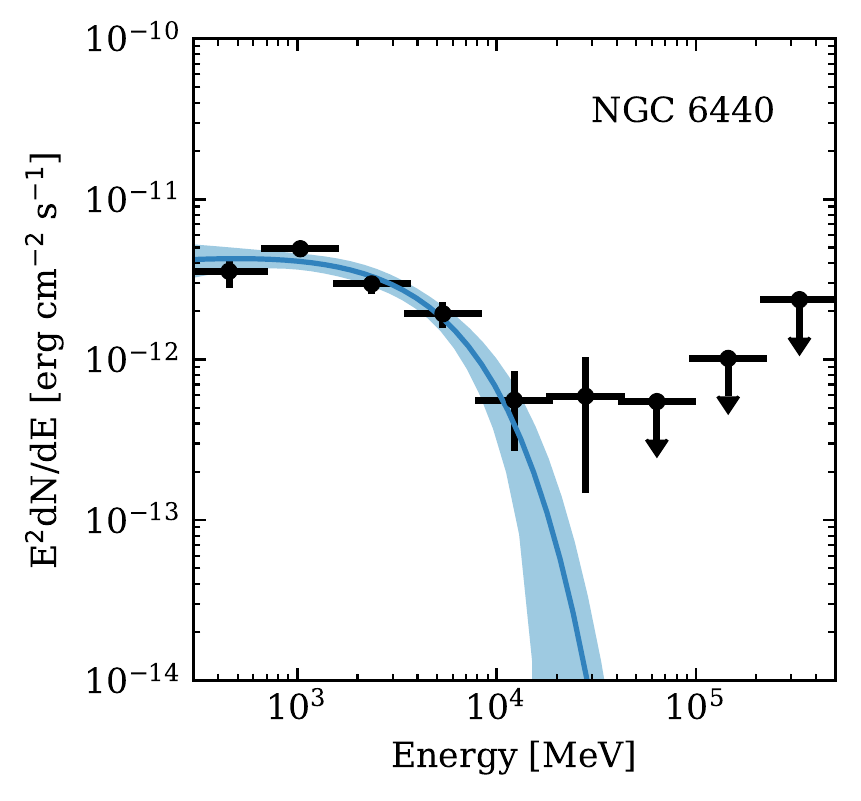}
\includegraphics[width=0.24\textwidth]{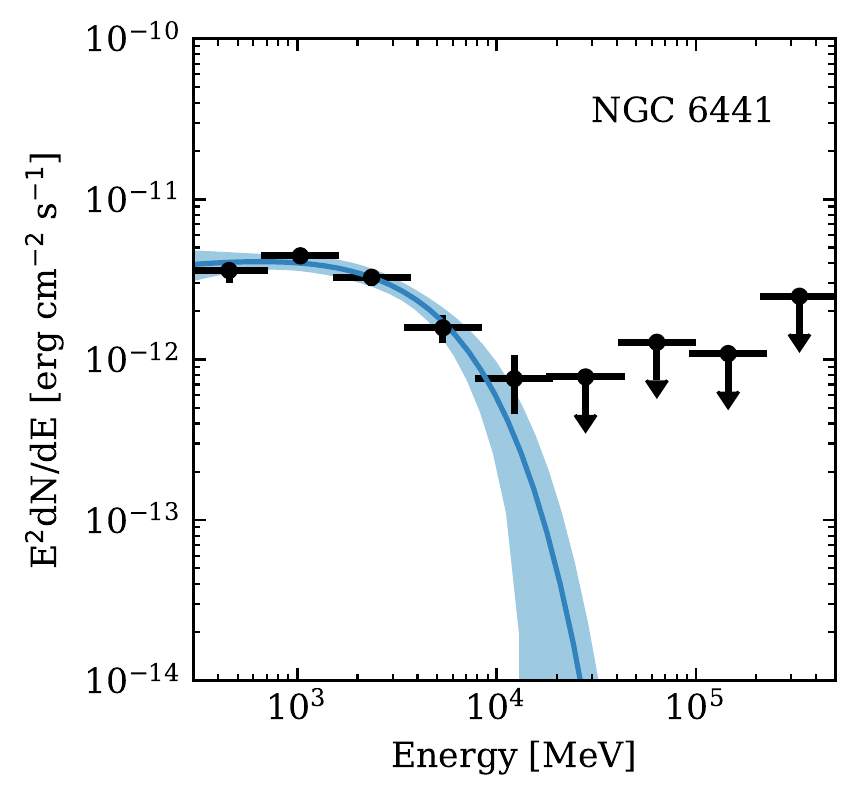}
\includegraphics[width=0.24\textwidth]{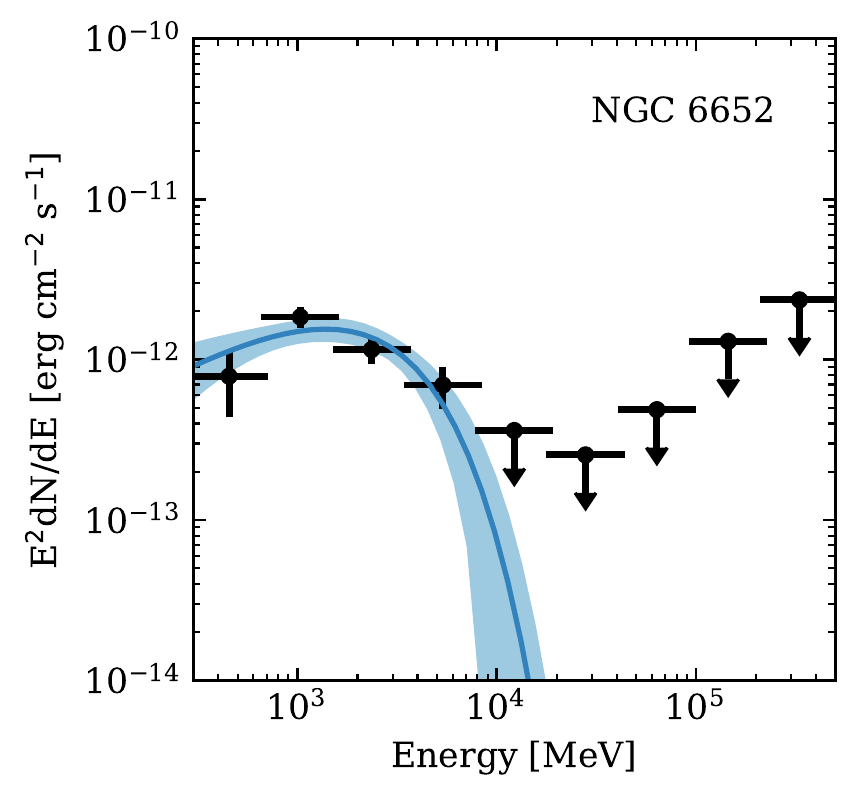}
\includegraphics[width=0.24\textwidth]{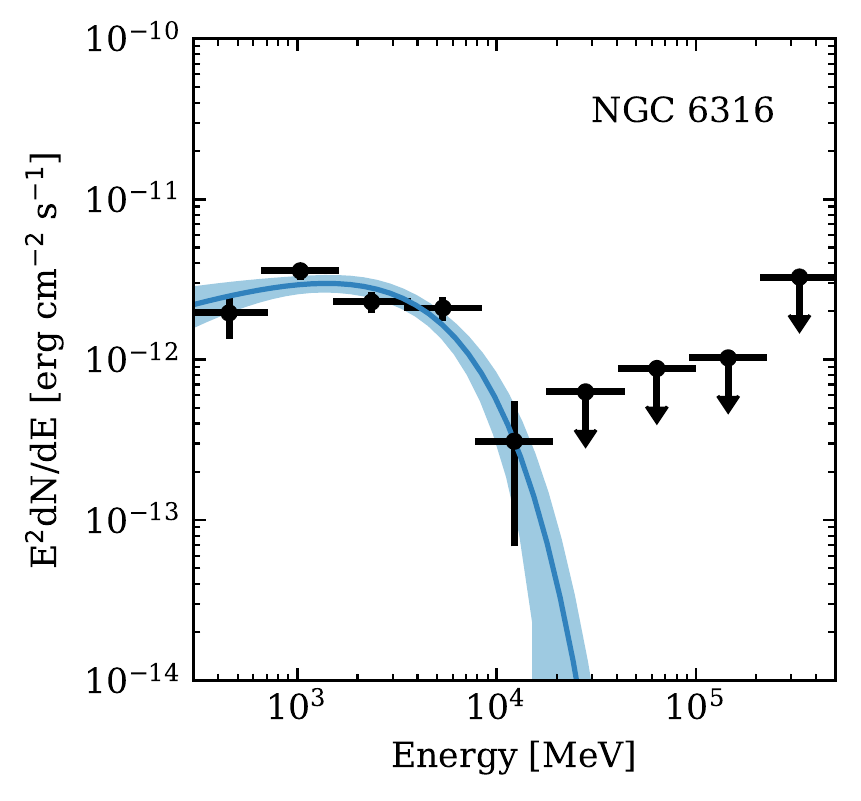}
\includegraphics[width=0.24\textwidth]{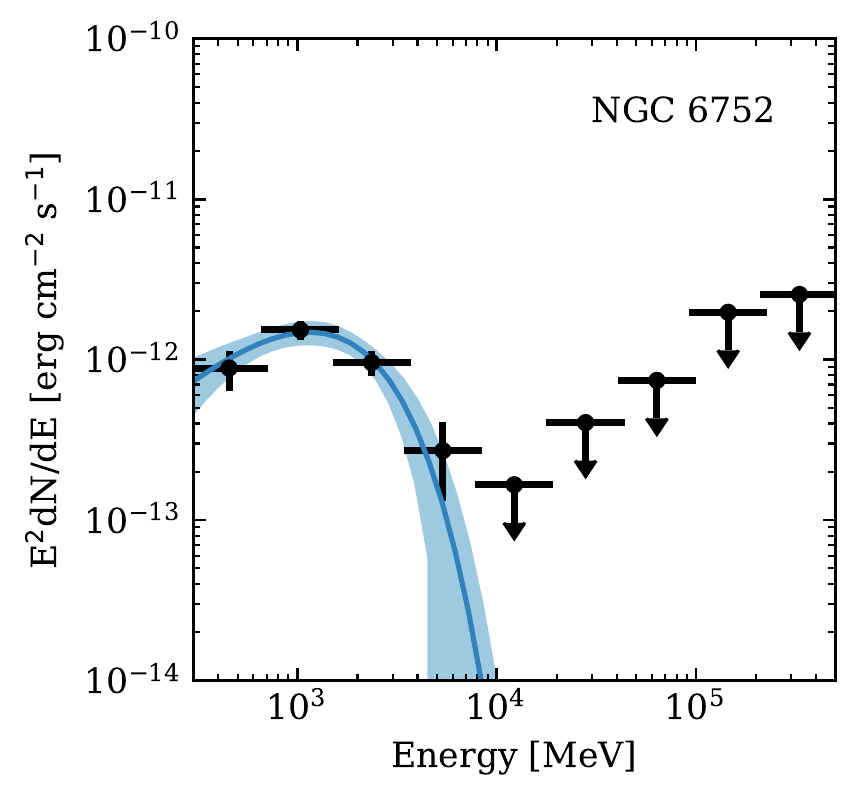}
\includegraphics[width=0.24\textwidth]{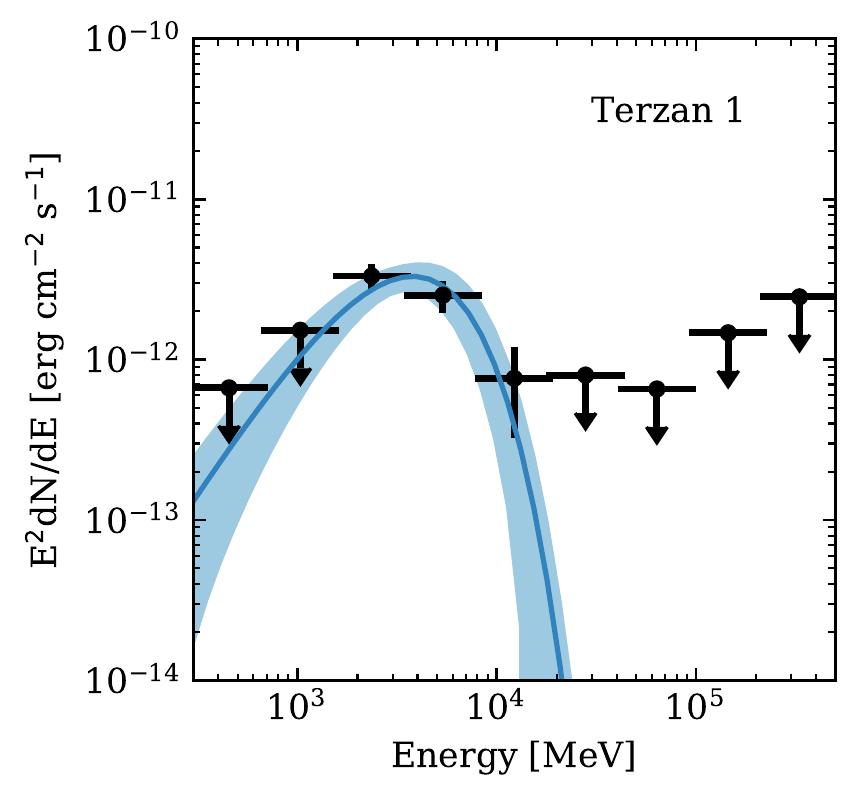}
\includegraphics[width=0.24\textwidth]{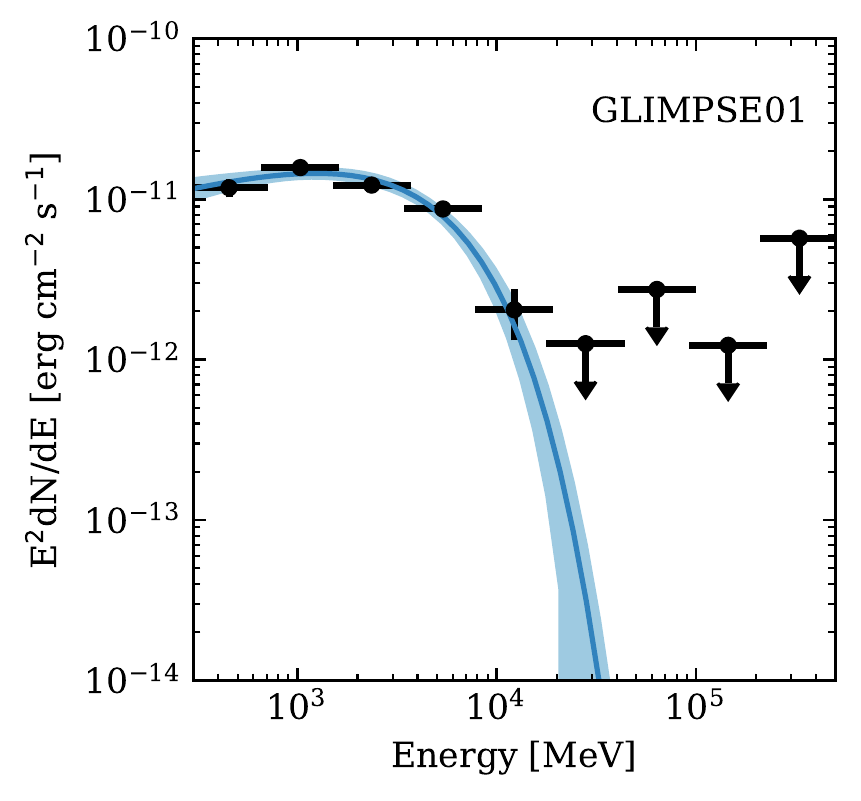}
\includegraphics[width=0.24\textwidth]{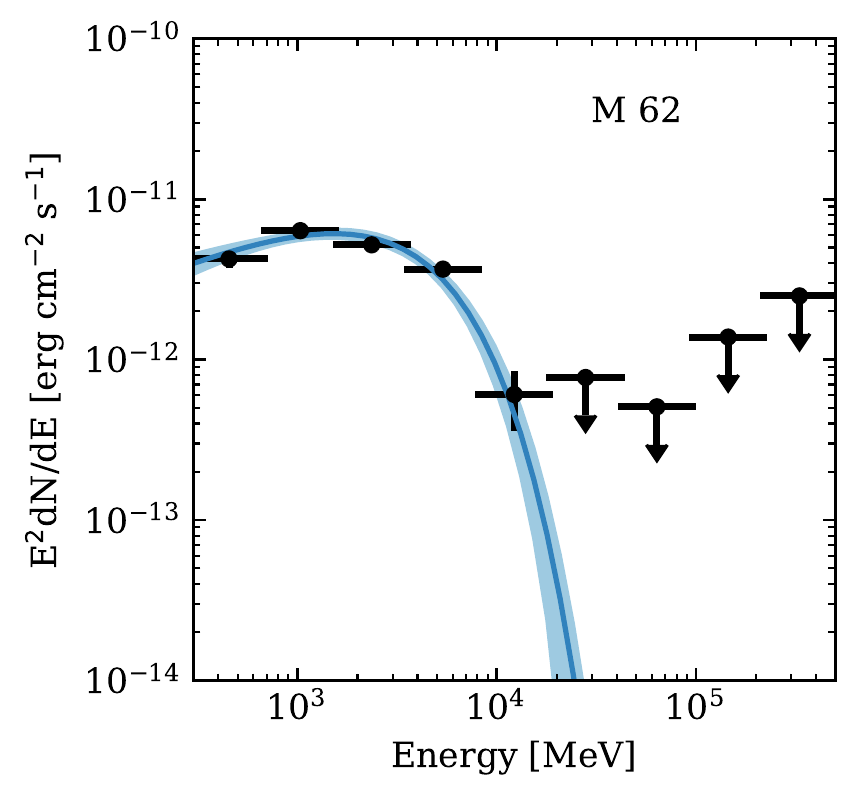}
\includegraphics[width=0.24\textwidth]{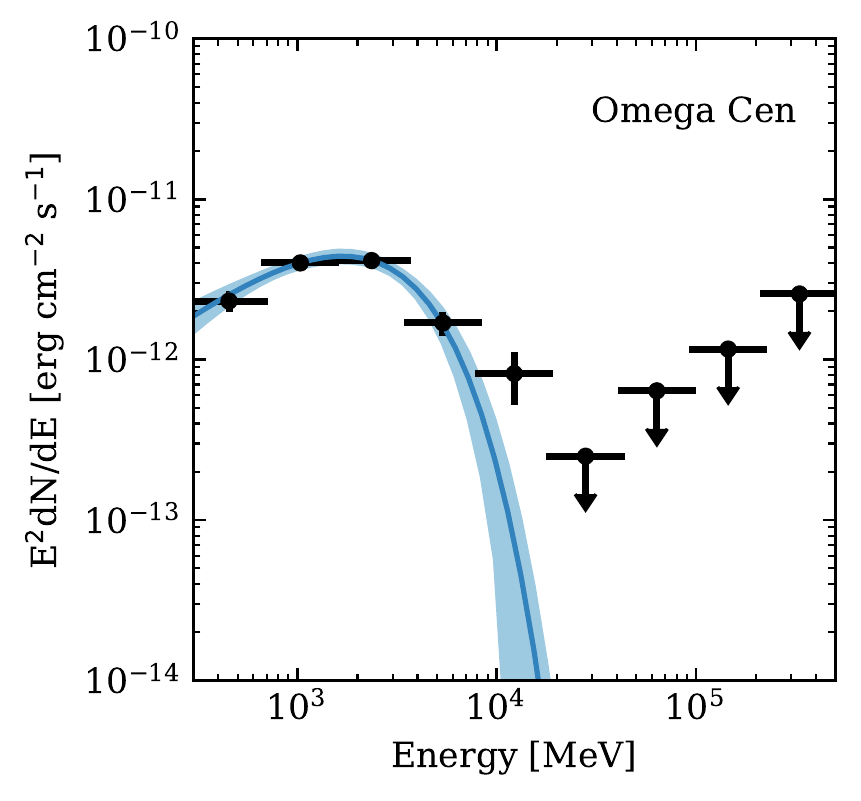}
\includegraphics[width=0.24\textwidth]{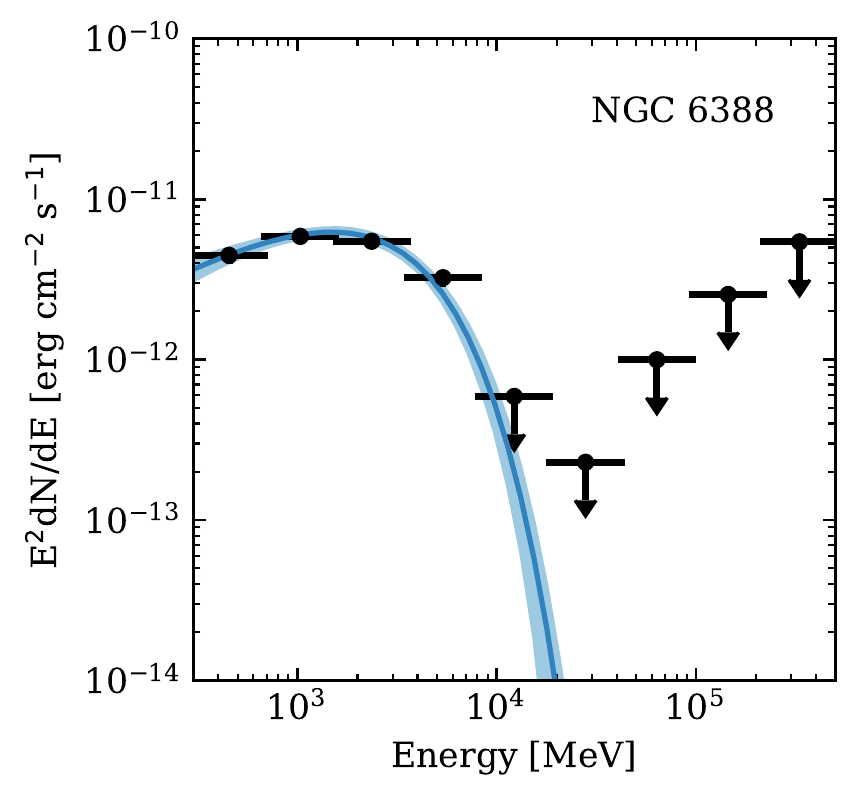}
\includegraphics[width=0.24\textwidth]{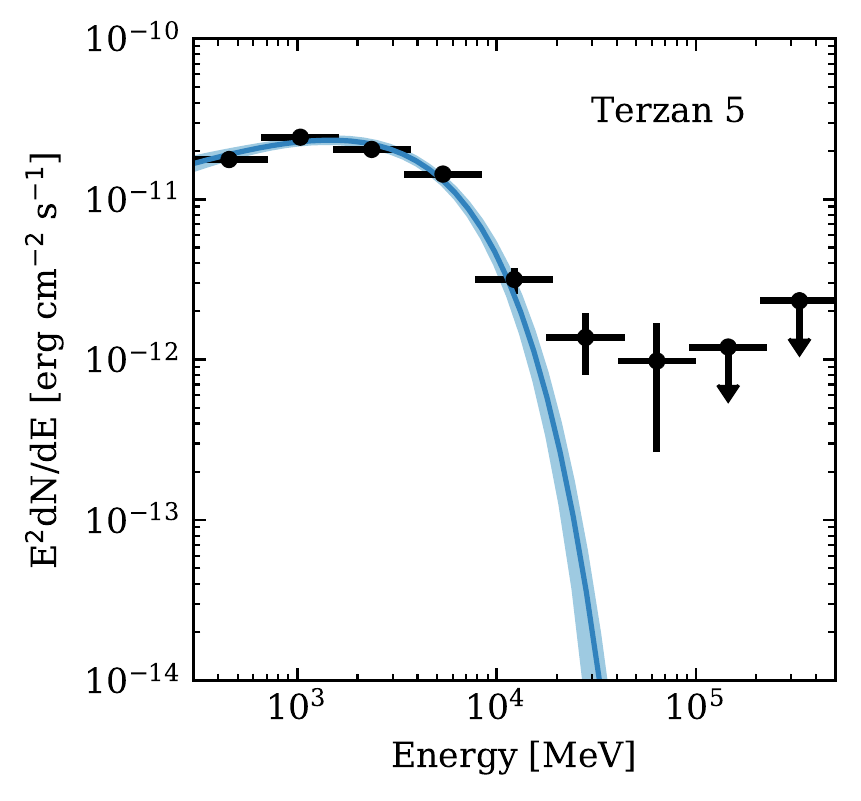}
\includegraphics[width=0.24\textwidth]{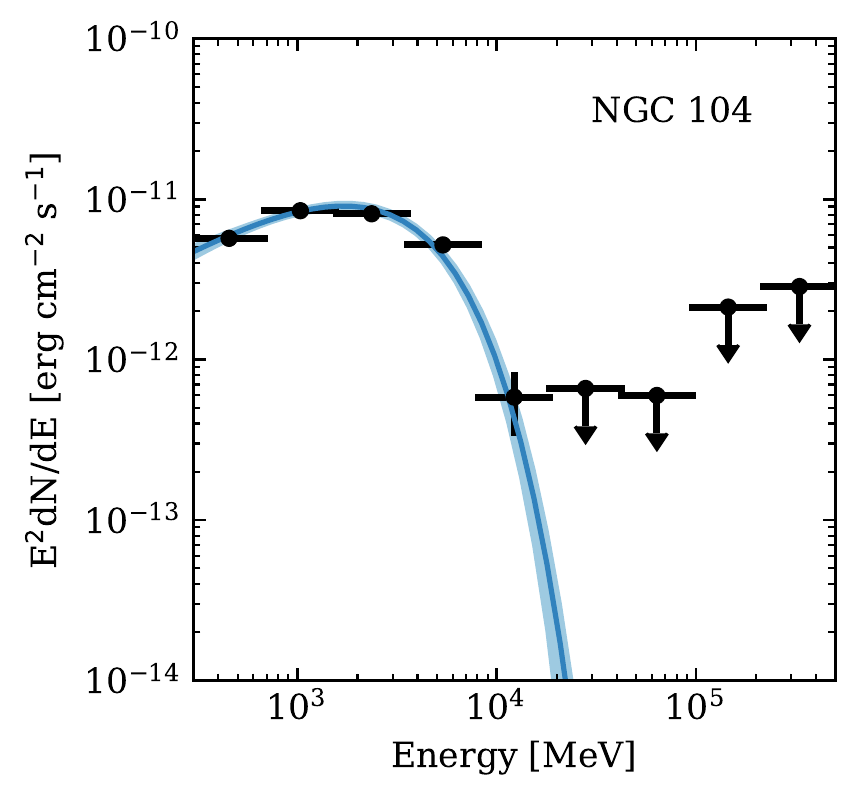}
\contcaption{}
\end{figure*}

\begin{figure*}
\centering
\includegraphics[width=0.24\textwidth]{figures/spectra/2comp_0.pdf}
\includegraphics[width=0.24\textwidth]{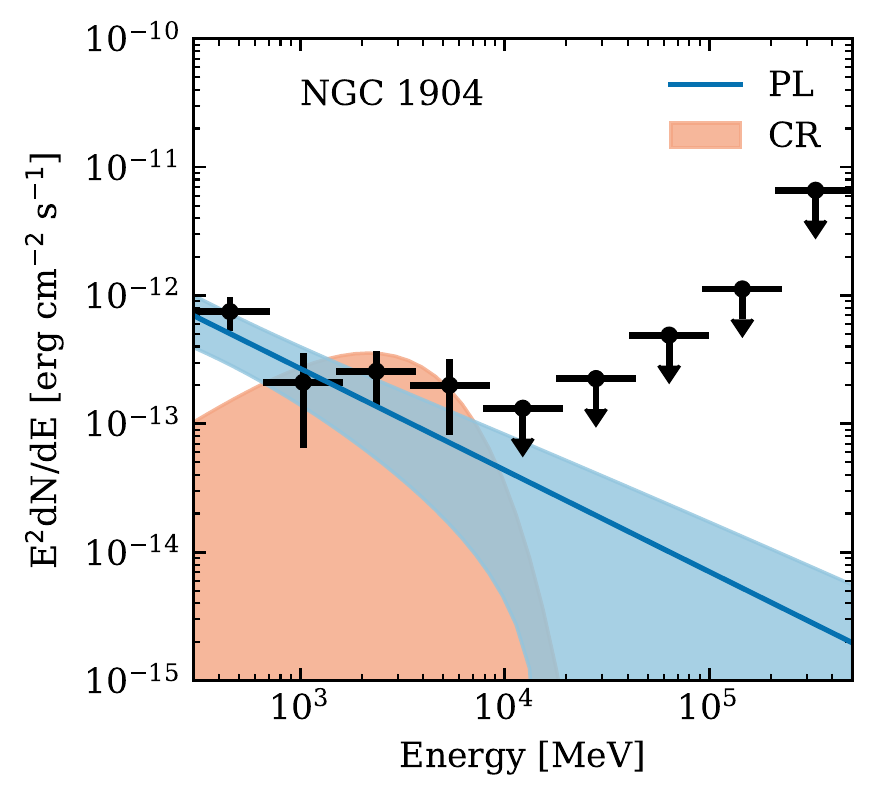}
\includegraphics[width=0.24\textwidth]{figures/spectra/2comp_16.pdf}
\includegraphics[width=0.24\textwidth]{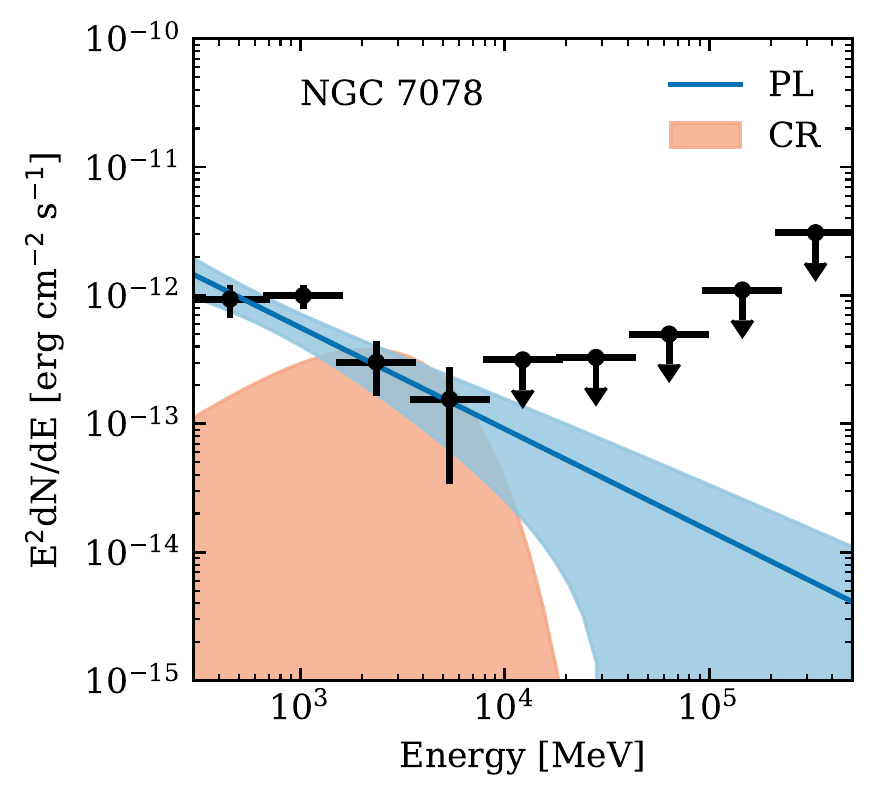}
\includegraphics[width=0.24\textwidth]{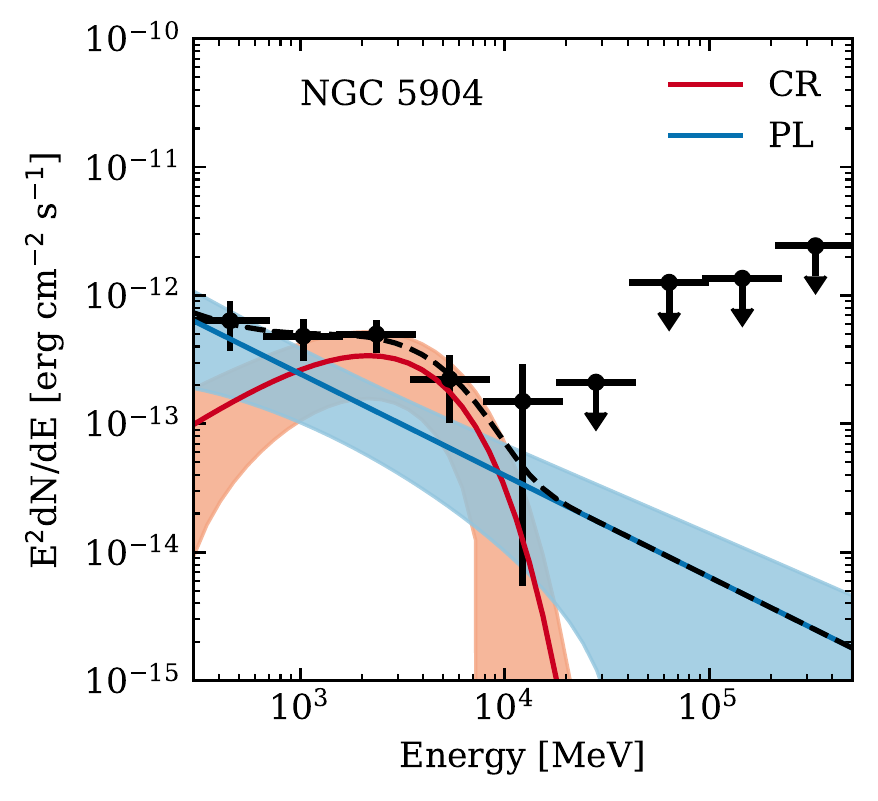}
\includegraphics[width=0.24\textwidth]{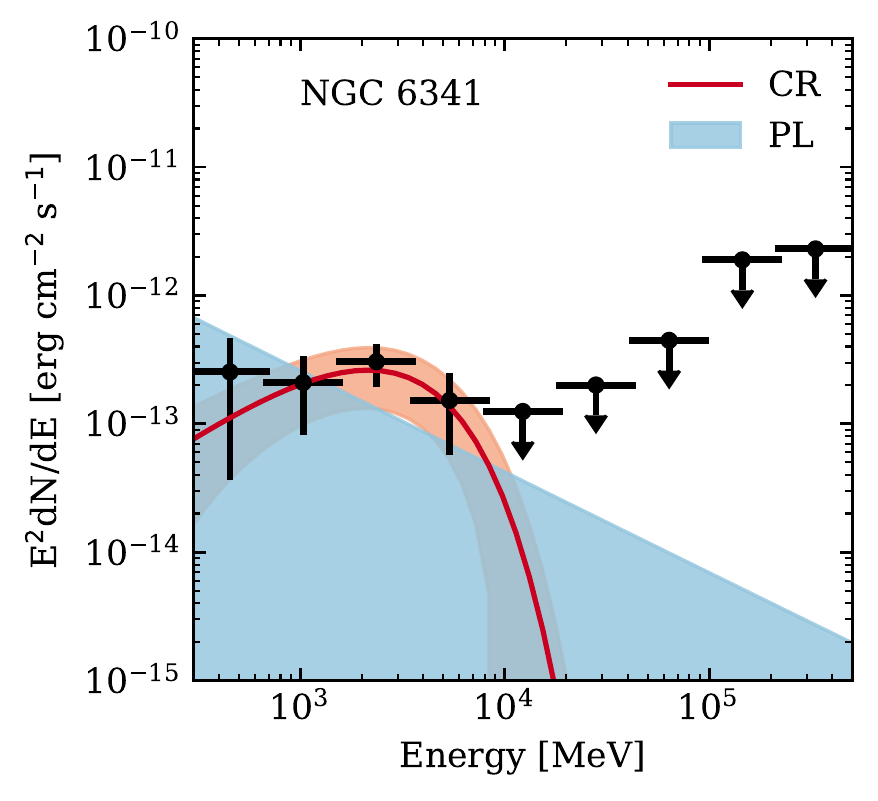}
\includegraphics[width=0.24\textwidth]{figures/spectra/2comp_21.pdf}
\includegraphics[width=0.24\textwidth]{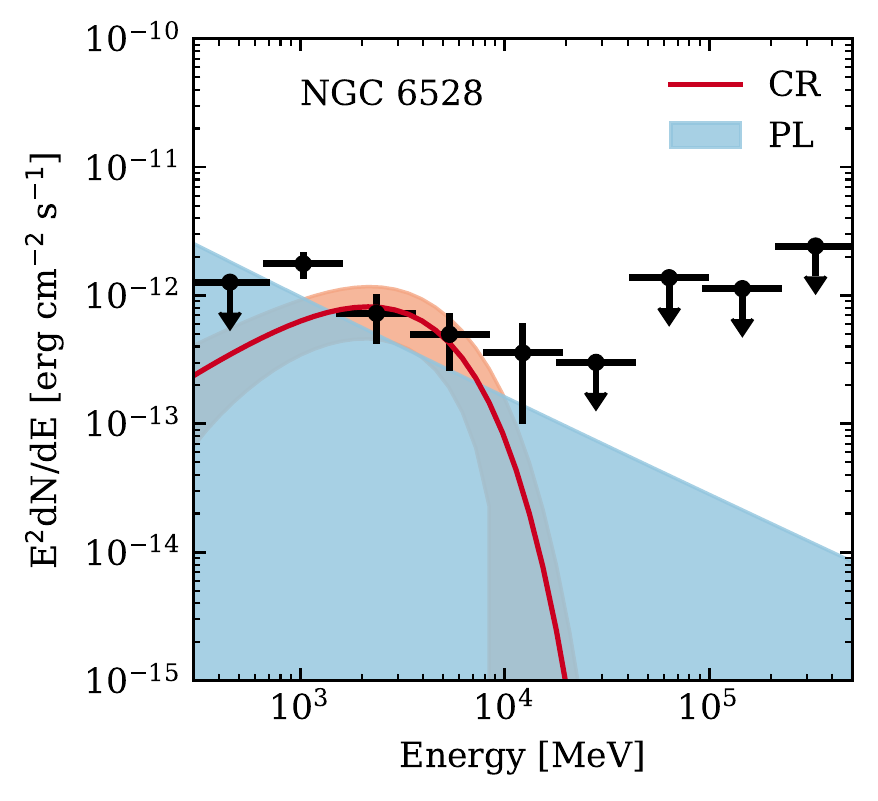}
\includegraphics[width=0.24\textwidth]{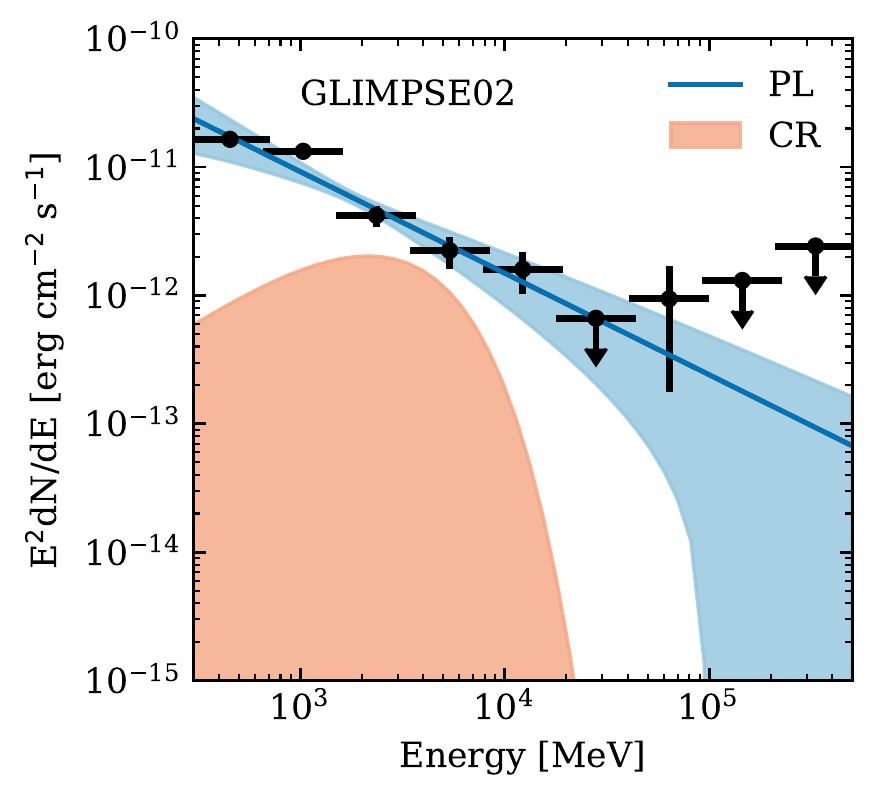}
\includegraphics[width=0.24\textwidth]{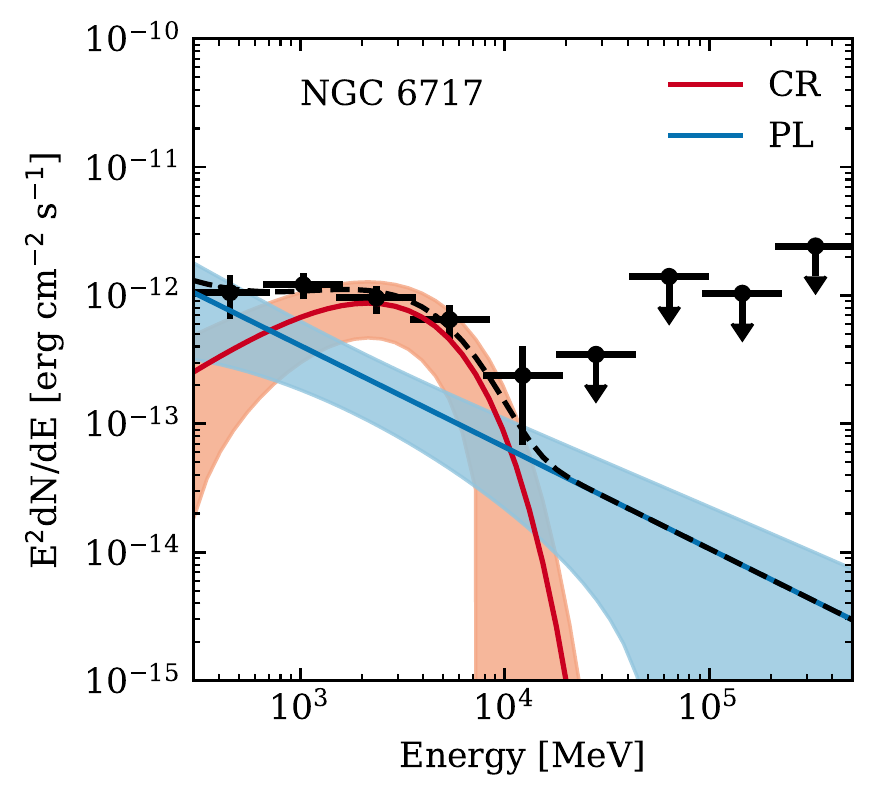}
\includegraphics[width=0.24\textwidth]{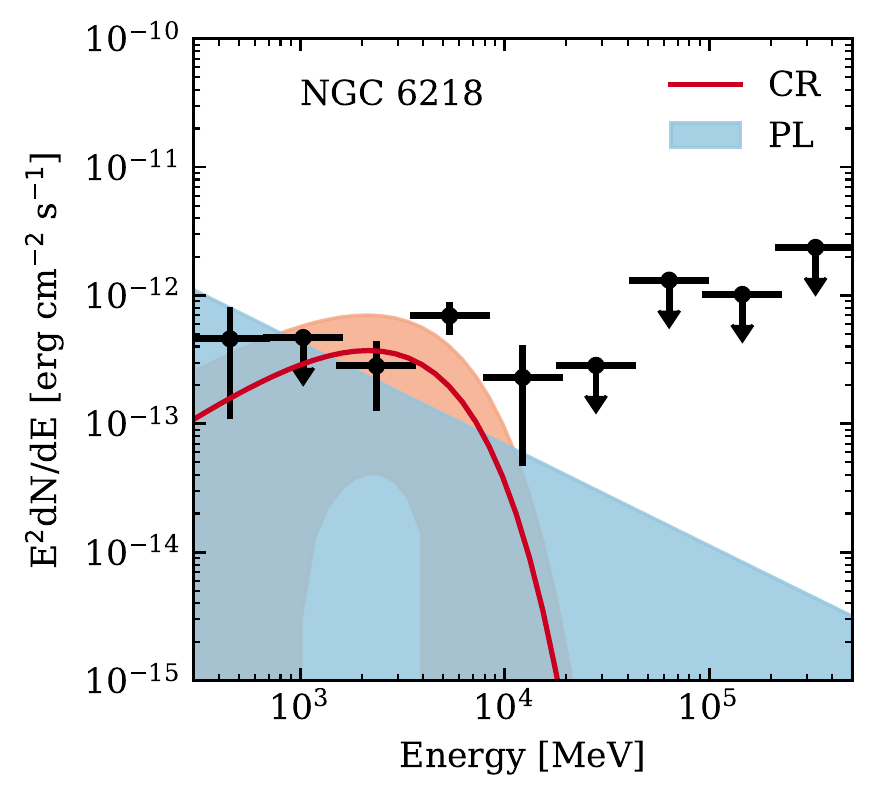}
\includegraphics[width=0.24\textwidth]{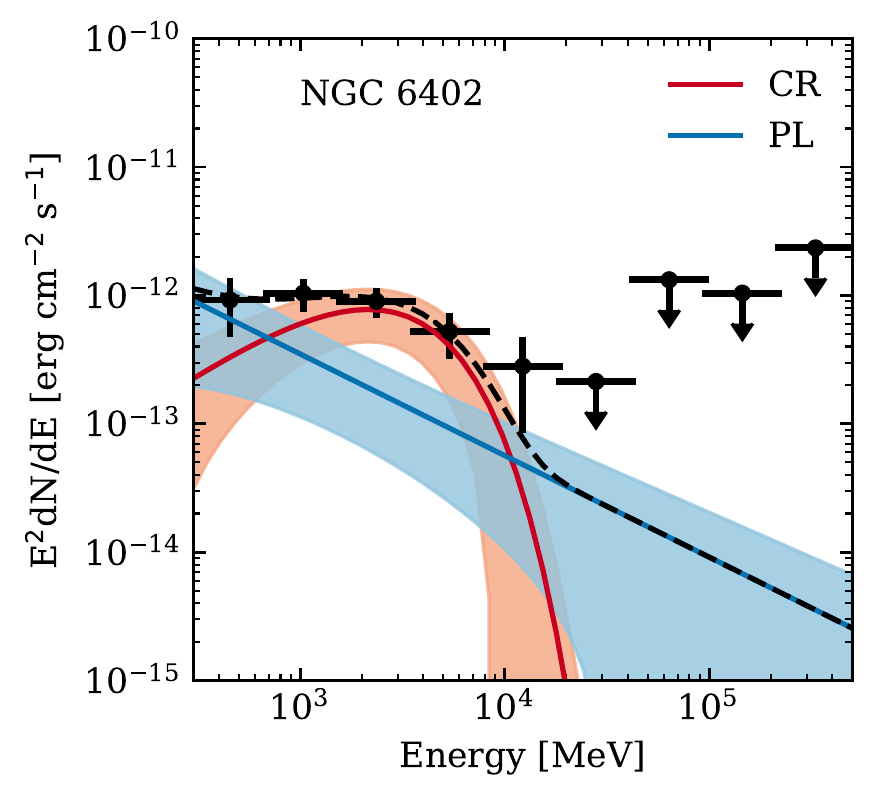}
\includegraphics[width=0.24\textwidth]{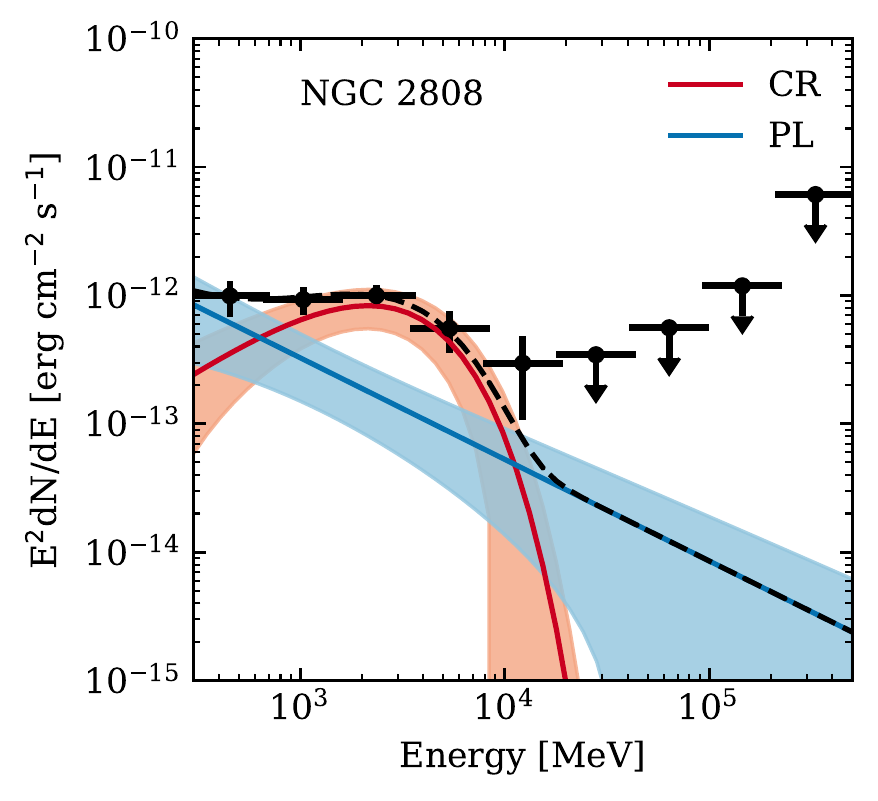}
\includegraphics[width=0.24\textwidth]{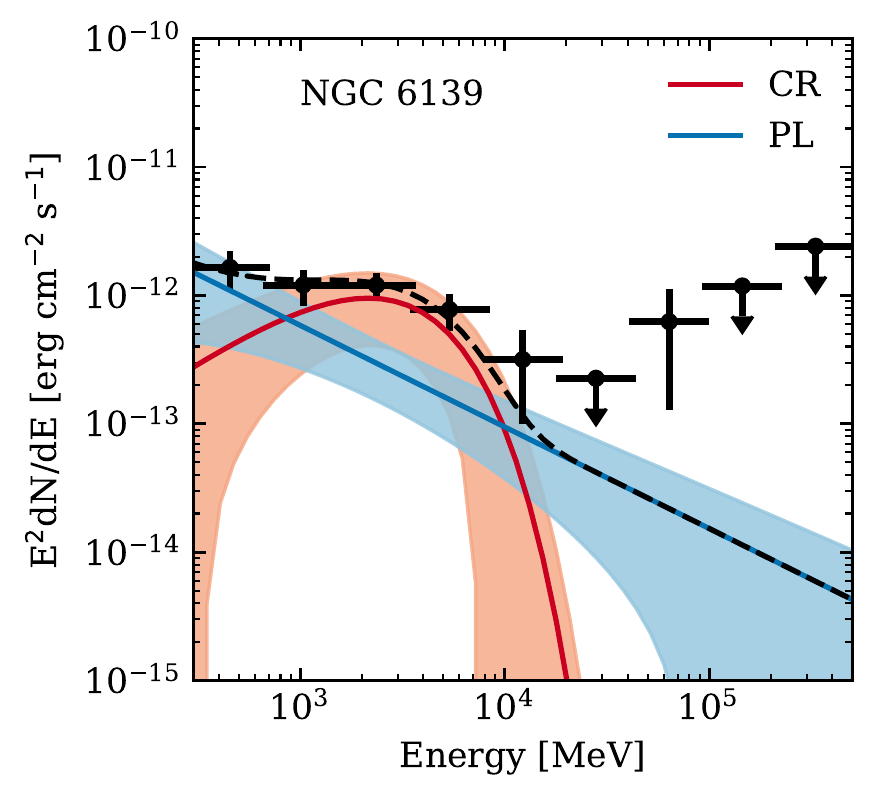}
\includegraphics[width=0.24\textwidth]{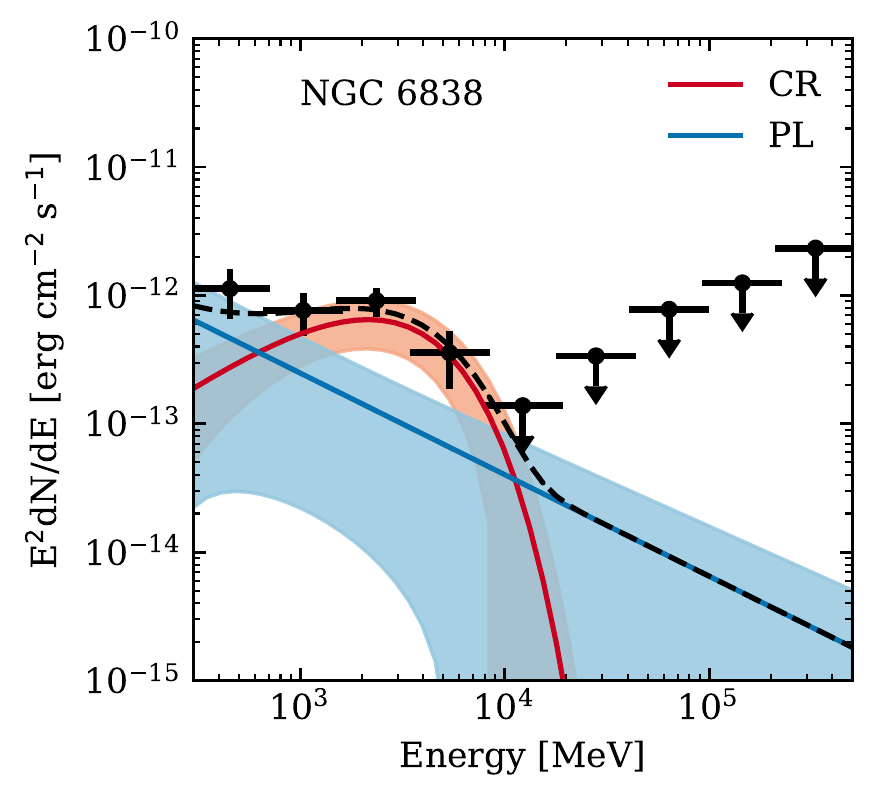}
\includegraphics[width=0.24\textwidth]{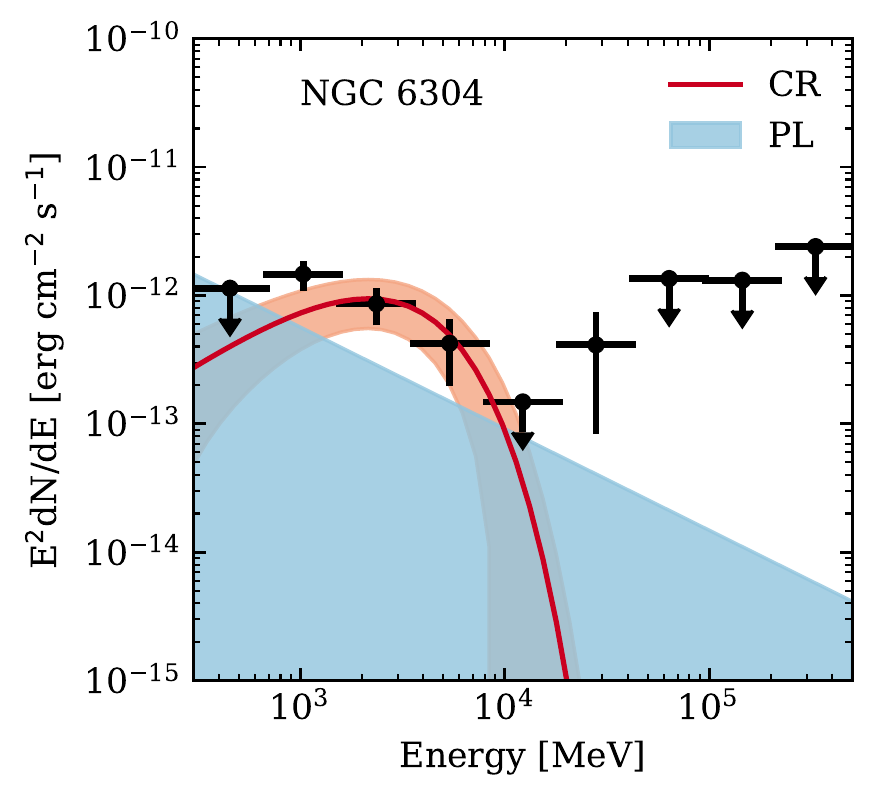}
\caption{Best-fit spectra and 1$\sigma$ uncertainty for 30 $\gamma$-ray-detected GCs from the universal two-component fits (see section~\ref{sec:spectra_global}). The best-fit parameters for the CR component (red line with shaded band) is $\Gamma_1 = 0.88 \pm 0.44$ and $\log(E_\mathrm{cut}/\mathrm{MeV})=3.28 \pm 0.16$. The best fit parameter for the {PL} component (blue line with shaded band) is $\Gamma_2 = 2.79 \pm 0.25$. 
The black dashed line shows the combined CR and {PL} components. When only one component detected, the 95\% C.L. upper limit for the CR ({PL}) component is shown by the red (blue) shaded area.}
\label{fig:2comp}
\end{figure*}

\begin{figure*}
\centering
\includegraphics[width=0.24\textwidth]{figures/spectra/2comp_8.pdf}
\includegraphics[width=0.24\textwidth]{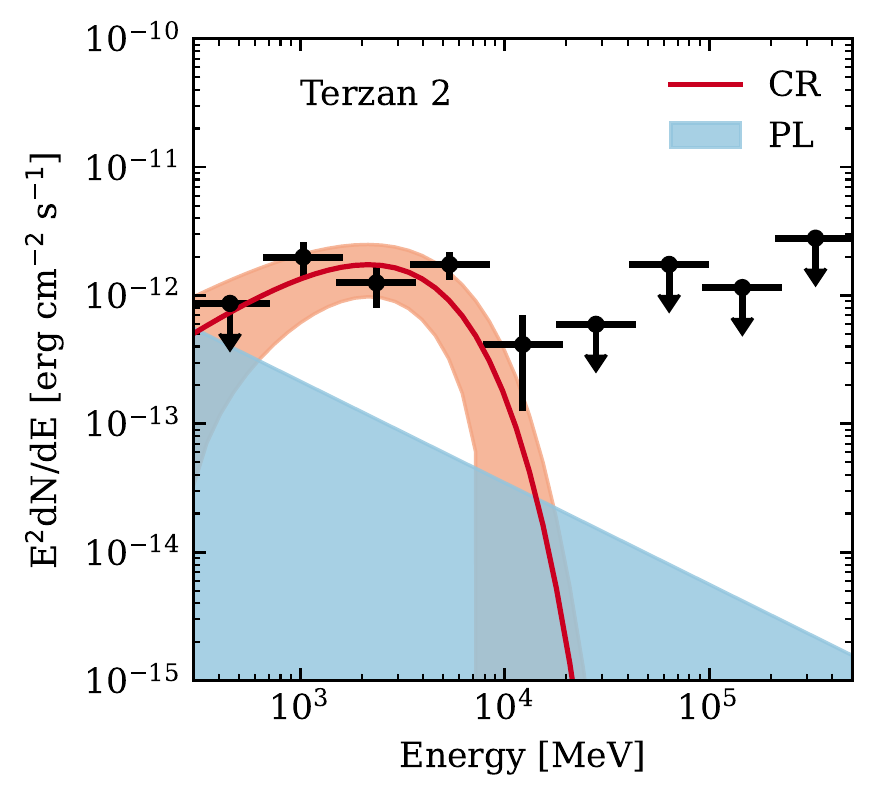}
\includegraphics[width=0.24\textwidth]{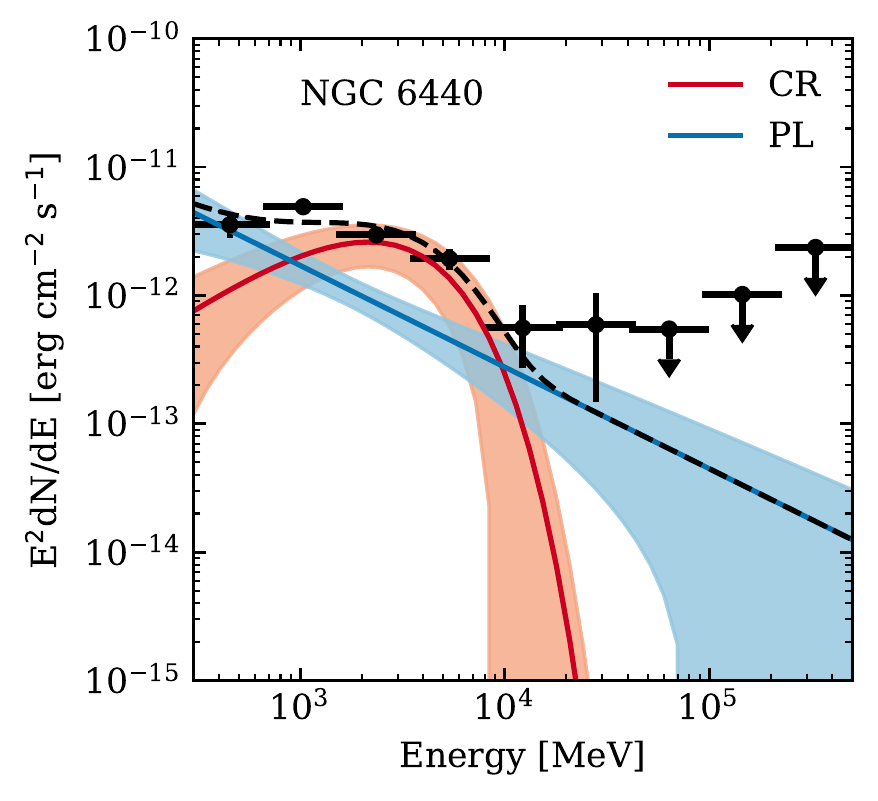}
\includegraphics[width=0.24\textwidth]{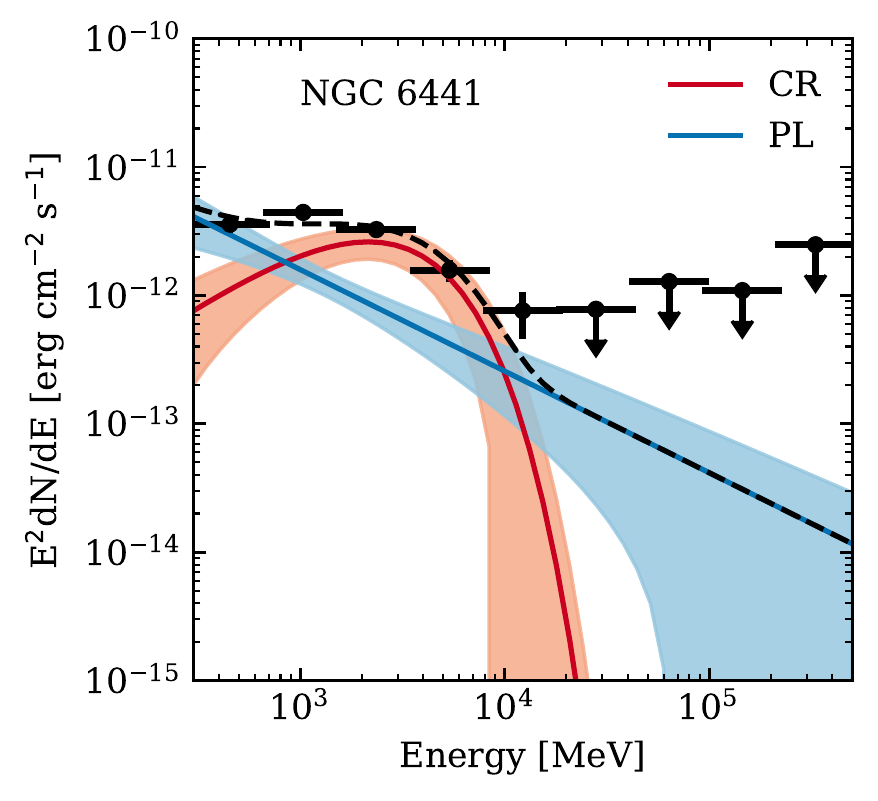}
\includegraphics[width=0.24\textwidth]{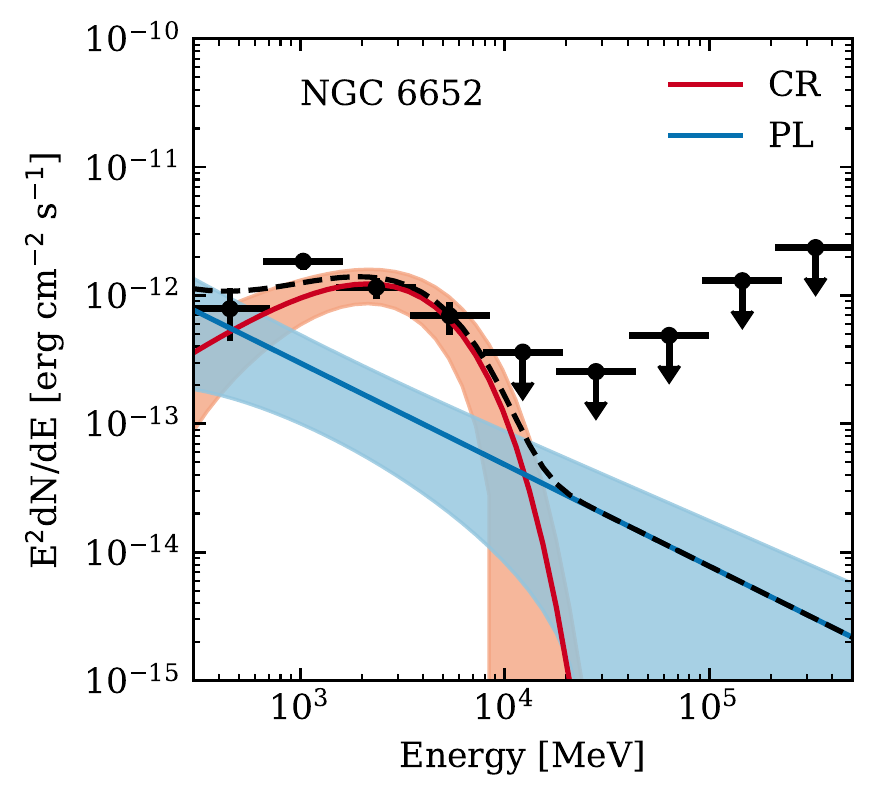}
\includegraphics[width=0.24\textwidth]{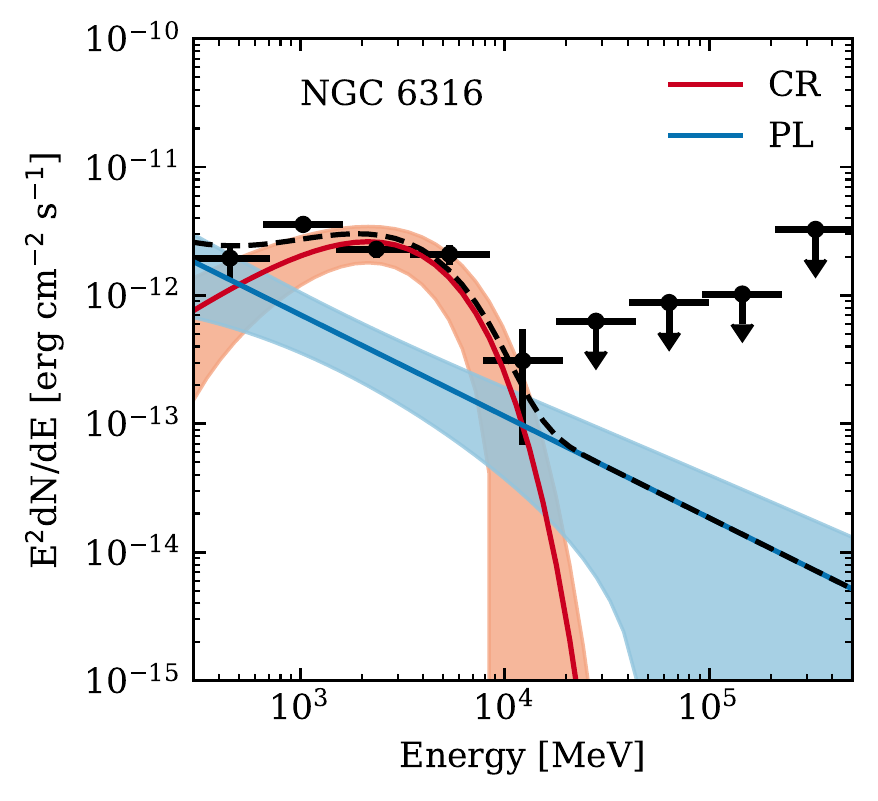}
\includegraphics[width=0.24\textwidth]{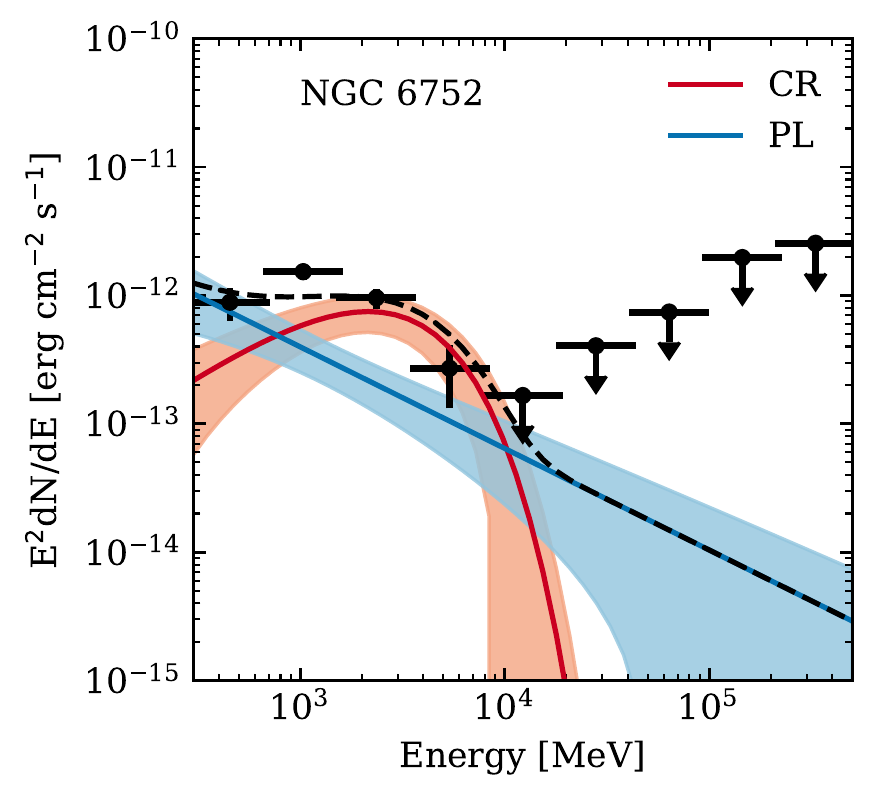}
\includegraphics[width=0.24\textwidth]{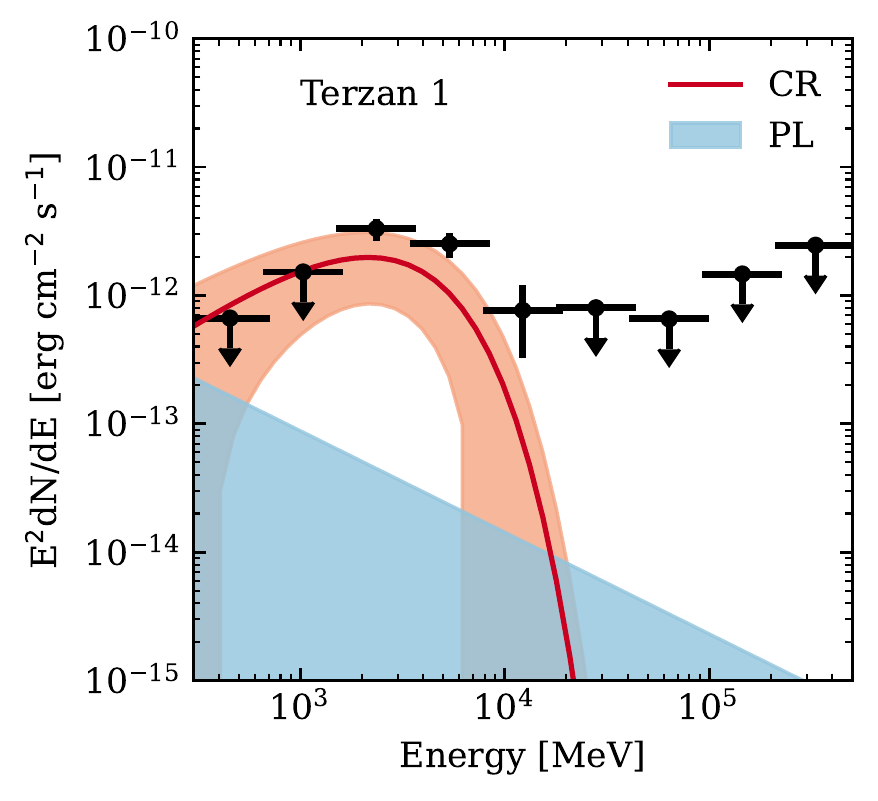}
\includegraphics[width=0.24\textwidth]{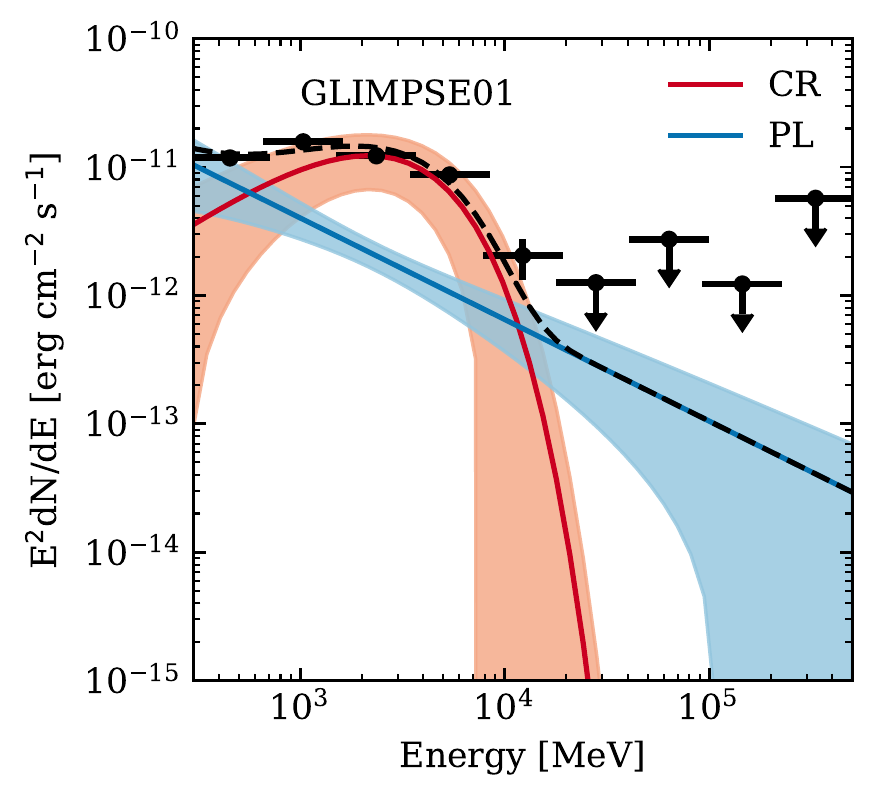}
\includegraphics[width=0.24\textwidth]{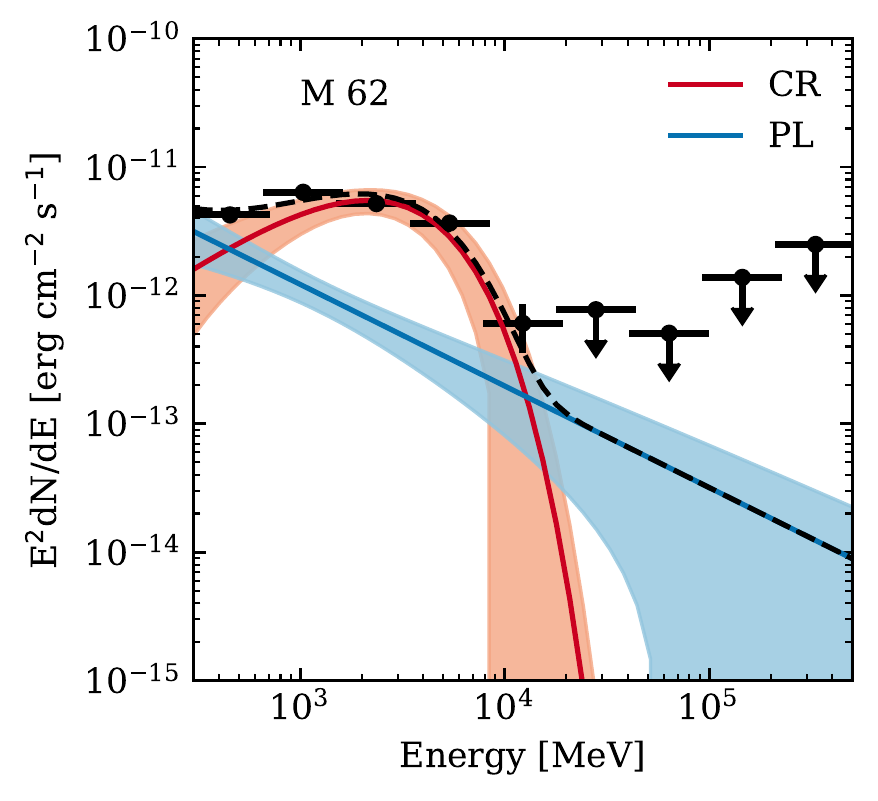}
\includegraphics[width=0.24\textwidth]{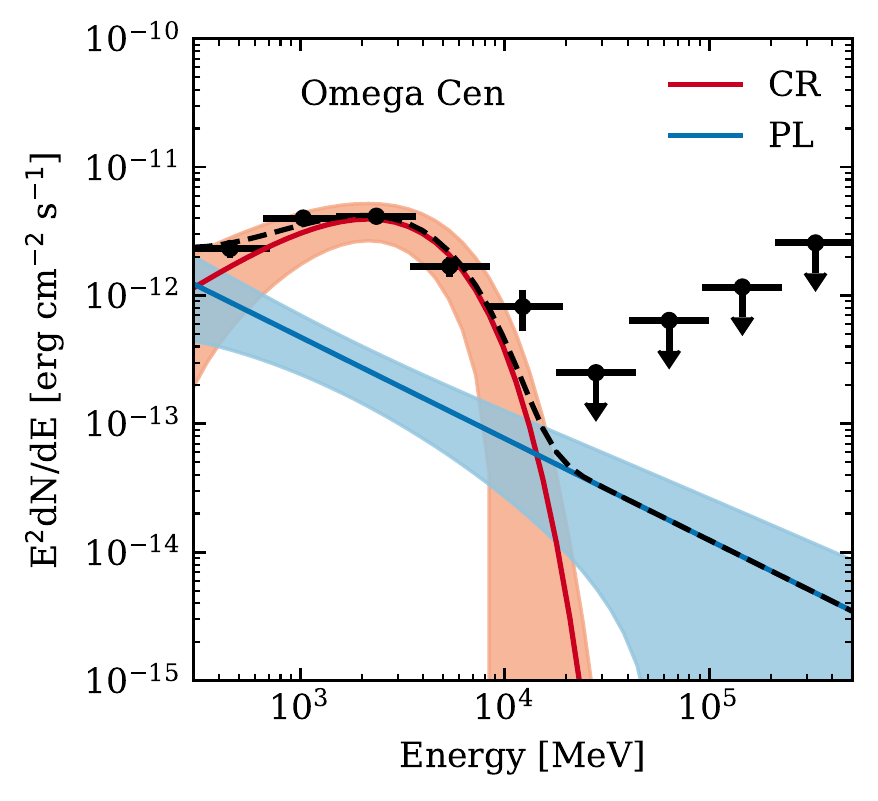}
\includegraphics[width=0.24\textwidth]{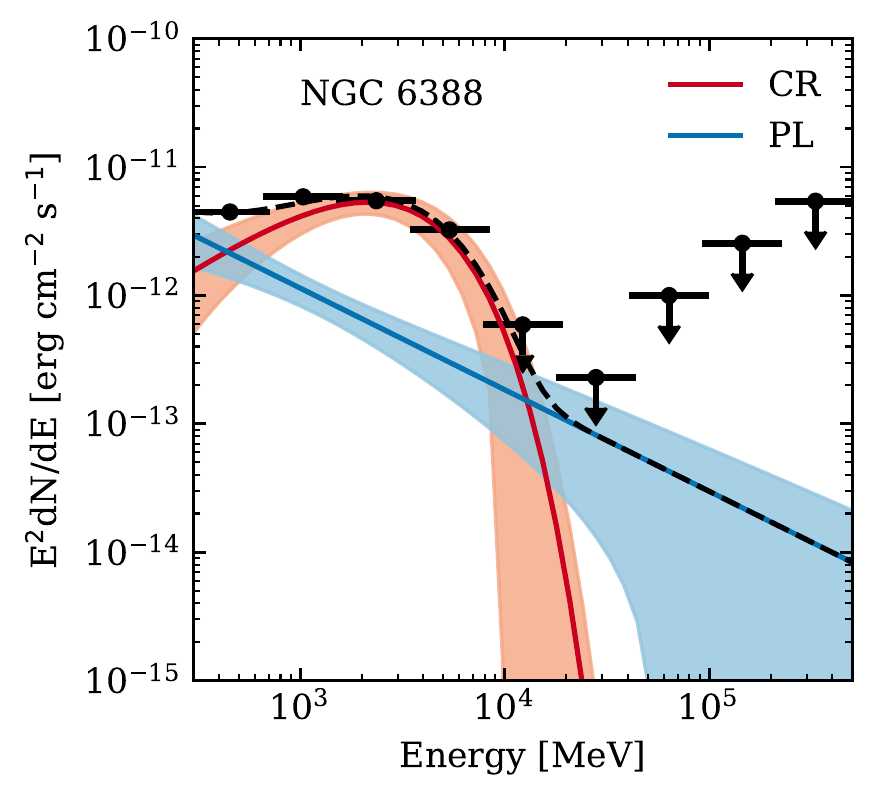}
\includegraphics[width=0.24\textwidth]{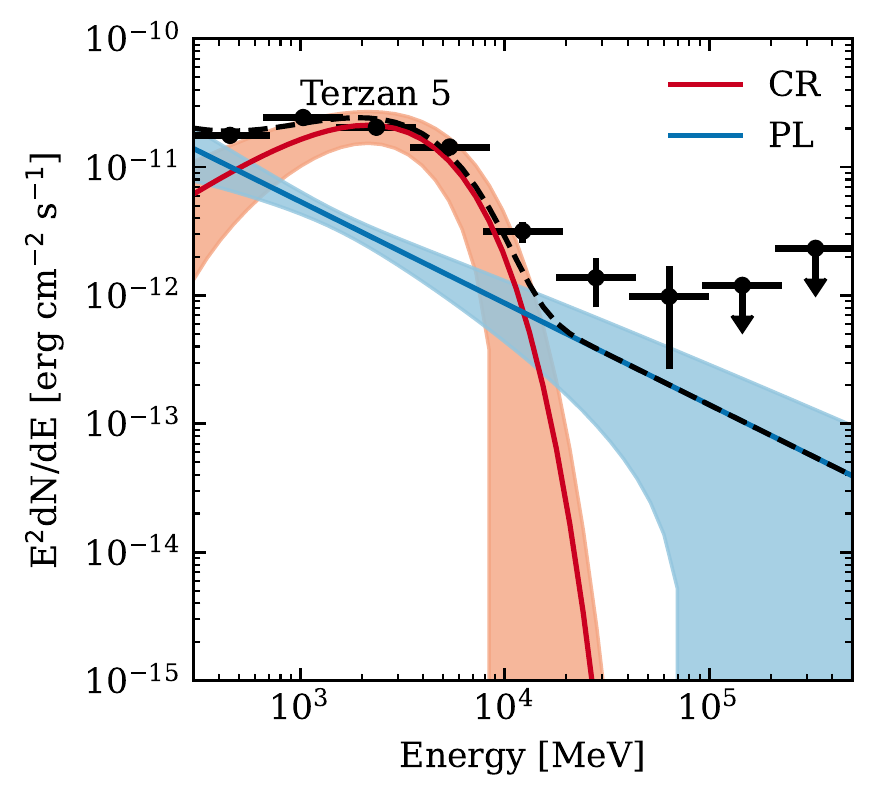}
\includegraphics[width=0.24\textwidth]{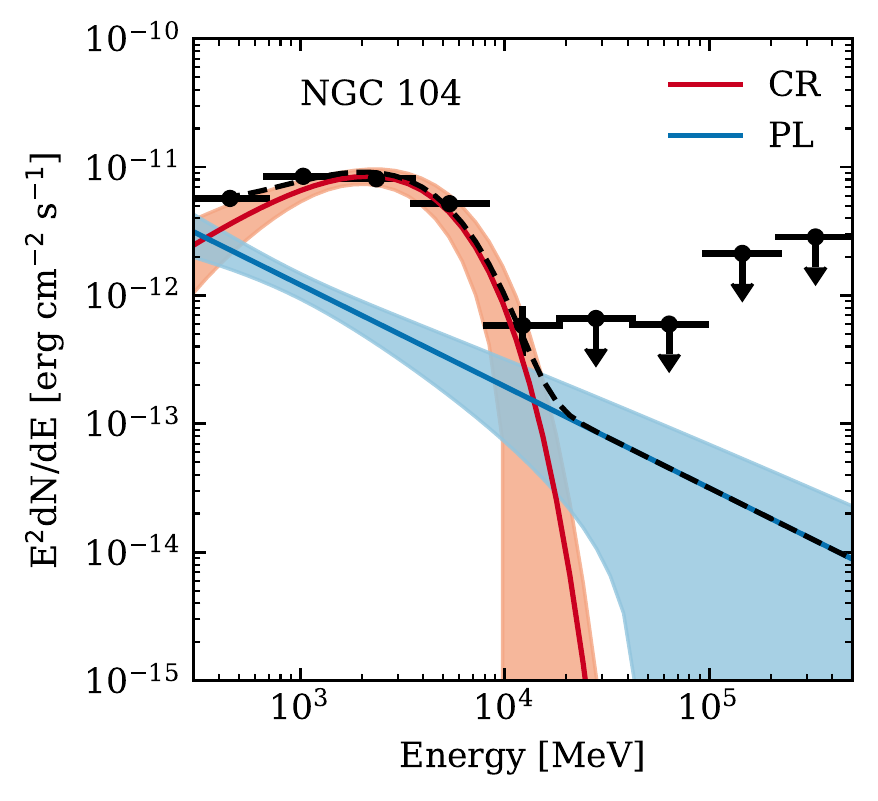}
\contcaption{}
\end{figure*}

\section{Systematic uncertainty on electron/positron luminosity }\label{appx:system_fe}

In Section~\ref{spectra_fe}, we discussed the implication for the $e^\pm$ injection efficiency from modelling the universal GC spectra. We argued that the $e^\pm$ luminosity from the MSPs in the GCs can be approximated by the luminosity of the IC component in our two-component model. 

\begin{figure*}
    \centering
    \includegraphics{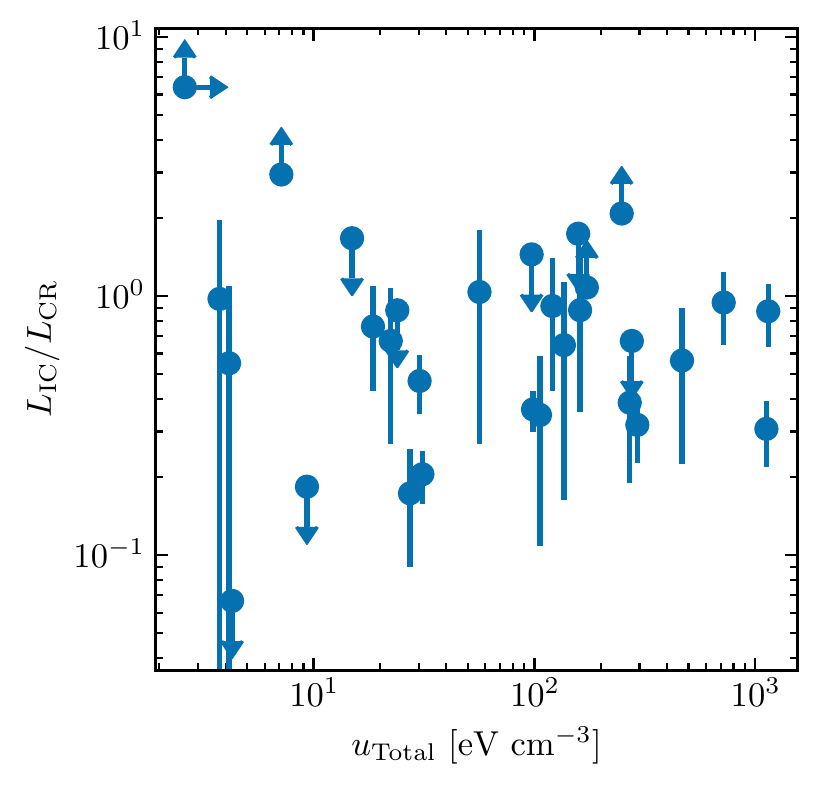}
    \includegraphics{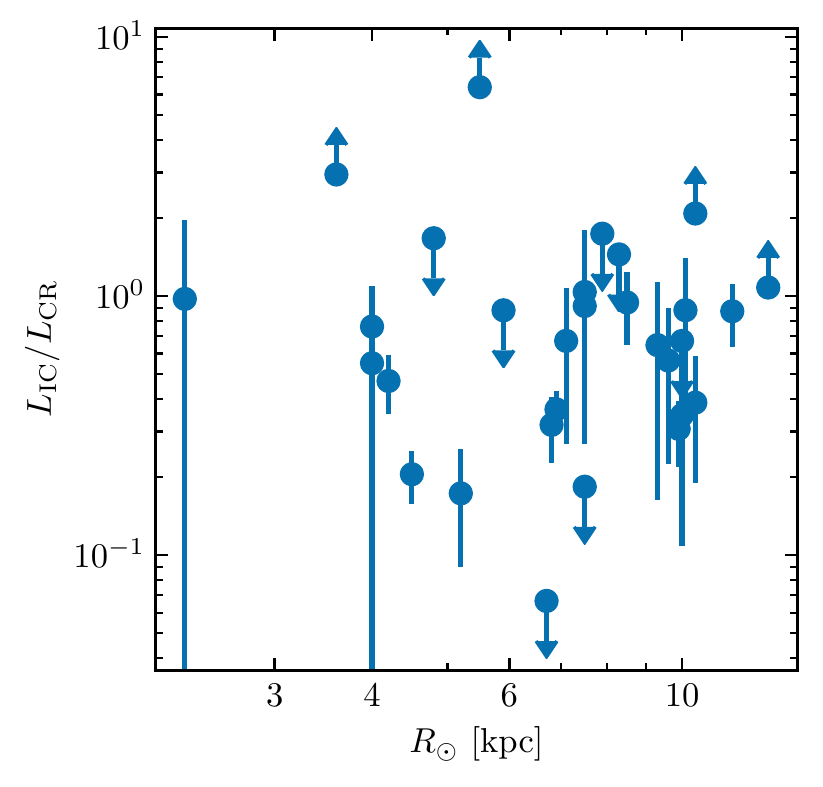}
    \caption{Distribution of the ratio $L_\mathrm{IC}/L_\mathrm{CR}$ compared with the total photon field energy density $u_\mathrm{Total}$ (left panel) and the distance from the Sun $R_\odot$ (right panel). The ratios are shown to be $\sim$ O(1) among GCs, with uncertainties presented. In both cases, the ratios show no obvious correlations with the other parameters.}
    \label{fig:fe}
\end{figure*}

Figure~\ref{fig:fe} plots the $L_\mathrm{IC}/L_\mathrm{CR}$ against the total radiation field $u_\mathrm{Total}$ of the GCs (left panel) and their distance $R_\odot$ from the Sun (right panel). It is shown that the ratios of $L_\mathrm{IC}/L_\mathrm{CR}$ are $\sim O(1)$. In both cases, the $L_\mathrm{IC}/L_\mathrm{CR}$ are scattered and show no obvious correlations with $u_\mathrm{Total}$ or $R_\odot$. This supports the idea that IC is the leading energy loss process for $e^\pm$ injected by MSPs diffusing in the GCs. The propagating $e^\pm$ eventually lose all their energy when interacting with the background photon field (IC) or magnetic field (synchrotron radiation). The $L_\mathrm{IC}$ would be suppressed for small $u_\mathrm{Total}$ when the synchrotron energy loss is comparable with the IC, which is not observed. 

For similar reason, we can rule out large uncertainties due to the point spread function effects of the \textit{Fermi}-LAT. In the present work, we are only considering the point-source luminosity of the GCs. For GCs closer to the Sun, the GC spatial extensions correspond to larger angular separations. The $e^\pm$ injected by MSPs could leak and carry away energy from the GCs. Should that happen, we may find a positive correlation between $L_\mathrm{IC}/L_\mathrm{CR}$ and $R_\odot$ since the apertures would only see partial IC from the $e^\pm$. Within uncertainties, this is not observed.

Given the above, it is a good approximation to assume that
\begin{equation}
    \dfrac{f_{e^\pm}}{f_\gamma} \simeq \dfrac{L_\mathrm{IC}}{L_\mathrm{CR}},
\end{equation}
as we have done in this study. 

\bsp	
\label{lastpage}
\end{document}